\documentclass[12pt]{article}

\usepackage{amsmath}
\usepackage{mathtools}
\usepackage[T1]{fontenc}
\usepackage{amssymb}
\usepackage{amsfonts}
\usepackage{amsthm}
\usepackage{mathrsfs}
\usepackage[all]{xy}
\usepackage{tensor}
\usepackage{stmaryrd}
\usepackage{bm}
\usepackage[labelfont={small, bf, sf}, textfont={small,sf}]{caption}

\usepackage[dvipsnames]{xcolor}
\definecolor{cadmiumred}{rgb}{0.89, 0.0, 0.13}
\definecolor{darkblue}{rgb}{0.2, 0, 0.8}
\definecolor{darkgreen}{rgb}{0.2, 0.71, 0}  
\definecolor{LinkGreen}{cmyk}{.9,0.05,.95,0.45}
\usepackage[colorlinks=true, 
citecolor=Blue,
linkcolor=LinkGreen,
urlcolor=Bittersweet]{hyperref}

\numberwithin{equation}{section} 

\usepackage{ragged2e} 

\usepackage[greek, english]{babel} 
\usepackage{teubner}  

\DeclareRobustCommand{\rchi}{{\mathpalette\irchi\relax}}
\newcommand{\irchi}[2]{\raisebox{\depth}{$#1\chi$}} 

\usepackage[vcentermath]{youngtab}

\usepackage{authblk}
\usepackage{fullpage}

\newcommand{\dd}{\mathrm{d}}
\newcommand{\lie}{\pounds}
\newcommand{\ep}{\epsilon}
\newcommand{\vep}{\varepsilon}

\newcommand{\beq}{\begin{equation}}
\newcommand{\eeq}{\end{equation}}

\newcommand{\Eth}{\text{\DH}}
\newcommand{\Ethb}{\text{\bf{\DH}}}
\DeclareMathOperator{\Dt}{D}
\DeclareMathOperator{\Db}{ \mathbf{D} }

\newcommand{\p}[1]{  {\bm{ #1} }   } 

\newcommand{\sg}{\varepsilon}

\newcommand{\un}[1]{ \underline{#1} }
\newcommand{\ind}[1]{\indices{#1}} 

\newcommand{\gim}{\gimel} 
\newcommand{\cn}{\gim} 

\newcommand{\out}{\mathcal{O}} 
\newcommand{\cat}{\mathcal{C}} 
\newcommand{\nout}{\mathcal{P}} 
\newcommand{\mix}{\mathcal{M}} 
\newcommand{\nix}{\mathcal{N}} 

\title{Geometrical tools for embedding fields, submanifolds, and foliations}
\author{Antony J. Speranza\thanks{asperanz@gmail.com}  }
\affil{\small \it 
Perimeter Institute for Theoretical Physics, 31 Caroline St. N, Waterloo, ON N2L 2Y5, Canada}
\date{}

\begin{document}

\maketitle

{\abstract
Embedding fields provide a way of coupling a background structure to
a theory while preserving diffeomorphism-invariance.  Examples of such background
structures include embedded submanifolds, such as branes; boundaries of local subregions,
such as the Ryu-Takayanagi surface in holography; and foliations, 
which appear in fluid dynamics and force-free electrodynamics.  
This work presents a systematic framework for computing geometric properties
of these background structures in the embedding field 
description.  An overview of the local geometric quantities associated
with a foliation is given, 
 including a review of  the Gauss, Codazzi, and Ricci-Voss
equations, which relate the geometry of the foliation to the ambient 
curvature.  Generalizations of these equations for
curvature in the nonintegrable normal directions are derived.  
Particular care is given to the question
of which objects are well-defined for single submanifolds, and which depend on the 
structure of the foliation away from a submanifold.  Variational formulas are provided for
the geometric quantities, which involve contributions both from the variation of the 
embedding map as well as variations of the ambient metric.  As an application of these
variational formulas, a derivation is given 
of the Jacobi equation, describing perturbations of extremal
area surfaces of arbitrary codimension.  The embedding field formalism is also 
applied to the problem of 
classifying boundary conditions for general relativity in a finite subregion that 
lead to integrable Hamiltonians.  
The framework developed in this paper will provide a useful set of tools for future
analyses of brane dynamics, fluid mechanics, and edge modes for finite subregions of 
diffeomorphism-invariant theories.  
}

\pagebreak

\renewcommand{\baselinestretch}{0.5}\normalsize
\tableofcontents
\renewcommand{\baselinestretch}{1.0}\normalsize

\section{Introduction}

A number of physical systems can be described in terms of a background
structure coupled to 
an otherwise generally covariant theory.  
One  example is a local subregion in a gravitational theory,
where the subregion boundary partially breaks the underlying 
diffeomorphism symmetry \cite{Donnelly2016F, Speranza2018}.  
Relativistic fluid dynamics provides another example, with the 
fluid rest frame determining the background structure
\cite{Friedman1978,Green2013,Andersson2007a,Dubovsky2012a}.  
Branes coupled to gravity  can  similarly
be viewed as a type of background structure \cite{Carter2001}, 
as can the preferred frame of Lorentz-violating
gravity theories \cite{Horava2009a,Blas2009a,Jacobson2001a,Jacobson2015c}.  

In each of these examples, the background structure introduces additional degrees of freedom
whose kinematics and dynamics require scrutiny.  Embedding fields 
provide a convenient formalism with which to perform this analysis.  An embedding
field is a diffeomorphism $X$ from a reference space $M_0$ into the 
dynamical spacetime manifold $M$.  
It provides a means for coupling fields living on the different spaces
by mapping them between the spaces via pullbacks. 
The background structure can  be defined as a fixed object on $M_0$, and then coupled 
to the dynamical fields on $M$ using the $X$ mapping.
Promoting $X$ to a dynamical object ensures that the coupling is covariant with respect
to diffeomorphisms of $M$.  
This restores  diffeomorphism symmetry to the theory,
and the $X$ field encapsulates the  additional degrees of freedom 
introduced by the background structure.

Fluid dynamics provides a canonical example of this procedure.  The reference space $M_0$ 
is the so-called Lagrangian frame, and the fixed structure is a congruence of curves describing
the flow of the fluid elements through time.  The $X$ field maps this congruence into the spacetime 
manifold $M$, which in this context is also referred to as the Eulerian frame.  $X$ contains
the information of how the fluid flow distorts as it evolves through spacetime, and hence
represents the fluid degrees of freedom.  This framework for describing relativistic fluids 
has found applications in forming covariant Lagrangians for fluid mechanics \cite{Friedman1978,
Andersson2007a}, describing the phase space for Einstein gravity coupled to a fluid
\cite{Iyer1997, Green2013}, and constructing effective field theories 
for fluid dynamics \cite{Dubovsky2012a, Haehl2018, Crossley2017a}.

In other applications, the background structure is a single embedded submanifold, rather
than a foliation.  Strings and branes are examples of this type of object, and the embedding fields 
define the map from the worldvolume into the target space $M$ \cite{Carter2001}.  
The boundary of a local
subregion is another example of a single-submanifold background
structure.  Although  such a boundary
should not define a dynamical object in the theory, being merely an arbitrary demarcation
for a subregion in a larger space, describing the location of the boundary
nevertheless introduces a background structure into the description.  The 
additional degrees of freedom that inevitably appear must be eliminated
during the process of gluing adjacent subregions to correctly reproduce the phase 
space of the larger region.  
This construction is particularly relevant when discussing entanglement entropy for 
a subregion in a gravitational theory, where the entangling surface serves as the 
subregion boundary.  The additional degrees of freedom represented by the embedding
fields contribute to the entropy of the subregion, and this edge mode entropy
is believed to be an important contribution to the entropy associated to horizons 
and black holes \cite{Donnelly2016F}.
A related construction by Isham and Kucha\v{r} 
in canonical general relativity extends the usual phase space 
by additional degrees of freedom representing the embedding of the Cauchy surface into
spacetime \cite{Isham1985, Isham1985a}.  Introducing the embedding fields in this 
context produces a constraint algebra that represents $\text{Diff}(M)$, as opposed 
to the field-dependent hypersurface deformation algebra that usually appears in 
the Hamiltonian analysis.

The entanglement wedge associated with  
a subregion of a holographic conformal field theory is another example where
embedding fields are used to describe a single submanifold.  The wedge is defined as 
the domain of dependence of a subregion bounded by a codimension-2 extremal surface
in the bulk.  This extremal surface is referred to as the RT surface,
in honor of the Ryu-Takayanagi formula, which expresses the 
entanglement entropy of the boundary CFT in terms of the area 
of this bulk surface \cite{Ryu:2006bv}.  In many applications, one is interested in the response of 
the RT surface to perturbations of the bulk geometry or the boundary subregion.  Examples 
include holographic proofs of the quantum null energy condition 
\cite{Koeller2016, 
Akers2017, 
Leichenauer2018},
discussions of bulk reconstruction from modular flow \cite{Lewkowycz2018a}
and derivations of the Einstein equation from entanglement beyond linear order 
\cite{Faulkner2017b}.  
Embedding fields again can be applied in these
situations, and give a way of cleanly separating perturbations induced by changing the geometry
from those coming purely from varying the embedding into spacetime.  

Embedding fields also find applications in situations where the background structure is a 
nondynamical metric.  One such application is the parameterized scalar field
\cite{Isham1985}, where the embedding fields impart diffeomorphism  invariance 
on an ordinary scalar field theory.  
Another is nonlinear massive gravity, where  a background metric appears
in the construction of the mass interactions for the massive spin two field \cite{DeRham2010a}.
Embedding fields are sometimes used as Stueckelberg fields to restore 
diffeomorphism invariance into the theory, which aids in establishing the absence of 
ghosts \cite{DeRham2011b}. 

Given the abundance of applications for embedding fields, it is desirable to have a 
systematic framework in which to perform calculations with them.  The present work seeks 
to develop this framework, placing specific emphasis on maintaining manifest covariance
in all expressions and derivations.  The embedding field $X$ is 
viewed as a specific type of dynamical field to be included in the theory, namely, a 
diffeomorphism from the reference space $M_0$ into spacetime $M$.  This 
means it has prescribed transformation properties under diffeomorphisms
of $M$.  
Section \ref{sec:overview} discusses these transformation properties of 
$X$, and describes generally how to couple $X$ to other dynamical fields
in a diffeomorphism-invariant way.   
Often embedding fields are viewed as a collection of scalar fields
representing the coordinates of the manifold, which particularly 
when performing concrete calculations.  In order to connect with this description, section 
\ref{sec:Xcoordexpr} discusses the coordinate expressions for the embedding fields.  

Following this, section \ref{sec:foliations} develops the local geometry associated with foliations
and submanifolds.  The fundamental object is the normal form to the submanifolds,
$\nu$, which is a differential form that annihilates all vectors tangent to the submanifolds.  
All geometric quantities associated with the submanifold and foliation
can be constructed in terms of $\nu$ and the metric $g_{ab}$. Section
\ref{sec:geom} performs these constructions explicitly, obtaining formulas for the normal
and tangential metrics as well as the extrinsic curvatures.  The normal and tangential covariant
derivatives are introduced in section \ref{sec:derivs}, and their associated curvatures are 
related to the spacetime curvature via the Gauss, Ricci-Voss, and Codazzi identities
in \ref{sec:gauss}.  In situations where one is only interested in a single submanifold, 
it is important to know which quantities constructed for a foliation
remain well-defined.  Section \ref{sec:invten} addresses this 
question, where it is shown that tensors associated with the intrinsic geometry such as the 
induced metric $h_{ab}$ and intrinsic curvature $\mathcal{R}\ind{^a_b_c_d}$ are invariants 
of a single submanifold, as are the extrinsic curvature tensor $K\ind{^a_b_c}$ and 
outer curvature $\out\ind{^a_b_c_d}$ (defined in (\ref{eqn:outabcd})), but notably
the normal extrinsic curvature $L\ind{_a_b^c}$ (defined in (\ref{eqn:Labc}))
is not invariant.  The transformation
properties of $L\ind{_a_b^c}$ under a change in foliation are derived in that section,
and its relation to the existence of a normal coordinate system near the submanifold 
is discussed.  

Section \ref{sec:variations} then combines the embedding fields with the submanifold calculus of 
section \ref{sec:foliations} in order to derive variational formulas 
for the foliation geometry.  Explicitly including the embedding
fields into the description allows one to separate out the variations induced by 
changing the metric from variations coming solely from a change in embedding.  
As a particular application,  section \ref{sec:Jacobi} uses the variational formulas for 
the mean extrinsic curvature $\p K_a$ to derive the Jacobi equation, which describes
perturbations of an extremal surface to a nearby extremal surface.  This equation
is of particular importance to holography and 
RT surfaces, and the geometric description provided in this section
will hopefully be relevant to future holographic calculations.  

The constructions of sections \ref{sec:foliations} and \ref{sec:variations} are performed for
foliations of arbitrary codimension.  To connect with more common constructions, 
section \ref{sec:special} analyzes the specific cases 
of hypersurfaces (codimension 1), congruences of curves (codimension $(d-1)$), 
and codimension-2 surfaces.  
In each of these cases, certain simplifications occur, and connections are 
made between the usual tensors defined for these foliations and the tensors associated
with the generic case.  Following this, section \ref{sec:hamiltonian}
gives an application of the formalism by discussing the boundary 
term than appears in the gravitational Hamiltonian for a local subregion.  
Section \ref{sec:conclusion} concludes with a discussion of possible generalizations
of the formalism to the case of null submanifolds,  and  points to a number of 
future applications.  

Finally, two appendices are included with additional geometric identities relevant
to the submanifold calculus of the paper.  Appendix  \ref{app:curvature}  generalizes the 
Gauss-Ricci-Voss-Codazzi identities associated with the tangential covariant derivative
$\Dt_a$ by deriving analogous statements for the normal covariant derivative $\Eth_a$, 
as well as identities associated with a mixed commutator between these two
derivatives.  Some of these identities have appeared before in the special case of 
foliations by one-dimensional curves 
\cite{C-G1961, C-G1963, 
Massa1974, Boersma1995c}; 
and general treatments of arbitrary codimension foliations has appeared before
in \cite[Ch. V, Sec. 7]{Schouten1954}, 
although appendix \ref{app:curvature} casts these identities in a modern
light and derives additional relations from them.  
Appendix \ref{app:coordexpr} gives numerous coordinate expressions for the 
geometric quantities associated with a foliation, which can be helpful in
concrete computations.  This appendix can be viewed as a generalization
of the  $3+1$ decomposition of spacetime to a $(d-p)+p$ decomposition for aribitrary codimension
foliations.  

The submanifold calculus developed in this work has been considered before by several
authors, and it is worth pointing out the similarities and differences between this paper
and previous work.  The work of Carter \cite{Carter1992, Carter2001} provides much of the 
basis for the formalism developed here.  A notable difference is that Carter was primarily concerned
with single submanifolds, as are applicable to the study of branes, and hence employs 
a formalism that does not require an extension of the submanifold to a local foliation.  
The advantage of working with a local foliation is that it canonically determines a
connection on the normal bundle to the surface through the normal extrinsic curvature 
tensor $L\ind{_a_b^c}$ (see section \ref{sec:derivs}).   As discussed in section \ref{sec:invten},
$L\ind{_a_b^c}$ is not an invariant tensor on the surface, since it depends on 
how the foliation is extended away from the surface.  This is just the statement that there
is no preferred connection on the normal bundle in the absence of a local foliation.  
In Carter's work, the analog of the arbitrariness in the foliation is the choice of an 
orthonormal basis on the normal bundle, needed 
in order to define the twist pseudotensor. 
Section \ref{sec:invten} shows that this twist pseudotensor coincides with the twist 
$\bar{F}\ind{_a_b^c}$ of a foliation whose normal directions are geodesic.
For this reason,  the acceleration tensor $A\ind{_a_b^c}$ never appears in Carter's formalism.  
Capovilla and Guven \cite{Capovilla1995, Capovilla1995a} and Armas and Tarr\'io \cite{Armas2017}
similarly employ an orthonormal basis to describe the normal geometry, and 
compute many of the variational formulas appearing in section \ref{sec:variations} 
of the present work.  The main difference between the present paper and the above 
is favoring a choice of local foliation over a choice of orthonormal normal basis,
since the goal is to have a formalism that applies both for single submanifolds and 
for  foliations.  
This is more in line with related treatments of submanifold variations employed
by Feng and Matzner \cite{Feng2017}, and Plebanski and Rozga \cite{Plebanski1989}.
A notable cosmetic difference between this paper and others is the treatment of the 
embedding field $X$ abstractly as a diffeomorphism, as opposed to using its coordinate
expression $\p X^\mu$, discussed in section \ref{sec:Xcoordexpr}.  This has some advantages in
leading to clean variational expressions in terms of the variational vector field $\rchi^a$
that appears in the pullback formula (\ref{eqn:pullbackX}), 
and we stress that the abstract calculus  provides
concrete computational tools through repeated use of this formula.  
Finally, Engelhardt and Fischetti's recent work \cite{Engelhardt2019}, which appeared concurrently
with the present one, presents a comprehensive, covariant treatment of submanifolds 
using distributional tensor fields, which do not require an extension to a local foliation.  
Their work therefore gives a complementary perspective on some of the results 
obtained in this paper.

\subsection{Notation}
Throughout this paper, Latin indices from the beginning of the alphabet, $a, b, c,\ldots $ are 
considered abstract indices \cite[Section~2.4]{Wald1984}, while Greek indices $\mu, \nu, \ldots$ are 
used as coordinate indices on $M$.  
Since tensor fields on both $M_0$ and on $M$ appear in the calculations below, it is helpful
to distinguish between them.  This is done using 
boldface font, $\p g_{ab}$, $\p K\ind{^a_b_c}$,
for tensors on the reference space $M_0$, and regular font, $g_{ab}$, $K\ind{^a_b_c}$
 for tensors on $M$.  The same letter is used for a bolded and unbolded tensor
when the two are related by a pullback by the embedding field, so that in general,
$
\p \psi = X^*\psi.
$
Finally, many expressions involve antisymmetrization or symmetrization over multiple indices
that are not adjacent.  Rather than using brackets and parentheses, which become cumbersome 
in these cases, we instead underline indices to denote antisymmetrization, and overline indices
for symmetrization, i.e.\ $T_{\un{ab}} \equiv T_{[ab]} = \frac12(T_{ab} - T_{ba})$ and 
$T_{\overline{ab}}\equiv T_{(ab)} = \frac12(T_{ab}+ T_{ba})$. The spacetime metric $g_{ab}$
is always assumed to have signature $(-,+,+,\ldots)$. 
All tensors are presented as tensors on spacetime, and as such, the same types of letters 
are used to describe both normal and tangential indices, in line with the philosophy of 
\cite{Carter2001}.  This has a slight disadvantage of 
requiring the reader to remember which indices are normal and tangential (i.e.\ $K\ind{^a_b_c}$
is normal on $a$ and tangential on $b$ and $c$), but has the advantage of 
making clear how objects such as the spacetime covariant derivative $\nabla_a$ act
on these tensors.

\section{Overview of embedding fields} \label{sec:overview}

This section develops the basic tools for computing with embedding fields.  
The starting point is a 
collection of dynamical fields, collectively denoted $\phi$, 
which are tensor fields on a spacetime manifold $M$.  
For this paper, the sole dynamical field will be the metric $g_{ab}$, but the basic operations
for coupling a generic field to the embedding fields is the same for fields  of any type.  
The diffeomorphism group $\text{Diff}(M)$ acts on these fields via pullbacks, sending
$\phi \mapsto Y^*\phi$ for $Y\in \text{Diff}(M)$.  
A theory in which all tensor fields needed to construct a covariant 
action are dynamical is diffeomorphism-invariant.  An example of a non-invariant theory 
is a massless scalar, since its Lagrangian density $\mathcal{L} = \eta^{ab}\nabla_a \varphi
\nabla_b \varphi$ involves a fixed metric $\eta^{ab}$ which does not transform under 
diffeomorphisms.  This metric therefore constitutes the background structure
of the theory that prevents diffeomorphism-invariance.  Similarly, a theory with 
an invariant Lagrangian defined in a finite subregion is still not fully diffeomorphism-invariant,
since the normal form to the boundary of the subregion is again a fixed structure
which does not transform.

\begin{figure}
\centering
\includegraphics[width=0.6\textwidth]{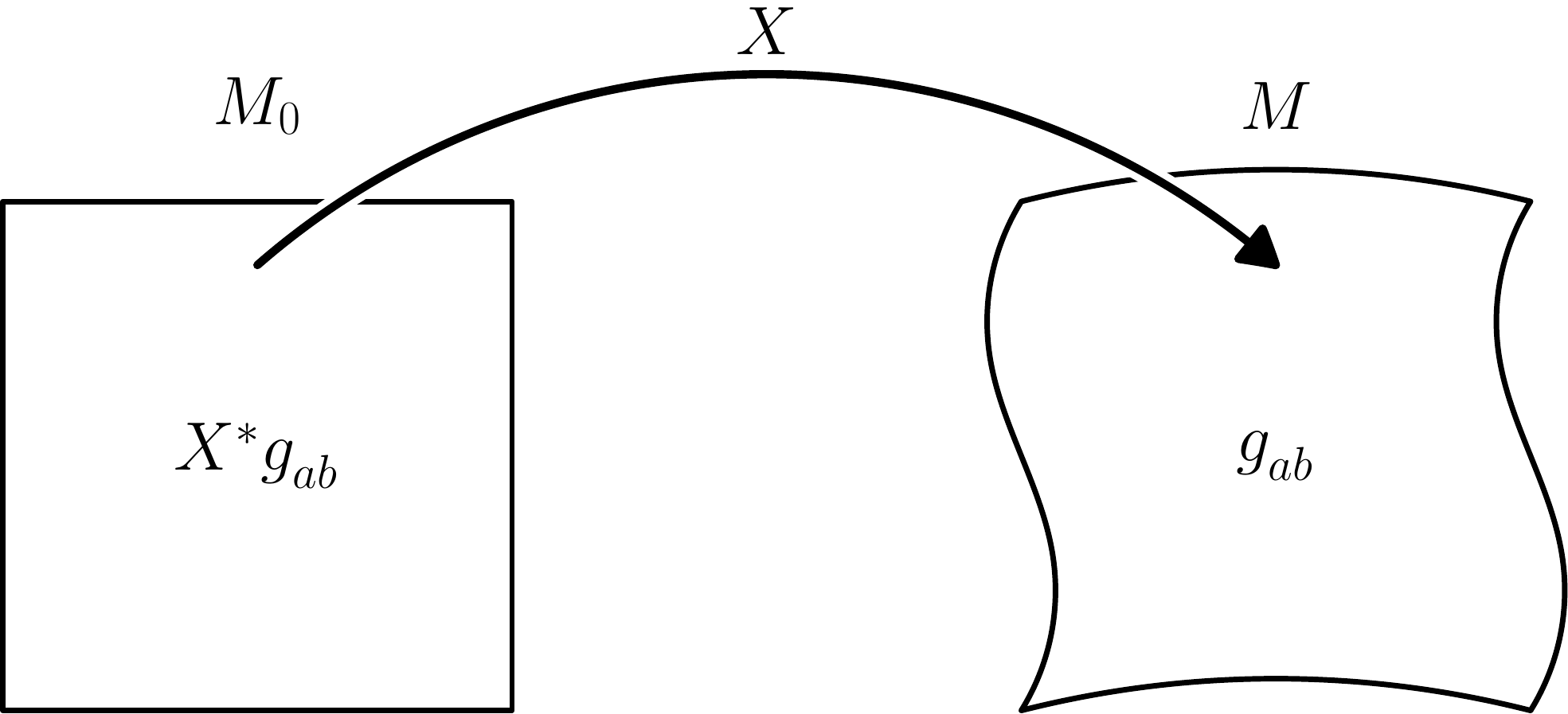}
\caption{ The embedding field $X$ is a map between a reference space $M_0$ and 
the spacetime manifold $M$.  The metric $g_{ab}$ and other dynamical fields live 
on $M$, and can be mapped to pulled-back fields $X^*g_{ab}$ on the reference 
space using the embedding map.  \label{fig:emb}}
\end{figure}

Embedding fields provide a way of modifying  theories with background structures
to make them invariant.  The embedding field $X$ is a diffeomorphism from a reference space
$M_0$ into $M$ (Figure \ref{fig:emb}).  
This reference space can simply be viewed as a second copy of the spacetime
manifold, although for calculations involving coordinate expressions
(section \ref{sec:Xcoordexpr}), it is often useful
to take it to be an open subset of $\mathbb{R}^d$. The embedding 
field transforms under diffeomorphisms of $M$ by the pullback
by $Y^{-1}$,
\beq \label{eqn:YinvX}
X \mapsto (Y^{-1})^* X = Y^{-1}\circ X.
\eeq
The natural way to couple the embedding field to the other dynamical fields is simply to form
the pulled back fields, $\p\phi\equiv X^*\phi$.  Throughout this work, bold
font will indicate tensors that live on the reference space $M_0$, and are related to 
their non-bold counterparts through a pullback by $X$.  
The fields $\p\phi$ have the property of being invariant under diffeomorphisms
of $M$, since 
\beq
(Y^{-1}\circ X)^* Y^*\phi = X^*(Y^{-1})^* Y^* \phi = X^*\phi.
\eeq
This invariance is intuitive from the perspective that $\p\phi$ live on a different manifold, $M_0$,
and hence should not transform under diffeomorphisms of $M$. 

A central goal of this work is to compute variations of objects constructed using $X$ and 
the dynamical fields.  For this, the following formula is of central importance,
\beq\label{eqn:pullbackX}
\delta X^*\phi = X^*(\delta\phi +\lie_{\rchi}\phi),
\eeq
where $\rchi^a$ is a vector field representing an arbitrary infinitesimal variation of 
the embedding map $X$ (see \cite{Donnelly2016F, Speranza2018} for derivations of this 
formula).  The coordinate expression for $\rchi^a$ is given in equation (\ref{eqn:chicoord}).  
A related formula involves mapping a field on $M_0$ to a field on $M$ with the inverse
pullback $(X^{-1})^*\equiv X_*$, and reads
\beq \label{eqn:delXpush}
\delta X_*\p\phi = X_*\delta \p \phi - \lie_{\rchi} \phi.
\eeq
 
Any expression involving $\delta$ should be taken to mean an arbitrary variation
of that quantity.  To denote a particular choice for the infinitesimal variation, we will
write $I_{\hat \Phi}$ acting on the expression, where $\hat \Phi$ then contains 
the information of which variation is being considered.  For example, we write 
$I_{\hat \Phi} \delta \phi = \Phi$ to denote the infinitesimal variation of the field $\phi$ given
by the specific field configuration $\Phi$.  This notation stems from viewing 
$\delta$ as the exterior derivative on the infinite-dimensional manifold of field 
configurations, in which case $I_{\hat \Phi}$ denotes contraction of a differential form with the 
vector field $\hat \Phi$ describing an infinitesimal change on this manifold \cite{Donnelly2016F, 
Speranza2018}.  

A particularly important class of variations are those induced by an infinitesimal
diffeomorphism of $M$ generated by a vector field $\xi^a$.  The contraction representing 
this transformation is denoted $I_{\hat\xi}$, so that $I_{\hat\xi}\delta\phi
= \lie_\xi\phi$, since the Lie derivative gives the transformation of a tensor field under
an infinitesimal diffeomorphism.  Note that since the pulled back fields $X^*\phi$ 
are diffeomorphism-invariant, they must satisfy $I_{\hat\xi} \delta X^*\phi = 0$, 
which implies via equation (\ref{eqn:pullbackX}) 
\beq\label{eqn:Ixichi}
I_{\hat{\xi}}\rchi^a = -\xi^a.  
\eeq
We can also consider diffeomorphisms of the reference space generated by a vector
$\p \xi^a$, and this transformation on the fields will be represented by $I_{\hat{\p\xi}}$. 
These diffeomorphisms of $M_0$ should not transform the dynamical fields which
live on $M$, which implies the relations for this contraction
\beq
I_{\hat{\p\xi}} \delta\phi = 0, \quad I_{\hat{\p\xi}} \rchi^a = X_* \p \xi^a = \xi^a.
\eeq

Occasionally we will also be interested in second variations of quantities.  
In these cases, when $\delta$ acts on a tensor that already involves a variation, we  
take it to mean the antisymmetric combination of the two variations.  This is consistent
with the interpretation of $\delta$ as the exterior derivative on the space of field
configurations.  In particular, it implies $\delta\delta \phi = 0$.
Also, any expression involving the product of two variational tensors
will similarly always be assumed to be antisymmetrized in the variations, so that 
$\delta\alpha \,\delta\beta = - \delta\beta\,\delta\alpha$.  With this interpretation, we note
 the useful formula for the variation of $\rchi^a$, which does not vanish, but rather
 satisfies \cite{Speranza2018}
\beq\label{eqn:delchi}
\delta \rchi^a = \delta X_*\p\rchi^a = X_*\left(\delta\p\rchi^a - \lie_{\p\rchi}\p\rchi^a\right)
 = -\frac12[\rchi,\rchi]^a.
\eeq

\subsection{Coordinate expressions} \label{sec:Xcoordexpr}

While the abstract definitions of the $X$ field and the vector $\rchi^a$ associated with its variation
suffice for formal manipulations, it is also useful to have coordinate expressions for these 
objects when performing direct calculations.  To obtain these,  introduce a coordinate 
system on $M$, which is defined by a collection of functions $y^\mu:M\rightarrow \mathbb{R}$, 
$\mu = 0,\ldots,d-1$, mapping each point in $x\in M$ to its coordinate values 
$y^\mu(x)$.\footnote{Since the following analysis deals only with local quantities,
it suffices to 
restrict attention to a single coordinate patch. }  The coordinate expression for the $X$ field
is simply obtained by pulling the coordinate functions back by $X$.  Taking $\mathbb{R}^d$
for the reference space $M_0$, this  leads to a 
collection of $d$ scalar functions $\p X^\mu:\mathbb{R}^d\rightarrow\mathbb{R}$, defined by
\beq
\p X^\mu = X^*y^\mu  = y^\mu \circ X.
\eeq
The functions $\p X^\mu$ then define a coordinate system on $\mathbb{R}^d$.
Promoting $X$ to a dynamical field in the theory therefore has the interpretation of giving dynamics 
to the coordinate system itself.  A consequence of $\p X^\mu$ defining a coordinate system is that
the gradients $\p \nabla_a \p X^\mu$ yield a basis for one-forms on $\mathbb{R}^d$.  This in turn
defines a basis $\p \partial_\mu^a$ for tangent vectors on $\mathbb{R}^d$ satisfying 
\begin{align}
\p \nabla_a \p X^\nu \p \partial_\mu^a  &= \delta^\nu_\mu \label{eqn:coordid}\\
\p \nabla_a \p X^\mu\p \partial_\mu^b &= \p \delta^b_a \label{eqn:id} \\
[\p \partial_\mu, \p \partial_\nu] &= 0.
\end{align}
These basis vectors are just the pullbacks of the coordinate basis vectors 
$\partial_\mu^a$ on $M$.  

The variations of the functions $\p X^\mu$ again define a collection of $d$ functions 
$\delta \p X^\mu :
\mathbb{R}^d\rightarrow \mathbb{R}$ interpretable as an infinitesimal change of coordinates.  
From the pullback formula (\ref{eqn:pullbackX}) and the requirement
that $\delta y^\mu=0$, this variation is given by
\beq
\delta\p X^\mu = X^*\lie_\rchi y^\mu = \p\rchi^a \p\nabla_a \p X^\mu = \p\rchi^\mu,
\eeq
so that $\delta\p X^\mu$ are simply the components of $\p\rchi^a$ in the coordinate system
on $\mathbb{R}^d$ defined by $\p X^\mu$,
\beq
\p \rchi^a = \delta \p X^\mu \p \partial_\mu^a.
\eeq 
This leads to the coordinate expression for 
$\rchi^a$,
\beq \label{eqn:chicoord}
\rchi^a = (X_* \delta \p X^\mu) \partial_\mu^a = (\delta \p X^\mu \circ X^{-1}) \partial_\mu^a,
\eeq
which was presented in \cite{Donnelly2016F}.

Because $\p \partial_\mu^a$ depends on $\p X^\mu$, the antisymmetric 
variation of 
the vector field $\p \rchi^a$ does not vanish; rather, it is given by
\begin{align}
\delta \p \rchi^a &= - \delta \p X^\mu \delta (\p\partial_\mu^a)  \nonumber\\
&= - \delta \p X^\mu \p \partial_\mu^b\, (\p \nabla_b \p X^\nu)
\delta(\p \partial_\nu^a) \nonumber \\
&=\p \rchi^b \p \nabla_b (\delta \p X^\nu) \p \partial_\nu^a \nonumber \\
&= \p \rchi^b (\nabla_b \p\rchi^a - \delta \p X^\nu \p\nabla_b\p\partial_\nu^a) \nonumber \\
&= \frac12[\p \rchi, \p \rchi]^a.
\end{align}
The second line follows after applying equation (\ref{eqn:coordid}), while the third line 
uses the identity $\p \nabla_a(\delta \p X^\mu) \p \partial_\mu^a
=-\p \nabla_a \p X^\mu \delta(\p \partial_\mu^a)$, obtained by acting with $\delta$ on 
equation (\ref{eqn:id}).  Finally, the fifth line uses that $\p \rchi^b\p\nabla_b\p\rchi^a 
= \frac12[\p \rchi, 
\p \rchi]^a$ due to the implied antisymmetrization 
between the two $\p\rchi^a$'s in this expression,
and also that  $\p \nabla_b$ can be taken 
to be the derivative operator associated with the $\p X^\mu$
coordinate 
system,
which annihilates the vectors $\p\partial_\nu^a$.\footnote{With 
any other derivative operator, one finds that the connection coefficients drop out
of this expression, so this choice is not  restrictive.}
 From this, we can also compute the variation
of the vector field $\rchi^a$ defined on $M$,
\beq
\delta \rchi^a = \delta X_* \p\rchi^a = X_*\delta \p \rchi^a - \lie_\rchi X_* \p\rchi^a 
= \frac12[\rchi,\rchi]^a - [\rchi,\rchi]^a = -\frac12[\rchi, \rchi]^a,
\eeq
giving a derivation of equation (\ref{eqn:delchi}).   

Finally, although this paper refrains from employing explicit coordinate expressions
for the pullback of the metric $\p g_{ab}=X^*g_{ab}$, it can be useful to have them on hand.  Letting 
$g_{\mu\nu} = g_{ab}\partial_\mu^a \partial_\nu^b$ denote the components of the metric
in the $y^\mu$ coordinate system, and $\p g_{\alpha\beta}$ the components of $\p g_{ab}$
in a coordinate system $x^\alpha$ on the reference space, they are related via
\beq
\p g_{\alpha\beta}(x) = 
\frac{\partial \p X^\mu}{\partial x^\alpha} \frac{\partial \p X^\nu}{\partial x^\beta}
 g_{\mu\nu}\big(X(x)\big).
\eeq

\section{Foliations by submanifolds} \label{sec:foliations}

The main application for
the embedding fields in this paper is in describing 
embedded submanifolds and foliations.  
This section discusses the basic construction of a foliation, and derives the local
geometric data associated
with it.  
The idea  is to fix a specific submanifold
or foliation
in the reference space $M_0$, and then use $X$ to map it into spacetime.  
Variations of geometric quantities under a change in the foliation are then parameterized by
variations of $X$. In addition, this setup cleanly separates the effects
 of varying the 
spacetime metric from those coming from varying the embedding.

A common way to define a submanifold of 
codimension $p$ on the reference space is to give $p$ functions $\p F^A(x)$, $A=1,\ldots, p$, 
which simultaneously vanish on the submanifold.\footnote{By assuming
that these functions are defined everywhere in a neighborhood of the submanifold, we are 
implicitly imposing that the normal bundle of the submanifold is trivial.  
This is not overly restrictive for many applicaitons; however, for the most
general class of submanifolds and foliations, the existence of such functions can only 
be established locally.  A simple example where global existence  fails is a circle
embedded in a Mobius strip, since the normal form would return to its negative after
going around the circle. } 
These functions also define a foliation 
via their simultaneous level sets.  One disadvantage of this description is that it is highly redundant:
any reparameterization of the form $\p G^A(x) = \p G^A(\p F^B(x))$ defines an equivalent foliation.
A way to avoid this redundancy is to instead work with  a differential $p$-form $\p{\nu}$, 
which vanishes when restricted to the submanifolds, i.e.\ the tangent vectors $\p{v}^a$
to the submanifolds
are precisely the ones which annihilate the $p$-form, $i_{\p{v}} \p{\nu}=0$.  
One can express $\p{\nu}$ in terms of the functions $\p F^A$ simply via
\beq
\p{\nu} = \dd \p F^1\wedge\ldots\wedge \dd \p F^p
\eeq
Under the reparameterization $\p G^A = \p G^A(\p F^B)$, $\p{\nu}$  changes by an overall
rescaling from the Jacobian.  Hence, $\p{\nu}$ and $\p f(x) \p{\nu}$ define equivalent foliations for
any positive function $f(x)$.  Working directly with $\p{\nu}$  reduces the redundancy 
of the description from arbitrary reparameterizations of $\p F^A$ to a single overall rescaling 
ambiguity in $\p{\nu}$.

\begin{figure}
\centering
\includegraphics[width=0.5\textwidth]{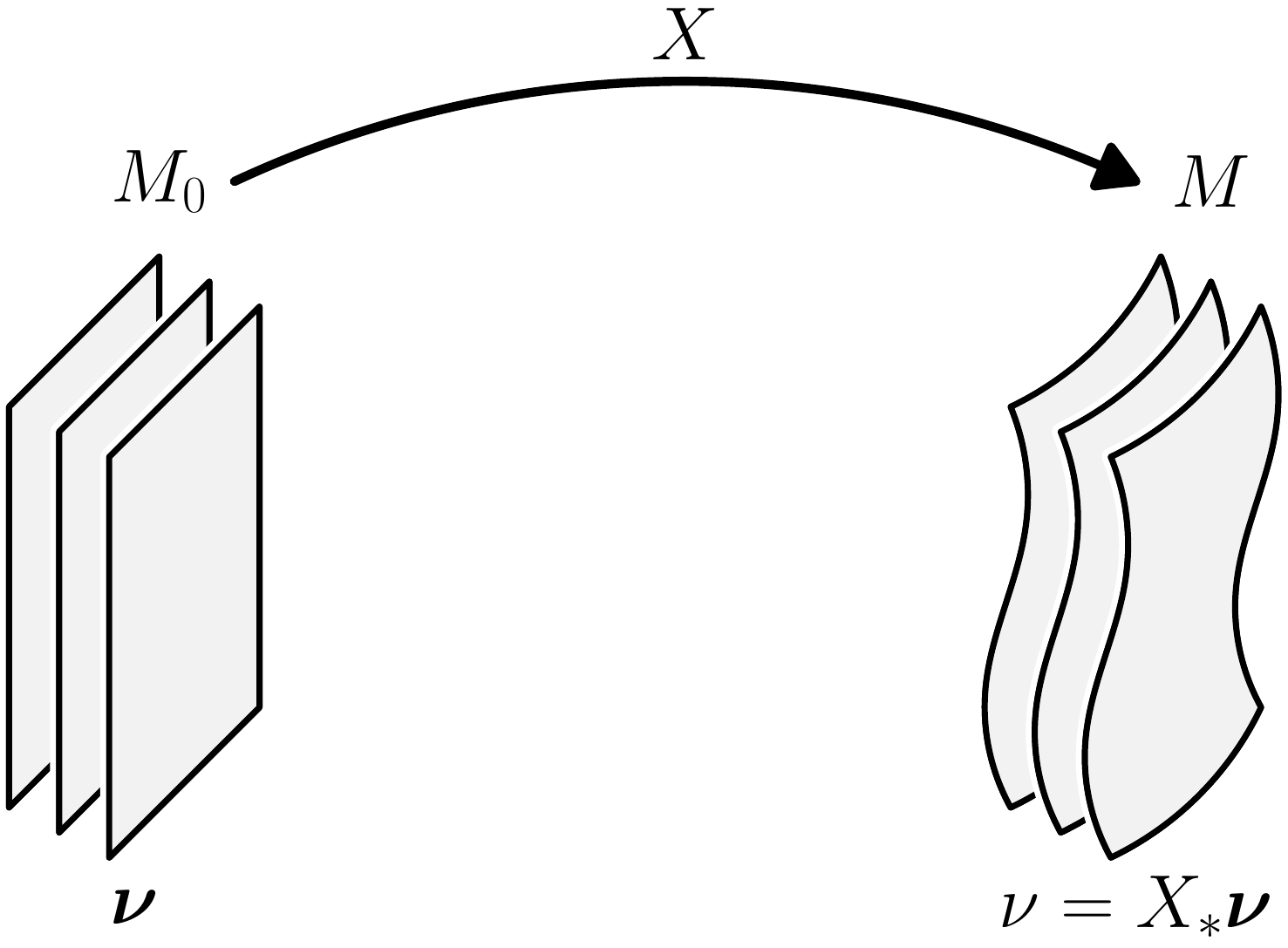}
\caption{\label{fig:foliation} The fixed foliation on $M_0$ is specified through
the normal form $\p\nu$ by requiring all tangent vectors to the foliation
annihilate it.  $X$ maps $\p\nu$ to the form $\nu = X_* \p \nu$,
which defines a foliation in $M$.  The $M$  foliation can be varied by changing 
the embedding map $X$.  }
\end{figure}

Going forward, we will forget entirely about the functions $\p F^A$ and instead work only with 
$\p{\nu}$.  It is important to note at this point that an arbitrary $p$-form is not suitable for 
defining the submanifolds; rather, it must satisfy two additional conditions.  The 
first condition is that the rank of $\p{\nu}$ must equal $p$, which means there exists
a set of one-forms $\p{e}^A$ with $A$ ranging from $1$ to $p$,
 in terms of which $\p{\nu}$ can be expressed as 
\beq \label{eqn:nusimpl}
\p{\nu} = \p \nu_{A_1\ldots A_p} \p{e}^{A_1}\wedge\ldots\wedge \p{e}^{A_p}.
\eeq
A generic $p$-form requires more than $p$ basis one-forms to express it, so this requirement
is that $\p{\nu}$ has the minimal possible rank, i.e.\ it is a simple form.   
This restriction
is an algebraic condition on $\p{\nu}$ that must hold pointwise.  
The reason for it is to ensure the dimension of the submanifold is precisely
$(d-p)$. If $\p{\nu}$ had rank greater than $p$, fewer vectors would annihilate it, 
and it would therefore define lower dimensional submanifolds.
The second condition is an integrability condition that ensures  that the vectors 
annihilating $\p{\nu}$ are tangent to submanifolds.  In terms of the vector fields, this means
that $[\p v,\p w]^a$ must annihilate $\p{\nu}$ whenever $\p v^a$ and $\p w^a$ do.  
The condition this imposes on $\p \nu$ is 
\beq \label{eqn:nuinteg}
\dd\p\nu = \p\rho \wedge\p\nu
\eeq
for some one-form $\p\rho$ \cite[Section~IV.C]{3-ladies}.  
Occasionally, the stronger condition $\dd\p\nu = 0$ is imposed, especially 
when $\p\nu$ has additional physical meaning, such as a conserved current.  
An example of this is ideal fluid mechanics, where $\p\nu$ can represent a 
conserved particle number or entropy current \cite{Andersson2007a}.  

Once $\p \nu$ satisfying the simplicity and integrability conditions has been chosen, it 
defines a fixed family of submanifolds in the reference space.  
It can be mapped to spacetime using the embedding fields,
\beq
\nu = X_* \p\nu,
\eeq
and $\nu$ 
defines a foliation in spacetime, since the pushforward preserves the conditions imposed on 
$\p \nu$ (see Figure \ref{fig:foliation}).  
Finally, we can require that $\p\nu$ is held fixed under all variations of dynamical
fields, $\delta\p\nu=0$, which fixes the variation of $\nu$ to be that induced by the variation
of the embedding fields $X$, 
\beq \label{eqn:delnu}
\delta \nu = - \lie_\rchi \nu.
\eeq
using equation (\ref{eqn:delXpush}).  
Note that this formula ensures that $\nu$ transforms as an ordinary tensor field under spacetime
diffeomorphisms, meaning that 
\beq \label{eqn:Ixidelnu}
I_{\hat\xi}\delta \nu = \lie_\xi\nu,
\eeq
by equation (\ref{eqn:Ixichi}), which guarantees that any tensor constructed from
$\nu$ and the dynamical fields will also have covariant transformation properties.  
Variations of objects constructed from $\nu$ are 
treated in detail in section \ref{sec:variations}.

\subsection{Intrinsic and extrinsic geometry} \label{sec:geom}
In some applications, such as descriptions of fluids, the unnormalized form $\p\nu$ is treated as 
a physical quantity that can be used to build actions and describe dynamics \cite{Andersson2007a,
Dubovsky2012a}.  
However, geometric invariants of the foliation should not involve $\p\nu$ 
directly, since it contains additional information beyond the embedding, due 
to the rescaling ambiguity.  Instead, the appropriate object to work with is 
the normalized unit normal.  Normalization of $\p\nu$ requires a metric, which is obtained from the 
dynamical metric on spacetime by an $X$ pullback,
\beq
\p g_{ab} = X^* g_{ab}.
\eeq
Alternatively, we can map the normal form to spacetime, giving $\nu = X_* \p \nu$, 
which then defines the foliation on $M$.  The various quantities constructed below from $\nu$ 
are  mapped to their counterparts on the reference space $M_0$ via $X^*$.  Hence, 
the formulas derived in this section are valid for 
both bolded and unbolded  
tensors.  

The unit normal $p$-form is defined as 
\beq \label{eqn:nNnu}
n_{a_1\ldots a_p} =  N  \nu_{a_1 \ldots a_{p}}, 
\eeq
which must satisfy the normalization condition
\beq \label{eqn:normnorm}
 n_{a_1\ldots a_p}  n^{a_1 \ldots a_p} = \sg\, p!,
\eeq
where $\sg = \pm 1$ depending on whether the foliation is spacelike $(-1)$ or 
timelike $(+1)$.\footnote{Null surfaces are excluded from this analysis.  The inability
to normalize the normal form is one of several reasons the 
geometric quantities considered in this section cannot be defined for a null surface.
A lengthier discussion of how one might adapt the formalism to null surfaces 
is given in section \ref{sec:null}. }  
$ N$ must therefore be given by
\beq \label{eqn:N}
N = \sqrt{p!} \left(\sg\, g^{a_1 b_1} \ldots  g^{a_p b_p} 
 \nu_{a_1 \ldots a_p}  \nu_{b_1 \ldots b_p}\right)^{-1/2}.
\eeq

The normal metric $ s_{ab}$  is constructed from the normal form as
\beq \label{eqn:sabdef}
 s_{ab} = \frac{\sg}{(p-1)!}  
 n_{a\, a_2\ldots a_p}  n_{b\, b_2 \ldots b_p}  g^{a_2 b_2} \ldots  g^{a_p b_p}.
\eeq
The normalization of this tensor comes from requiring that $ s\indices{^a_b}$ be a normal 
projector.  
One can verify using (\ref{eqn:sabdef}) that if $ \alpha_a$ is a normal $1$-form,
then 
\beq
 s\ind{^a_b}  \alpha\ind{_a} = \sg (-1)^{p-1} (\star \star \alpha)_b =   \alpha_b,
\eeq
where $ \star$ is the Hodge dual on the normal space. 
This is the point where the simplicity  of $\nu$ is used: 
the normalization condition (\ref{eqn:normnorm}) and the 
projector requirement  $s\ind{^a_b}
s\ind{^b_c} = s\ind{^a_c}$ cannot both hold unless $\nu$ is simple.  
More generally, one can form the projector
onto the space of normal $q$-forms by contracting $(p-q)$ indices of a pair of unit normals,
\beq \label{eqn:qformproj}
 s\ind{^{a_1\ldots a_q}_{b_1\ldots b_q} } =  s\ind{^{{a_1}} _{\un{b_1} }}  
\ldots  s\ind{^{{a_q}} _{ \un{b_q}}} = \frac{\sg}{q!(p-q)!}  n^{a_1\ldots a_q e_{q+1}\ldots e_p}
 n_{b_1\ldots b_q e_{q+1}\ldots e_p},
\eeq
and its status as a projector also follows by expressing its action on a $q$-form in terms of Hodge
duals.  

 The tangential metric $ h_{ab}$ 
is then 
constructed via
\beq
 h_{ab} =  g_{ab} -  s_{ab},
\eeq
and the tangential projector $ h \indices{^a_b}$ is obtained by simply raising the index with the 
metric.  When pulled back to the  submanifolds, 
$ h_{ab}$ defines the induced metric that determines the intrinsic geometry.  The 
volume forms $ \mu_{b_1\ldots b_{d-p}}$ for the surfaces are obtained by contracting 
the spacetime volume form $\ep_{a_1\ldots a_d}$ with the 
unit normal form, giving the equation
\beq \label{eqn:mu}
 \mu_{b_1\ldots b_{d-p}}  = 
-\frac{\vep}{p!}  n^{a_1\ldots a_p}\ep_{a_1\ldots a_p b_1\ldots b_{d-p}}.
\eeq  
This ensures that the normal form and induced volume form combine into the full spacetime
volume form through the equation\footnote{The minus depends on the choice of 
orientation for both spacetime and the submanifolds.  The choice made here
is most convenient for spacelike submanifolds with a future-pointing 
timelike normal.} 
\beq
 \ep = - n\wedge \mu.
\eeq

Additional geometric information is contained in the covariant derivative of $ s\indices{^a_b}$.  
It is straightforward to see that 
$ \nabla_a  s\indices{^b_c}$ vanishes when
$b$ and $c$ are both projected in the normal direction, or both tangentially.  Of the remaining
components, the ones with $a$ and $c$ tangential (or equivalently, $a$ and $b$) 
encode the extrinsic geometry of the embedding through the extrinsic curvature tensor,
defined as
\beq \label{eqn:Kabc}
 K\indices{^a_b_c} =  h\indices{^d_b}  h\indices{^e_c}\nabla_d  s\indices{^a_e} = 
- h\indices{^d_b}  h\indices{^e_c} \nabla_d  h\indices{^a_e},
\eeq
which is manifestly tangential on $b$ and $c$ and normal on $a$.  
An important property of the extrinsic curvature tensor $ K\indices{^a_b_c}$ is that it is symmetric
on its tangential indices, $b$ and $c$, which follows from the 
integrability condition (\ref{eqn:nuinteg})
for the normal
form $ \nu$.  Integrability implies that $ \nabla\ind{_{   \un{  a}} }
 n_{c\, \un{c_2\ldots c_p}}  = 
{\tilde\rho}\ind{_{\un{a} } }
 n_{c \, \un{c_2\ldots c_p}}$ 
for some one form ${\tilde\rho}_a$, and so in particular, $\dd n$ has 
at most one tangential component.  We can then write out the condition that $\dd n$ vanishes 
when projected tangentially on two indices to derive
\begin{align}
0&=(p+1) h\indices{^a_b}  h\indices{^c_d} \nabla_{\un{a \vphantom{c_p}}} 
 n_{c\, \un{c_2\ldots c_p}} \nonumber\\
&= %
 h\indices{^a_b}  h\indices{^c_d} \left( \nabla_a  n_{c\,c_2\ldots c_p} - \nabla_c
 n_{a\, c_2 \ldots c_p} + (p-1)  \nabla_{\underline{c_2}}  n_{ac\,\underline{c_3\ldots c_p} }\right)
\nonumber \\
&= %
2  h \indices{^a_{\un b}}  h\indices{^c_{\un d}}  \nabla_{a} n_{c\, c_2 \ldots c_p}.
\end{align}
From this, the symmetry of $ K\indices{^a_b_c}$ follows straightforwardly,
\beq
 K\indices{^a_b_c}  = \frac{\sg}{(p-1)!}  h\indices{^d_b}  h\indices{^e_c}  n^{a\,c_2\ldots c_p}
 \nabla_d  n_{e\, c_2 \ldots c_p}  
=  %
\frac{\sg}{(p-1)!}  h\indices{^d_c}  h\indices{^e_b}  n^{a\,c_2\ldots c_p}
 \nabla_d  n_{e\, c_2 \ldots c_p}
= %
 K\indices{^a_c_b}
\eeq

The components of $ \nabla_a  s\indices{^b_c}$ with $a$ and $c$ normal 
(equivalently, $a$ and $b$) encode the extrinsic geometry of the normal planes to the surface
through the normal extrinsic curvature tensor,
\beq \label{eqn:Labc}
 L \indices{^a_b_c} = - s\indices{^d_b} s\indices{^e_c} \nabla_d  s\indices{^a_e}
=   s\indices{^d_b} s\indices{^e_c} \nabla_d  h\indices{^a_e},
\eeq
which is manifestly normal on $b$ and $c$ and consequently tangential on $a$.  Unlike 
$ K\indices{^a_b_c}$, the normal extrinsic curvature $ L\indices{^a_b_c}$ is not symmetric
on its lower indices, since in general the normal planes are not  integrable.    
Hence, $ L\indices{^a_b_c}$ further decomposes into its antisymmetric and 
symmetric parts
\begin{align}
 F\indices{^a_b_c} &= -2  L\indices{^a_{\un{bc} }} \label{eqn:F} \\
 A\indices{^a_b_c} &=  L\indices{^a_{\overline{b c}}  } \label{eqn:A}.
\end{align}
$ A\indices{^a_b_c}$ is called the generalized acceleration tensor, because 
its contraction with a normal vector $W^b$ gives
\beq \label{eqn:WWA}
W^b W^c A\ind{^a_b_c} = -h\ind{^a_c}W^b\nabla_b W^c.
\eeq 
The second expression is the tangential component of the acceleration of a flow
along $W^a$.  Since a symmetric tensor is entirely determined by repeated
contractions of this form, knowledge of the accelerations in all possible normal 
directions completely determines $A\ind{^a_b_c}$.

The twist 
tensor $ F\indices{^a_b_c}$ measures the  non-integrability of the normal
planes.  This follows from viewing the foliation  as a fiber bundle, with
the leaves of the foliation comprising the fibers.  The tangential directions
to the leaves coincide with the vertical directions of the fiber bundle, and hence the 
tangential projector $ h\indices{^a_b}$ defines a connection on the fiber bundle.  Its curvature
is given by the Fr\"olicher-Nijenhuis bracket of $ h\indices{^a_b}$ with itself, and measures the 
obstruction to integrability of the horizontal (i.e.\ normal) planes.  This bracket can be
related to $F\indices{^a_b_c}$, by noting first that $[ h, h]=
[ \delta - s,  \delta - s] = [s,s]$ since the identity 
$ \delta\indices{^a_b}$ has vanishing bracket with everything, and then computing
(see \cite[Section~II.8]{Michor})
\beq \label{eqn:FNB}
\frac12 [ h ,  h]\indices{^a_b_c} =\frac12[ s,  s]\indices{^a_b_c} =
2\left(  s\indices{^d_{\un b} } \nabla _d 
 s\indices{^a_{\un c}} -  s\indices{^a_d}  \nabla_{\un b}  s\indices{^d_{\un c} } \right)
\eeq
Projecting $a$ onto the normal direction is seen to give something something proportional to 
$ K\indices{^a_{\un{bc}} }$ which vanishes, hence the bracket must be tangential on $a$, giving
\beq
\frac12[ h,  h]\indices{^a_b_c} = 2  h\indices{^a_e} s\indices{^d_{\un b}} 
 s\indices{^f_{\un c}}  \nabla_d  s\indices{^e_f} =  F\indices{^a_b_c}.
\eeq
Another way to see that $F\ind{^a_b_c}$ measures normal integrability obstruction 
is to contract with two normal vector $W^b$ and $U^c$, which gives
\beq \label{eqn:Fcomm}
W^b U^c F\ind{^a_b_c} = h\ind{^a_c}[W,U]^c.
\eeq
Hence when $F\ind{^a_b_c}$ is nonvanishing, the commutator of normal vectors 
fails to remain normal, signaling a lack of integrability.  

A final note on the normal extrinsic curvature $L\ind{^a_b_c}$ is that although it is a globally 
well-defined tensor on the surface, it can become singular if the foliation crosses itself at
caustics.  Such behavior is inevitable unless the normal bundle is topologically trivial,
and we will comment below on the effects of this singular behavior when it occurs.

\subsection{Covariant derivative operators} \label{sec:derivs}

Although the induced metric $h_{ab}$ is defined as a spacetime tensor, 
restricting its action to vectors tangent to the submanifolds of the foliation naturally defines
a metric on these submanifolds.  It therefore is fruitful to discuss the covariant derivative operator
compatible with this induced metric.  This tangential covariant derivative operator
is denoted $\Dt_a$, and its action on a tangential vector $ V^b$ is the tangential projection of
the spacetime covariant derivative of $ V^b$,
\beq
\Dt_a  V^b =  h\indices{^c_a} h\indices{^b_d} \nabla_c  V^d = h\indices{^c_a}\nabla_a V^b + 
K\indices{^b_a_d}V^d,
\eeq
where the second equality follows from applying equation (\ref{eqn:Kabc}).  The action of 
$\Dt_a$ on covectors and multi-index tensors is defined similarly by projecting the covariant 
derivative of the tensor tangentially on all indices.  A consequence of this definition is that
$h_{ab}$ is annihilated by $\Dt_a$, since 
\beq \label{eqn:Dhab}
\Dt_a h_{bc} = h\indices{^d_a} h\indices{^e_b} h\indices{^f_c}\nabla_d h_{ef}
= h\indices{^d_a}(h\indices{^e_b}\nabla_d h_{ec} - h\indices{^f_b}\nabla_d h_{fc}) = 0.
\eeq
Hence, $\Dt_a$ is the unique covariant derivative 
compatible with the induced metric on the submanifolds.  This fact leads to the usual coordinate
expressions for the connection coefficients of $\Dt_a$ in terms of derivatives of $h_{ab}$, 
written in detail in appendix \ref{app:conncoefs}.  

It is also interesting to consider derivatives of normal vectors along the tangential directions of the 
submanifold.  The covariant derivative associated with vectors in the normal bundle will also
be denoted by $\Dt_a$, and its action on a normal vector $W^b$ is obtained by a different projection
of the spacetime covariant derivative, namely,
\beq
\Dt_a W^b = h\indices{^c_a}s\indices{^b_d}\nabla_c W^d = h\indices{^c_a}\nabla_cW^b 
- K\indices{_d_a^b}
W^d,
\eeq
where again the second equality follows from equation (\ref{eqn:Kabc}). This definition
guarantees that $\Dt_a W^b$ remains normal on the $b$ index.  $\Dt_a$ can  be extended
to act on tensors with multiple normal indices  by projecting all indices in the normal direction
after acting with $h\indices{^b_a} \nabla_b$.  Hence, using an argument similar to 
(\ref{eqn:Dhab}), one finds that $\Dt_a s_{bc} =0$.  
Appendix \ref{app:conncoefs} 
derives the coordinate expressions for the normal connection coefficients, 
and it is here that one finds that $L\indices{_a_b^c}$ is associated with these coefficients.  
This follows from the equation
\beq \label{eqn:Lconnect}
\Dt_a W^b = s\indices{^b_d}h\indices{^c_a}\partial_c W^d + L\indices{_a_c^b}W^c
\eeq
where $\partial_c$ is a coordinate derivative for a coordinate system compatible with the foliation, 
(see appendix \ref{app:coordexpr}).  

The appearance of $L\ind{_a_c^b}$ as connection coefficients suggests  a
modified connection on the normal bundle, $\tilde\Dt_b$, which acts on normal vectors and 
covectors as
\beq \label{eqn:Dtilde}
\tilde \Dt_a W^b = \Dt_a W^b - L\ind{_a_c^b}W^c, \qquad \tilde \Dt_a W_b = \Dt_a W_b+
L\ind{_a_b^c}W_c.
\eeq
This connection simply subtracts off the contribution from $L\ind{_a_c^b}$ in (\ref{eqn:Lconnect}),
and so acts like a coordinate derivative on the components of the normal vectors.  
It appears in a number of variational formulas in section \ref{sec:variations}.  
$\tilde\Dt_b$ 
annihilates the tangential metric $h_{ab}$ 
because it agrees with $\Dt_b$ acting on tangential tensors; however, it is not compatible with the 
normal metric $s_{ab}$, 
\beq
\tilde\Dt_a s_{bc} = 2 A_{abc}.
\eeq 
This equation is directly related to the coordinate expression for the tensor $A_{abc}$ 
derived in appendix \ref{app:extrinsic}.  If $L\ind{_a_c^b}$ has singularities due to 
caustics in the foliation, the modified connection $\tilde\Dt_a$ is only well-defined 
away from these singular points.  In particular, $\tilde \Dt_a$ can be globally defined
only for certain normal bundle topologies, including, but not limited to, trivial
normal bundles.  This is in contrast to $\Dt_a$,
which is well-defined for any normal bundle topology.   

The action of $\Dt_a$ on tensors with indices of 
both normal and tangential type is given by first acting with $h\indices{^b_a}\nabla_b$, and 
then projecting tangentially all indices that were originally tangential, and projecting
normally all indices that were originally normal.  For example, the extrinsic curvature
tensor $K\indices{^a_b_c}$ is normal on $a$ and tangential on $b$ and $c$, so its 
tangential covariant derivative is
\beq
\Dt_a K\indices{^b_c_d} = h\indices{^m_a} s\indices{^b_n} h\indices{^p_c} h\indices{^q_d}
\nabla_m K\indices{^n_p_q}.
\eeq
For a tensor with indices that are not definitely tangential or normal, the action of $\Dt_a$ is
defined by first decomposing the tensor into tangential and normal pieces, and then acting according
to the above definitions.  A straightforward application of this rule shows that 
the spacetime metric $g_{ab}$ is  annihilated
by $\Dt_a$ since
\beq
\Dt_a g_{bc} = \Dt_a(h_{bc} + s_{bc}) = \Dt_a h_{bc} + \Dt_a s_{bc} = 0.
\eeq

When working with foliations of submanifolds, it is important to also consider derivatives in the 
normal direction.  Analogous to the tangential covariant derivative, we define the normal 
covariant derivative $\Eth_a$ acting on a normal vector $W^b$ by
\beq
\Eth_a W^b = s\indices{^c_a} s\indices{^b_d} \nabla_c W^d = s\indices{^c_a}\nabla_c W^b
+ L\indices{^b_a_d}W^d,
\eeq
and on a tangential vector $V^b$ by
\beq
\Eth_a V^b = s\indices{^c_a}h\indices{^b_d}\nabla_c V^d = s\indices{^c_a}\nabla_c V^b
-L\indices{_d_a^b}V^d.
\eeq
Its action on multiple indices of mixed tangential and normal type are defined completely analogously
to the tangential covariant derivative $\Dt_a$.  
The normal covariant derivative $\Eth_a$ has many properties analogous $\Dt_a$, including
annihilating the normal and tangential metric, $\Eth_a s_{bc} = \Eth_a h_{bc} = \Eth_a g_{bc} = 0$.
A notable difference arises from the fact that the tangent planes are not necessarily integrable,
which means that there are no submanifolds on which $\Eth_a$ restricts to 
a genuine affine connection.  This means that $\Eth_a$ behaves somewhat like a connection
with torsion.  This can be quantified by computing the commutator $[\Eth_a, \Eth_b]$ acting on
a scalar function,
\beq \label{eqn:Ethtors}
(\Eth_a \Eth_b - \Eth_b \Eth_a)f = 2s\indices{^c_{\un a} }s\indices{^d_{\un b}}
\nabla_c(s\indices{^e_d} \nabla_e f)= F\indices{^e_a_b} \Dt_e f.
\eeq
Here, the negative twist tensor $-F\indices{^e_a_b}$ is acting 
like a torsion for the connection $\Eth_a$.  
However, it differs from a usual torsion tensor in that the derivative appearing on the 
right hand side of (\ref{eqn:Ethtors}) is the tangential derivative $\Dt_e$, rather than $\Eth_e$.
This means the torsion points in a tangential direction rather than a normal direction. 
The tensor $-F\ind{^e_a_b}$ 
is sometimes called the deficiency of the connection $\Eth_a$, to distinguish it from
genuine torsion \cite{Perjes1993, Boersma1995c}.  
Additional discussion of this interpretation of $-F\indices{^e_a_b}$ as a type of torsion
for $\Eth_a$ is provided in appendix \ref{app:normalcurv}.  

There is also a modification of $\Eth_b$ acting on tangent vectors that appears in many of the 
variational formulas.  This modified connection $\tilde \Eth_b$ acts on tangential vectors 
and covectors as
\beq \label{eqn:Ethtilde}
\tilde\Eth_a V^b = \Eth_a V^b - K\ind{_a_c^b}V^c; \qquad \tilde \Eth_a V_b = 
\Eth_a V_b +K\ind{_a_b^c}V_c.
\eeq
Similar to $\Dt_a$, this modified connection is compatible with the normal metric $s_{ab}$, but
not the tangential metric,
\beq
\tilde\Eth_a h_{bc} = 2K_{abc}.
\eeq

\subsection{Gauss, Codazzi, and Ricci-Voss identities} \label{sec:gauss}

This section describes the curvature tensors associated with the tangential covariant 
derivative $\Dt_a$ and their relationships to the spacetime curvature tensors.  
The derivation of these equations as well as their generalizations to the normal connection
$\Eth_a$ are given in appendix \ref{app:curvature}.
Since the tangential connection $\Dt_a$ is  compatible with the induced metric on the 
submanifolds, its curvature gives information about their intrinsic geometry.  
This intrinsic curvature $\mathcal{R}\indices{^a_b_c_d}$ tensor
is defined through the commutator of two covariant derivatives
acting on a tangential vector $V^a$ according to the equation
\beq
(\Dt_c \Dt_d - \Dt_d \Dt_c)V^a = \mathcal{R}\indices{^a_b_c_d}V^b.
\eeq
Its relationship to the ambient spacetime curvature and extrinsic geometry of the surface
is encoded in the Gauss equation, which reads
\beq \label{eqn:Gauss}
\mathcal{R}_{abcd} = h\indices{^m_a}h\indices{^n_b} h\indices{^p_c} h\indices{^q_d}
R_{mnpq} + K\indices{^e_a_c}K\indices{_e_b_d} - K\indices{^e_a_d} K\indices{_e_b_c},
\eeq
showing that the tangential components of the spacetime curvature tensor and 
the extrinsic curvature determine the intrinsic curvature of the submanifolds.

$\Dt_a$ also defines a  connection on the normal bundle, and there is an  
outer curvature tensor $\out\indices{^a_b_c_d}$ associated with this connection.  
It is defined by the commutator $[\Dt_c, \Dt_d]$ acting on a normal vector $W^a$ via
\beq
(\Dt_c \Dt_d - \Dt_d \Dt_c) W^a = \out\indices{^a_b_c_d}W^d.
\eeq
It also satisfies an equation relating it to the spacetime curvature and extrinsic curvature known
as the Ricci-Voss equation,
\beq \label{eqn:RVmain}
\out_{abcd} = s\indices{^m_a} s\indices{^n_b} h\indices{^p_c} h\indices{^q_d}R_{mnpq}
+ K\indices{_b^e_d} K\indices{_a_c_e} - K\indices{_b^e_c} K\indices{_a_d_e}
\eeq
Since $\out_{abcd}$ is normal and antisymmetric in its first two indices, and tangential and 
antisymmetric in its second two indices, it is trivially traceless on all indices.  One can 
also straightfowardly see that the traces on the right hand side of the equation drop out as well, so
it can instead be expressed in terms of the spacetime Weyl tensor $C_{abcd}$ and the 
traceless extrinsic curvature $\tilde{K}_{abc} = K_{abc} - \frac1{d-p} K_a h_{bc}$ \cite{Carter1992}, 
\beq
\out_{abcd} = s\indices{^m_a} s\indices{^n_b} h\indices{^p_c} h\indices{^q_d} C_{mnpq}
+ \tilde{K}\indices{_b^e_d} \tilde{K}\indices{_a_c_e} - \tilde{K}\indices{_b^e_c} 
\tilde{K}\indices{_a_d_e}.
\eeq
From equation (\ref{eqn:Lconnect}), we know that the tensor $L\indices{_a_b^c}$ fulfills the role
of the normal bundle connection coefficients, and hence there should also be an expression
for $\out_{abcd}$ in terms of $L\indices{_a_b^c}$.  This alternative equation for 
the outer curvature is derived in appendix \ref{app:mixedcurv}, and reads
\beq\label{eqn:outabcd}
\out_{abcd} = \Dt_c L_{dba} - \Dt_d L_{cba} + L\indices{_c_b^e}L\indices{_d_e_a}
-L\indices{_d_b^e}L\indices{_c_e_a}.
\eeq
One can further  decompose this expression by separating $L_{abc}$ into its symmetric and 
antisymmetric parts according to equations (\ref{eqn:F}) and (\ref{eqn:A}), which leads to the 
final expression for the outer curvature,
\beq \label{eqn:outdFAA}
\out_{abcd} = \Dt_{\un c} F_{\un{d}\,ab} 
+\frac12 F\indices{_{\un c}_b^e} F\indices{_{\un d}_e_a}
+ 2 \tilde{A}\indices{_{\un c}_b^e}
\tilde{A}\indices{_{\un d}_e_a},
\eeq
where only the traceless part of the acceleration tensor $\tilde{A}_{cea} = A_{cea} - \frac1p A_c
s_{ea}$ appears.  

The modified connection $\tilde\Dt_a$ introduced in (\ref{eqn:Dtilde}) in principle also has an
associated outer curvature, but 
appendix \ref{app:modifiedcurv} demonstrates that this curvature 
vanishes.  This is consistent with the interpretation of $\tilde\Dt_a$ treating normal
vectors as a collection of scalars, and hence acting as a tangential partial derivative.
Note that 
for this flat connection to be globally defined, the normal bundle must satisfy certain toplogical 
restrictions.  In particular, it requires any topological invariants constructed from the outer curvature, 
such as the Euler number, to vanish.  

Finally, there is an identity associated with the requirement that $(\Dt_c \Dt_d - \Dt_d \Dt_c)V^a$
is tangential on $a$ for $V^a$ tangential, 
which according to the definition of $\Dt_c$ is trivially true.  However, expressing this condition
in terms of the spacetime curvature leads to the Codazzi equation,
\beq \label{eqn:Codazzimain}
h\indices{^m_a}h\indices{^n_b} h\indices{^p_c} s\indices{^q_d}R_{mnpq} = 
\Dt_a K_{dbc} - \Dt_{b} K_{dac}.
\eeq
This same equation arises from requiring that $(\Dt_c \Dt_d - \Dt_d \Dt_c)W^a$ is normal on $a$
for $W^a$ normal.  

There are similar curvature quantities and identities associated with the normal covariant 
derivative $\Eth_a$, which are discussed in more detail in appendix
\ref{app:normalcurv}.

\subsection{Invariant tensors of a submanifold} \label{sec:invten}

For some applications, one is interested only in the properties of a single submanifold of a foliation.
In these cases, it is important to know which quantities are 
independent of how the foliation is extended away from the submanifold.  
These quantities will be called invariant tensors of the submanifold.
This section argues that the invariant tensors consist of the tangential metric $h_{ab}$,
the normal metric $s_{ab}$, the tangential extrinsic curvature $K\ind{^a_b_c}$, 
and any quantities constructed from tangential covariant
derivatives $\Dt_a$ of invariant tensors, which include the intrinsic curvature 
$\mathcal{R}\ind{^a_b_c_d}$ and outer curvature $\out\ind{^a_b_c_d}$.  Of course,
tensors that are defined without reference to the embedding, such as the spacetime
Riemann tensor $R\ind{^a_b_c_d}$, are also invariant. 
Notably, the normal extrinsic curvature $L\ind{_a_b^c}$ is not an invariant tensor, and 
its transformations under a change in the foliation away from the submanifold is 
derived below.  Although it is always possible to set $L\ind{_a_b^c}$ to zero at a point, 
the fact that the invariant outer curvature is constructed from $L\ind{_a_b^c}$ 
means that it cannot vanish everywhere, as is typical 
of connection coefficients in the presence of curvature.  Nevertheless, there always exists
a choice of foliation such that the symmetric piece $A\ind{^a_b_c}$ vanishes, and 
this coincides with extending the foliation away from the submanifold by flowing radially along 
normal geodesics.  
With this choice of foliation, there is still additional freedom to adjust 
the antisymmetric part $F\ind{^a_b_c}$, which can be used to
impose gauge conditions on this tensor.  

The analysis of invariant tensors begins by 
noting that the unit normal $n$ is invariant since it depends algebraically on the 
normal form $\nu$, and normalization ensures it does not change when $\nu$ is rescaled.  
Any tensors constructed algebraically from $n$ are then also invariant, and these include
the normal metric $s_{ab}$, tangential metric $h_{ab}$, and induced volume form $\mu$.  

Next we show
that the tangential covariant derivative $\Dt_a$ acting on an invariant
tensor produces another invariant tensor.  
This follows immediately from the fact that if 
two tensors $T\ind{^{a\ldots}_{b\ldots}}$ and $U\ind{^{a\ldots}_{b\ldots}}$
agree on a submanifold $\Sigma$, 
then the gradient of their difference 
is tensorial and normal to the surface.  This is because 
$\nabla_e(T\ind{^{a\ldots}_{b\ldots}}-U\ind{^{a\ldots}_{b\ldots}})
= \partial_e(T\ind{^{a\ldots}_{b\ldots}}-U\ind{^{a\ldots}_{b\ldots}})$ since the contributions 
for connection coefficients all involve $T\ind{^{a\ldots}_{b\ldots}}-U\ind{^{a\ldots}_{b\ldots}}$
undifferentiated at the surface, which vanishes.  Furthermore, contracting with a tangent vector 
on $e$ also gives a vanishing result since $T\ind{^{a\ldots}_{b\ldots}}-U\ind{^{a\ldots}_{b\ldots}}$
vanishes everywhere on the surface.  Now if  $T\ind{^{a\ldots}_{b\ldots}}$ 
and $U\ind{^{a\ldots}_{b\ldots}}$ are taken 
to be invariant tensors for different foliations that agree at 
$\Sigma$, their difference must vanish on $\Sigma$, and 
\beq
\Dt_eT\ind{^{a\ldots}_{b\ldots}}-\Dt_eU\ind{^{a\ldots}_{b\ldots}}
= h\ind{^d_e}\nabla_d(T\ind{^{a\ldots}_{b\ldots}}-U\ind{^{a\ldots}_{b\ldots}}) = 0.
\eeq
Hence, $\Dt_eT\ind{^{a\ldots}_{b\ldots}}$ defines an invariant tensor.  
This also implies that the curvatures $\mathcal{R}\ind{^a_b_c_d}$ and $\out\ind{^a_b_c_d}$
are invariant, since they can be expressed in terms of $\Dt_a$ acting on
invariant tensors.  

The invariance of the extrinsic curvature $K\ind{^a_b_c}$ follows from a similar argument.  
Note that $s\ind{^a_b}$ is invariant, and let $\bar{s}\ind{^a_b}$ denote the normal projector
for a different foliation that agrees at $\Sigma$.  Then $\nabla_e(s\ind{^a_b}-\bar{s}\ind{^a_b})$ is
normal on $e$, and so
\beq
K\ind{^a_b_c}- \bar{K}\ind{^a_b_c}=h\ind{^e_b}h\ind{^d_c}\nabla_e(s\ind{^a_b}
-\bar{s}\ind{^a_b}) = 0.
\eeq

We now turn to the transformation properties of $L\ind{_a_b^c}$ under a change in foliation.  
First, consider the expression for the covariant derivative of the unit normal, 
$\nabla_e n_{a_1\ldots a_p}$.  This has no component that is normal on all the $a_i$ indices,
since this would be proportional to $n_{a_1\ldots a_p}$ due to antisymmetry, but
$n^{a_1\ldots a_p}\nabla_e n_{a_1\ldots a_p} = 
\frac12\nabla_e(n^{a_1\ldots a_p}n_{a_1\ldots a_p}) = 0$.  Hence, $\nabla_e n_{a_1\ldots a_p}$
is tangential on at least one $a_i$ index, and by projecting tangentially we derive that
\beq
\nabla_e n_{a_1\ldots a_p} = p\left(K\ind{^b_e_{\,\un{a_1}}} - L\ind{_{\un{a_1}}_e^b} \right)
n\ind{_b_{\,\un{a_2\ldots a_p}}}
\eeq
Now take $\bar{n}$ to be the unit normal of a different foliation that agrees with $n$ at $\Sigma$.  
From the above formula, we find that their normal gradients differ according to
\beq\label{eqn:delen}
\nabla_e(\bar{n}_{a_1\ldots a_p} - n_{a_1\ldots a_p}) = 
p\, l\ind{_{\un{a_1}}_e^b} n\ind{_b_{\,\un{a_2\ldots a_p}}}
\eeq
with
\beq\label{eqn:Lshift}
l\ind{_a_e^b} = L\ind{_a_e^b}-\bar{L}\ind{_a_e^b}.
\eeq
Hence, the tensor $L\ind{_a_e^b}$ shifts under a change in the foliation  by $-l\ind{_a_e^b}$,
which characterizes the first order change in the unit normal when moving off of $\Sigma$ 
by equation (\ref{eqn:delen}).  

The transformation of $L\ind{_a_e^b}$ in (\ref{eqn:Lshift}) suggests that it could be set to 
zero by choosing $l\ind{_a_e^b}$; however, not all tensors $l\ind{_a_e^b}$ define a valid 
change in the foliation.  The new unit normal $\bar{n}$ must remain normalized 
according to (\ref{eqn:normnorm}) and satisfy
the simplicity and integrability constraints arising from
(\ref{eqn:nusimpl}) and (\ref{eqn:nuinteg}). After ensuring $\bar{n}$ is normalized, the 
latter two conditions are equivalent to  
\begin{align}
\bar s\ind{^a_b}\bar s\ind{^b_c} 
&= \bar s\ind{^a_c} \\
\bar s\ind{^a_d}[\bar s,\bar s]\ind{^d_b_c} &= 0, \label{eqn:FNbracket}
\end{align}
where the second condition involves the Fr\"olicher-Nijenhuis bracket
(see equation (\ref{eqn:FNB})).   
Taking gradients of these equations with 
$\bar s\ind{^a_b}$ expressed in terms of $\bar n$ leads to additional constraints on 
the gradients of $(\bar n - n)$.  In particular, 
the gradient of (\ref{eqn:FNbracket}) gives a differential restriction on $l\ind{_a_e^b}$.  
A quicker way to derive this restriction is to note that the outer curvature $\out_{abcd}$
is an invariant tensor, and hence should be the same whether computed using 
$L\ind{_a_b^c}$ or $\bar{L}\ind{_a_b^c}$.  
Using equation (\ref{eqn:outabcd}), this implies that
\beq
\Dt_{\un c} l_{\un{d} ba} + L\ind{_{\un c}_b^e}l\ind{_{\un d}_e_a}
+ l\ind{_{\un c}_b^e}L\ind{_{\un d}_e_a} -l\ind{_{\un c}_b^e}l\ind{_{\un d}_e_a} = 0,
\eeq
which can be simplified by using  the modified connection 
$\tilde \Dt_a$ (\ref{eqn:Dtilde}), 
\beq
\tilde \Dt_{\un c} l_{\un{d} ba}+2l\ind{_{\un c}_b^e}A\ind{_{\un d}_e_a} 
- l\ind{_{\un c}_b^e}l\ind{_{\un d}_e _a} = 0
\eeq
or equivalently
\beq \label{eqn:lflat}
\tilde \Dt\ind{_{\un c}} l\ind{_{\un d}_b^a} - l\ind{_{\un c}_b^e}l\ind{_{\un d}_e ^a} = 0.
\eeq

(\ref{eqn:lflat}) looks like a condition of vanishing curvature for a connection, and 
so it is easy to write
down solutions.  Choose any tensor $m\ind{_e^a}$ with normal indices and require that it be
invertible in the sense that $(m^{-1})\ind{_b^e} m\ind{_e^a} = s\ind{_b^a}$ for some tensor
$(m^{-1})\ind{_b^e}$.  Then 
\beq \label{eqn:lsoln}
l\ind{_d_b^a} = -(m^{-1})\ind{_b^e}\tilde\Dt\ind{_d} m\ind{_e^a}
\eeq
solves (\ref{eqn:lflat}), which is seen using the identity $\tilde\Dt\ind{_c}(m^{-1})\ind{_b^e}
= - (m^{-1})\ind{_b^f} \tilde\Dt\ind{_c}m\ind{_f^g} (m^{-1})\ind{_g^e}$, and the fact that 
the outer curvature of $\tilde\Dt_c$ vanishes, as shown in appendix \ref{app:modifiedcurv}.  
A particularly useful class of solutions are those in which $l_{d\,\overline{ba}} = A_{dba}$, 
since this transforms to a new foliation in which $\bar{A}_{dba}=0$ on $\Sigma$.  
Expressing this condition in terms of $m\ind{_e^a}$ gives
\begin{align}
2 A_{dba} &= 
\tilde \Dt_d s_{ba} = -(m^{-1})\ind{_b^e} s\ind{_a_c}\tilde \Dt\ind{_d} m\ind{_e^c} 
-(m^{-1})\ind{_a^e}s\ind{_b_c}\tilde\Dt\ind{_d} m\ind{_e^c} 
\nonumber \\
0&= 
m\ind{_b^e}m\ind{_a^c}\tilde\Dt\ind{_d}s\ind{_e_c} 
+ (\tilde \Dt\ind{_d}m\ind{_b^e})m\ind{_a^c} s\ind{_e_c}
+ m\ind{_b^e}(\tilde\Dt\ind{_d} m\ind{_a^c})s\ind{_e_c}
 = \tilde\Dt\ind{_d}\left(m\ind{_b^e}m\ind{_a^c}s\ind{_e_c}\right)
\end{align}
Hence, the matrix $m\ind{_b^e}$ must transform $s_{ec}$ 
into a metric $\eta_{ab} = m\ind{_a^c}m\ind{_b^e} s_{ec}$ 
that is compatible with the modified connection $\tilde\Dt_d$.  
Since all metrics of the same signature are related by some $GL(p)$ transformation,
such a matrix can always be found pointwise, but there can be topological 
obstructions to choosing it smoothly globally. 
Hence, $\bar{A}_{dba}$ can be set to zero everywhere on the surface only for special
topologies, although for many applications the topology is sufficiently trivial that 
the acceleration can be set to zero globally.  

The tensor $m\ind{_b^e}$ is not unique; the additional
freedom in choosing it consists of transformations that leave $\eta_{ab}$ invariant.  
These comprise an orthogonal group, $SO(p-1,1)$ for $\sg = -1$ or $SO(p)$ for $\sg = +1$,
and they can be used to change $F\ind{^a_b_c}$ while leaving $A\ind{^a_b_c}$ invariant.  
Since $F\ind{^a_b_c}$ is antisymmetric on its normal indices $b$ and $c$, it can be viewed 
on $\Sigma$ as a vector valued in the Lie algebra of the appropriate orthogonal group. 
On the other hand, $l\ind{_a_b^c}$ in (\ref{eqn:lsoln})
is a gradient of a group element, so can be used to cancel a scalar degree of freedom from
$F\ind{^a_b_c}$. 
 For example, one might use this remaining freedom to impose an axial-like 
gauge
$V_aF\ind{^a_b_c} = 0$ for some fixed tangential covector $V_a$, or a Coulomb-like gauge
$\tilde\Dt_a F\ind{^a_b_c} = 0$.  

A tensor $m\ind{_a^b}$ that sets the acceleration to zero is closely related to 
a vielbein for the normal space.  The inverse vielbein is $m\ind{_A^b}$, obtained from 
$m\ind{_a^b}$ by expressing its lower index in a coordinate basis. It satisfies
$\eta^{AB}m\ind{_A^a}m\ind{_B^b} = s^{ab}$, where  $\eta^{AB}$ are the components
of the inverse metric compatible with the flat normal bundle connection $\tilde\Dt_a$,  
which can be taken to be constant. 
One can further show that the modified normal extrinsic curvature can be
expressed in terms of $m\ind{_a^b}$ as
\beq \label{eqn:vieltwist}
\bar{L}\ind{_a_b^c} = (m^{-1})\ind{_b^A}\Dt_a m\ind{_A^c},
\eeq
which only involves the twist term $-2 \bar{F}\ind{_a_b^c}$ when $m\ind{_A^c}$ 
defines an inverse vielbein.   The use of vielbeins is common in 
Carter's treatment of embedded submanifolds \cite{Carter1992, Carter2001},
where an analogous formula to (\ref{eqn:vieltwist}) is used to define the twist tensor.  
The above discussion shows that the acceleration tensor will never appear when 
vielbeins are employed, which explains why such an object is not discussed in 
Carter's work.  Allowing for an acceleration tensor represents a more general choice
of parameterization of the normal bundle, and is indispensable when considering a
foliation as opposed to single submanifold.  This acceleration tensor can also encode
topological information about the normal bundle, since it must be nonvanishing somewhere
when there is a topological obstruction to choosing $m\ind{_a^b}$ smoothly.  In particular, 
topological invariants constructed the outer curvature $\out\ind{^a_b_c_d}$, such as the 
normal bundle Euler number, can be expressed as global integrals in terms of the original
$F\ind{_a_b^c}$ and $A\ind{_a_b^c}$, but must be generally 
constructed in local patches when
working with an orthonormal normal basis and  $\bar{F}\ind{_a_b^c}$.

The coordinate expression (\ref{eqn:AiAB}) for $A_{abc}$ shows that choosing the foliation
away from $\Sigma$ to set $A_{abc}$ to zero is equivalent to setting the components $s_{AB}$
of the normal metric to constant values on $\Sigma$.  
Furthermore, the tensor $A_{abc}$ measures the acceleration of  
 flows normal to $\Sigma$ due to equation (\ref{eqn:WWA}).
Hence, a foliation in which neighboring surfaces are reached by following a normal geodesic
necessarily will have $A_{abc}=0$ on $\Sigma$.  
These foliations therefore define a class of normal coordinates adapted to the surface.  
In such a coordinate system, one would like to define the transverse basis vectors to be tangent
to a family of normal geodesics; however, this is generally not possible when $F\ind{^a_b_c}$
is nonzero \cite{Carter1992, Carter2001}.  
This is because $F\ind{^a_b_c}$ measures the tangential component of the commutator
of normal vectors according to equation (\ref{eqn:Fcomm}).  
Since coordinate basis vectors must commute, generically one cannot choose the coordinate
basis vectors  to be normal to the surface.  
At best, the freedom to shift $F\ind{^a_b_c}$ can be used to set it to zero at a point, and the 
basis vectors can be chosen to be normal there.
Away from the point where $F\ind{^a_b_c}$ vanishes, some other prescription must be 
given to determine the transverse basis vectors, and the local shift freedom in $F\ind{^a_b_c}$
will lead to different available choices in this prescription.  Hence, unless $F\ind{^a_b_c}$ vanishes
everywhere, which is only possible if the outer curvature $\out_{abcd}$ vanishes, 
there is no unique set of normal coordinates.

\section{Variational formulas} \label{sec:variations}

A central motivation for working with the embedding field $X$ comes from considering
variations of geometric quantities of  surfaces.  
Variations of the spacetime metric $g_{ab}$ and the embedding map
both lead to changes in the geometry, and the $X$ field
allows one to cleanly separate the contributions coming from each type of variation.  
This section works out a number of formulas for these variations.

The key property that determines the variations is that the normal form
$\p \nu$ is fixed in the reference space, $\delta\p\nu=0$.  
As mentioned in equation (\ref{eqn:delnu}), this sets the variation of the normal form on 
spacetime to be $\delta\nu = - \lie_\rchi \nu$.  It receives no contribution from the variation
of the metric, which is expected since the foliation is defined without reference to a metric.  
$\nu$ transforms as a tensor under spacetime diffeomorphisms by 
equation (\ref{eqn:Ixidelnu}), and this ensures any object built from $\nu$ and spacetime
tensor fields will also transform covariantly.  On the other hand, $\p\nu$ is \emph{not}
covariant under diffeomorphisms of the reference space, since $I_{\hat{\p \xi}}
\delta\p\nu = 0$, which generally differs from 
$ \lie_{\p \xi}\p\nu$.  This means 
the geometric quantities constructed from
$\p\nu$ will not  transform as tensors on the reference space, so some care 
must be taken when computing their variations under a change in the embedding $X$
induced by a reference space diffeomorphism.
Nevertheless, it will be shown below that the unit normal $\p n$ is covariant 
with respect to 
foliation-preserving transformations of the references space,  generated
by vectors 
with a covariantly constant normal component with respect to $\tilde{\Db}_a$
defined in equation (\ref{eqn:Dtilde}).  These include, in particular, purely tangential vectors.  
Since all other geometric tensors are constructed from $\p n$ and $\p g_{ab}$, their
variation under a tangential diffeomorphism will always be given by a Lie 
derivative.  The variations with respect to normal diffeomorphisms are generally more involved,
and require a careful analysis to determine.

Before doing any calculations, there are a few general rules that help simplify the 
computations.  First, since the restriction $\delta\p \nu=0$ is imposed in the reference space, 
it often is easier to compute variations of quantities $\p\phi$ in the reference space first, and from
this infer the variation of the corresponding spacetime quantity using 
$\delta \phi = X_*\delta\p\phi - \lie_\rchi \phi$.  Second, variations of tensors with mixed 
tangential and normal indices tend to be simplest if all normal indices are taken to be 
covariant (downstairs) and all tangential indices taken to be contravariant (upstairs).  
This is related to the fact that being tangential for a contravariant 
index depends only on $\p\nu$ and 
not the metric, and so the variation will remain tangential.  Being normal for covariant indices
is also metric-independent, so these also remain normal when varied.  
The variations of  tensors with different index placements can then
be straightforwardly computed using $\delta \p\phi_a = \delta \p g_{ab} \p\phi^b + \p g_{ab}
\delta \p \phi^b$ and $\delta \p \psi^a = -\delta \p g_{cb}\p g^{ca}\p\psi^b + \p g^{ab}\delta \p\psi_b$.

\subsection{Intrinsic geometry}
We begin with the unit normal $\p n$ defined in equation (\ref{eqn:nNnu}).  Its variation comes 
entirely from the variation of the norm $\p N$ from equation (\ref{eqn:N}), which 
is calculated as follows
\begin{align}
\delta \p N &= -\frac12 \sqrt{p!} (\sg \p \nu_{c_1\ldots c_p} \p \nu^{c_1\ldots c_p})^{-3/2}
\sg p\, \delta\p g^{a_1 b_1} \p\nu_{a_1a_2\ldots a_p} \p \nu\indices{_{b_1}^{a_2 \ldots}^{a_p}}
\nonumber \\
&=%
\frac12 \frac{\p N^3}{p!} \sg p\,\delta \p g_{cd}\,\p \nu^{c a_2\ldots a_p}\p\nu\ind{^d_{a_2\ldots a_p}}
\nonumber\\
&=%
\frac12 \p N \p s^{cd}\delta \p g_{cd},
\end{align}
where the definition (\ref{eqn:sabdef}) was used.  This immediately leads to the formula for the 
variation of the unit normal,
\beq \label{eqn:deln}
\delta \p n = \frac12 \p s^{cd}\delta \p g_{cd} \p n.
\eeq
It is interesting to single out the contribution to this variation coming purely from a change
in the embedding, encapsulated by the terms involving $\p\rchi^a$.  They appear 
implicitly in (\ref{eqn:deln}) through $\delta \p g_{cd}$ since $\delta \p g_{cd} = X^*\delta
g_{cd}+\lie_{\p \rchi} \p g_{cd}$.  The terms involving $\p\rchi^a$ in $\delta\p n$ will be denoted
$L_{\hat{\p\rchi}} \p n \equiv I_{\hat{\p \rchi}} \delta \p n $, employing
the notation of section \ref{sec:overview} for this variation.  
It is helpful to decompose $\p\rchi^a=\p\tau^a +\p\sigma^a$ in to its
tangential $\p\tau^a$ and normal $\p\sigma^a$ pieces, 
\beq
\p \tau^a = \p h\ind{^a_b}\p\rchi^b,\quad \p\sigma^a = \p s\ind{^a_b}\p\rchi^b.
\eeq
Then (\ref{eqn:deln}) leads immediately to
\beq \label{eqn:Lchin}
L_{\hat{\p\rchi}} \p n = \p s^{cd}(\p\nabla_c \p\rchi_d) \p n = \left(\Ethb_c\p\sigma^c 
+\p \tau^c\p A_c\right)\p n.
\eeq
It is helpful to compare this expression to $\lie_{\p \rchi} \p n$.   The $\p\tau^a$
contribution remains proportional to $\p n$, since
$\lie_{\p\tau}\p n = i_{\p \tau}\dd \p n  = \p\tau^c(\Db_c \log \p N + \p\rho_c) \p n$
with $\p\rho_c$ defined by the integrability condition (\ref{eqn:nuinteg}) for $\p\nu$.
By expanding out the Lie derivative and contracting with $\frac{\sg}{p!} \p n^{b_1\ldots b_p}$,
a different expression for the $\p n$ coefficient is derived, 
\beq
\frac{\sg}{p!}\p n^{b_1\ldots b_p} \left(\p\tau^e\p\nabla_e \p n_{b_1\ldots b_p}
+p (\p\nabla\ind{_{\un{\vphantom{b_p} b_1} }} \p \tau^e) \p n_{e\un{b_2\ldots b_p}}\right)=\p s\ind{^{b_1}_e}\p\nabla_b
\p\tau^e = \p\tau^c\p A_c,
\eeq
which matches the term appearing in (\ref{eqn:Lchin}).  Incidentally, it also gives a relation for 
$\p\rho_c$,
\beq
\p\rho_c = \p A_c - \Db_c \log \p N.
\eeq
For the $\p\sigma^a$ contribution, there will be a piece proportional to $\p n$, and a term with
one tangential index.  The purely normal term is isolated as before by contracting with 
$\frac{\sg}{p!} \p n^{b_1\ldots b_p}$, which is seen to give $\Ethb_b \p\sigma^b$.  The term
with a tangential index is 
\beq
\frac{\sg}{(p-1)!} 
\p n^{c\,b_2\ldots b_p} \p h\ind{^b_d} \left(\p\sigma^e\p\nabla_e \p n_{b\,b_2\ldots b_p}
+ p ( \p\nabla\ind{_{\un{ b} \vphantom{b_p} } } \p\sigma^e)\p n_{e\,\un{b_2\ldots b_p} }\right) 
= \Db_d\p\sigma^c - \p L\ind{_e_d^c}\p\sigma^e = \tilde\Db_d\p\sigma^e.
\eeq
Hence the normal Lie derivative is
\beq
\lie_{\p\sigma} \p n_{b_1\ldots b_p} = (\Ethb_c\p\sigma^c) \p n_{b_1\ldots b_p} 
+ p(\tilde{\Db}_{\un{b_1}}\p \sigma^c) \p n_{c\,\un{b_2\ldots b_p}},
\eeq
and only matches (\ref{eqn:Lchin}) if $\tilde{\Db}_b\p\sigma^c=0$, which is the condition 
for the diffeomorphism to preserve the foliation.\footnote{As an aside,
it is also interesting to consider when the Lie derivative of the unnormalized form $\p\nu$
agrees with its variation under a change in embedding, which
simply vanishes $L_{\hat{\p\rchi}} \p\nu = 0$. 
For tangential vectors, $\lie_{\p\tau} \p\nu = \p\tau^c\p\rho_c \p\nu$.  
This vanishes automatically if $\dd \p\nu = 0$, which is relevant 
in applications where $\p\nu$ represents a conserved quantity, 
such as the entropy current in fluid dynamics.  The normal Lie derivative 
is $\lie_{\p\sigma} \p \nu_{b_1\ldots b_p}
= \Ethb_c\left(\frac{\p\sigma^c}{\p N}\right)\p n_{b_1\ldots b_p} 
+ p (\tilde{\Db}_{\un {b_1} }
\p\sigma^c) \p\nu_{c\,\un{b_2\ldots b_p}}$, so 
for this to vanish, $\p\sigma^c$ must both be foliation-preserving
and satisfy $\Ethb_c\left(\frac{\p\sigma^c}{\p N}\right)=0 $, which is interpreted as a 
volume-preserving condition in the normal directions.  
Volume-preserving diffeomorphism symmetry is often used as an organizing principle
in treatments of finite-temperature fluid mechanics 
\cite{Dubovsky2012a, Haehl2018, Crossley2017a}.}

Next, applying the definition (\ref{eqn:sabdef}) for the normal metric, its variation is
\begin{align}
\delta \p s_{ab} &= \frac{\sg}{(p-1)!} 
(\p s^{cd}\delta \p g_{cd} \p n_{a e_2\ldots e_p} \p n\ind{_b^{e_2\ldots e_p}}  
-(p-1) \delta \p g_{cd} \p n\ind{_a^c^{e_3\ldots e_p} } \p n\ind{_b^d_{e_3\ldots e_p}} ) \nonumber \\
&= \p s^{cd}\delta \p g_{cd} \p s_{ab} -\frac{\sg}{(p-1)!} (p-1) \delta \p g_{cd} \,\sg (p-2)!
(\p s_{ab} \p s^{cd} - \p s\ind{^d_a} \p s\ind{^c_b}) \nonumber \\
&= \p s\ind{^d_a} \p s\ind{^c_b}\delta \p g_{cd}, 
\label{eqn:delsab}
\end{align}
where the second line applied the identity (\ref{eqn:qformproj}) with $q=2$.  Hence the variation
of the normal metric is simply the normal projection of the full metric variation.  The inverse metric
variation is a bit more complicated because it has contravariant normal
indices,
\beq \label{eqn:delscontra}
\delta \p s^{ab} = \delta \p g_{cd}(\p s^{ca}\p s^{db} -\p s^{ca} \p g^{db} - \p g^{ca} \p s^{db})
=-\delta  \p g_{cd} (\p s^{ca} \p s^{db} + \p s ^{ca} \p h^{db} + \p h^{ca} \p s^{db})
\eeq
which involves both normal-normal and normal-tangential components of the metric variation.  

The normal and tangential projector variations are closely related since $\delta\p s\ind{^a_b}
 = \delta(\p \delta\ind{^a_b} - \p h\ind{^a_b}) = -\delta \p h\ind{^a_b}$.  Explicitly, this 
 variation is
 \beq
 \delta \p s\ind{^a_b} = \delta \p g_{cd}(-\p g^{ac}\p s\ind{^d_b} + \p s\ind{^a^c}\p s\ind{^d_b})
 = -\p h\ind{^a^c} \p s\ind{^d_b}\delta \p g_{cd}
 \label{eqn:delmixed}
 \eeq
 The inverse tangential metric also has a simple variation, given by
 \begin{align}
 \delta \p h^{ab} &= \delta \p g^{ab} - \delta (\p g^{ac}\p g{^db} \p s_{cd}) \nonumber \\
 &= %
 \delta \p g_{cd}(-\p g^{ac} \p g^{bd} + \p g^{ac} \p s ^{db} + \p s^{ac} \p g^{db} -\p s^{ac}\p s^{db})
 \nonumber \\
 &= %
 -\p h^{ac} \p h^{db} \delta \p g_{cd}.
 \label{eqn:delhab}
 \end{align}
This involves only the tangential components of the metric variation.  For the tangential 
metric with covariant indices, there are additional contributions coming from normal-tangential
components of $\delta \p g_{ab}$, similar to how they arose in (\ref{eqn:delscontra}),
\beq \label{eqn:delhcov}
\delta \p h_{ab} = \delta \p g_{cd}(\p h\ind{^c_a} \p h{^d_b} + \p h\ind{^c_a}\p s\ind{^d_b}
+ \p s\ind{^c_a} \p h\ind{^d_b}).
\eeq
Equations (\ref{eqn:delsab}), (\ref{eqn:delmixed}), and (\ref{eqn:delhab}) cover all of the 
independent components of the metric variation, and determines its decomposition into 
normal and tangential pieces to be
\beq \label{eqn:delgdecomp}
\delta \p g_{de} = \delta\p s_{de} + \p h_{dm}\delta \p h\ind{^m_e} + \p h_{en}
\delta \p h\ind{^n_d} - \p h_{dm}\p h_{en}\delta \p h^{mn}.
\eeq
We also need the variation of the induced volume form $\p\mu$.  From
its definition (\ref{eqn:mu}) we have (suppressing the $(d-p)$ tangential indices)
\begin{align}
\delta\p\mu 
&=
-\frac{\sg}{p!}(\delta \p n_{a_1\ldots a_p}\p\ep\ind{^{a_1\ldots a_p}  }
-p \,\delta \p g_{cd}\p n^{c\, a_2\ldots a_p}\p\ep\ind{^d_{a_2\ldots a_p}}
+ \p n^{a_1\ldots a_p} \delta \p \ep_{a_1\ldots a_p } ) 
\nonumber \\
&=
\delta \p g_{cd}\left(
\p s^{cd}\p\mu \left(\frac12-\frac{\sg}{(p-1)!}\sg(p-1)! +\frac12\right)
+\frac12 \p h^{cd}\p \mu 
+\frac{\sg}{(p-1)!} \p n^{c\, a_2\ldots a_p} \p h^{de} \p \ep_{e\, a_2\ldots a_p} 
\right)
\nonumber \\
&= \label{eqn:delmu1}
\delta \p g_{cd}\left(\frac12 \p h^{cd} \p\mu
+\frac{\sg}{(p-1)!} \p n^{c\, a_2\ldots a_p} \p h^{de} \p \ep_{e\, a_2\ldots a_p} \right)
\end{align}
The second term can be simplified somewhat by recalling that 
$\p \ep = -\p n \wedge\p \mu$
and employing a useful formula for the wedge product,
\beq
(\alpha\wedge \beta)_{a_1\ldots a_p b_1\ldots b_q}
= \sum_{n=0}^{\min (p,q)} (-1)^n \binom{p}{n} \binom{q}{n} 
\alpha\ind{ _{\un{a_1\ldots a_n \vphantom{a_q} }\,} _{\un{\un{b_{n+1} \ldots b_p}} } }
\beta\ind{_{\un{\un{ b_1 \ldots b_n\vphantom{b_p} } } \,} _{\un{a_{n+1} \ldots a_q}}} 
\eeq
where the antisymmetrization occurs separately for the single-underlined and 
double-underlined indices.  Applying this to the second term in (\ref{eqn:delmu1})
gives
\beq \label{eqn:delmu2}
\delta \p \mu_{b_1\ldots b_{d-p}}
=
\delta \p g_{cd}\left(
\frac12 \p h^{cd} \p \mu_{b_1\ldots b_{d-p}}
+(d-p) \p s\ind{^c_{\un {b_1\vphantom{b_p} }}} \p \mu\ind{^d_{\un{b_2\ldots b_{d-p}}}}
\right).
\eeq
The first term gives the expected expression for the variation of the intrinsic volume
form on the surfaces, and is the only term that remains when $\delta \p \mu$  is 
restricted to the surface.  
The second term tracks the component that must be added to the original volume form 
after the variation to keep it tangential, since tangential covectors do not remain tangential
when the metric is varied.  

\subsection{Tangential extrinsic curvature}
Next we turn to the variational formulas for the extrinsic curvatures.  These feature
prominently when considering perturbations to extremal surfaces and lead to a derivation
of the Jacobi equation for such surfaces.  We begin with 
the tangential extrinsic curvature tensor defined by (\ref{eqn:Kabc}).  As usual, the variation
is simplest with the tangential indices up and normal index down, 
and applying (\ref{eqn:delsab}), (\ref{eqn:delmixed}), and (\ref{eqn:delhab}) leads to
\begin{align} \label{eqn:delKabcmaster}
\delta \p K\ind{_a^b^c} &= \delta(\p s_{ad} \p h^{be} \p h^{cf}\p \nabla_e \p s\ind{^d_f}) \nonumber \\
&= \delta \p g_{de}(\p s\ind{^d_a} \p K^{ebc} 
- \p h^{bd} \p K\ind{_a^e^c} - \p h^{cd}\p K\ind{_a^b^e}) - \delta \p \Gamma^f_{de} \p s_{af}
\p h^{bd} \p h^{ce}
\end{align}
where the variation of the Christoffel symbol is 
\beq
\delta \p\Gamma^f_{de} = \frac12 \p g^{fm}(\p \nabla_d  \delta \p g_{me}+\p\nabla_e\delta\p g_{md}
-\p\nabla_m \delta\p  g_{de}).
\eeq
The contribution of $\p\rchi^a$ to this expression is of particular interest.  Since 
$\p K\ind{_a^b^c}$ is constructed from $\p n$ and $\p g_{ab}$, the contributions from
$\p\tau^a = \p h\ind{^a_b}\p\rchi^b$ will be simply given by $\lie_{\p \tau}\p K\ind{_a^b^c}$.  
Hence, we focus on the normal contribution involving $\p\sigma^a = \p s\ind{^a_b}\p\rchi^b$,
denoted $L_{\hat{\p\sigma}}\p K\ind{_a^b^c}$. 
First, consider the $\delta\p\Gamma^f_{de}$ term of (\ref{eqn:delKabcmaster}),
\begin{align}
-(L_{\hat{\p\sigma}} \p\Gamma^f_{de}) \p s_{af}\p h^{bd}\p h^{ce} &=
-\p h^{bd} \p h^{ce} \p s\ind{^f_a}(\p\nabla_d \p\nabla_{\overline f} \p\sigma_{\overline e}
+\p \nabla_e\p\nabla_{\overline f}\p\sigma_{\overline d} 
- \p\nabla_f\p\nabla_{\overline d}\p\sigma_{\overline e}) 
\nonumber \\
&=
\p h^{bd} \p h^{ce} \p s\ind{^f_a}\left(\p R\ind{^m_{\overline {ed}}_f}\p\sigma_m
-\p\nabla_{\overline d}\p\nabla_{\overline e} \p\sigma_f\right).
\label{eqn:LchiGamma}
\end{align}
The second term here can be expanded by employing the identity (\ref{eqn:DDWdeldelW})
derived in appendix \ref{app:tangcurv} to give
\begin{align}
-\p h^{bd}\p h^{ce}\p s\ind{^f_a} \p\nabla_{\overline d} \p\nabla_{\overline e} \p \sigma_f
= &\; 
-\Db^{\overline b} \Db^{\overline c}\p\sigma_a -\p K^{ebc}\Ethb_e\p\sigma_a
+\p K\ind{_a^{\overline b}^e} \p K\ind{_d_e^{\overline c} } \p\sigma^d 
\label{eqn:hhsdeldelchi}
\end{align}
The remaining terms in (\ref{eqn:delKabcmaster}) involving $\delta \p g_{de}$ are straightforward to 
evaluate, giving
\begin{align}
2\p\nabla\ind{_{\overline d}}\p\sigma\ind{_{\overline e}}
(\p s\ind{^d_a} \p K^{ebc} 
- \p h^{bd} \p K\ind{_a^e^c} - \p h^{cd}\p K\ind{_a^b^e}) 
&=
\p K^{dbc} (\Ethb_d\p\sigma_a +\Ethb_a\p\sigma_d) 
- 4\p K\ind{^d^{\overline b}_e}\p K\ind{_a^e^{\overline c} }\p\sigma_d
\label{eqn:dsigK}
\end{align}
Combining (\ref{eqn:hhsdeldelchi}) 
and (\ref{eqn:dsigK}), 
we arrive at the final expression for the change in $\p K\ind{_a^b^c}$ under a change in
the embedding,
\begin{align}
L_{\hat{\p\sigma}} \p K\ind{_a^b^c} = &\;
\p K\ind{_d^b^c}\Ethb_a \p \sigma^d 
-\Db^{\overline b}\Db^{\overline c}\p\sigma_a
+\left(\p h^{\overline{b} m} \p h^{\overline c e}\p s\ind{^f_a}\p R\ind{^d_e_m_f}
-3\p K\ind{^d^{\overline{b}}_e} \p K\ind{_a^e^{\overline c}}\right)\p\sigma_d
\label{eqn:LchiKabc}
\end{align} 

The full extrinsic curvature variation (\ref{eqn:delKabcmaster}) also receives contributions 
from the spacetime metric variation in the form $X^*\delta g_{de}$.  The expression involving
these terms can be simplified by 
decomposing the metric variation into its normal and tangential
components as in (\ref{eqn:delgdecomp})
and computing each contribution separately.  The first thing to notice is that 
the terms in (\ref{eqn:delKabcmaster}) involving $\delta \p s_{de}$ drop out,
\beq
\delta \p s_{de}\p s\ind{^d_a}\p K^{ebc} 
- \p h^{d\overline b} \p h^{e\overline c}\p s\ind{^f_a} \p\nabla_d \delta \p s_{ef} = 0.
\eeq
The terms coming from $\delta\p h\ind{^c_a}$ are
\beq
-\frac12\left(\Db^b\delta \p h\ind{^c_a} + \Db^c\delta\p h\ind{^b_a} 
+\p L\ind{^c_a^e}\delta \p h\ind{^b_e} + \p L\ind{^b_a^e}\delta \p h\ind{^c_e}\right)
 = -\tilde\Db\vphantom{\p h}\ind{^{\overline b}}\delta \p h\ind{^{ \overline{c \vphantom b}} _a}.
\eeq
Finally, the remaining pieces from $\delta \p h^{bc}$ are 
\beq
\delta \p h^{be}\p K\ind{_a_e^c} + \delta \p h^{ec}\p K\ind{_a^b^c} 
-\frac12\left( \p K\ind{_a^b_e}\delta\p h^{ce} + \p K\ind{_a^c_e}\delta \p h^{eb} 
+ \Ethb_a\delta \p h^{bc} \right)
= -\frac12\tilde\Ethb_a\delta \p h^{bc}.
\eeq

Of course, these expressions are simply a rewriting of (\ref{eqn:delKabcmaster}), hence still
contain contributions from $\p \rchi^a$; however, it is straightforward to separate off
the spacetime metric variation.  Denoting this contribution $X^*\kappa\ind{_a^b^c}$,
we have 
\beq
\kappa\ind{_a^b^c}  = \frac12 \tilde\Eth_a(h^{bd} h^{ce}\delta g_{de}) 
- \tilde \Dt\vphantom{h}^{\overline b}(h^{\overline cd} s\ind{^e_a}
\delta g_{de}).
\label{eqn:kappa}
\eeq
Thus, we have the general formula that $\delta \p K\ind{_a^b^c} = 
 X^*\kappa\ind{_a^b^c} + L_{\hat{\p \sigma}}
\p K\ind{_a^b^c} + \lie_{\p\tau}\p K\ind{_a^b^c} $, 
with the individual contributions 
given by (\ref{eqn:LchiKabc}) and (\ref{eqn:kappa}).

The formula for $\delta \p K_a$ then follows from those for $\delta\p K\ind{_a^b^c}$
since $\delta \p K_a = \p h_{bc}\delta \p K\ind{_a^b^c} + \p K\ind{_a^b^c}\delta\p h_{bc}$.
Applying (\ref{eqn:delKabcmaster}) then gives
\beq
\delta\p K_a = \delta \p g_{bc} \p K\ind{_a^b^c} - \delta \p \Gamma^d_{bc}\p s_{ad} \p h^{bc}.
\eeq
The contribution involving $\p\sigma^a$ follows in a similar manner from (\ref{eqn:LchiKabc})
giving
\begin{align}
L_{\hat{\p\sigma}} \p K_a = 
\p K_d \Ethb_a\p\sigma^d - \Db_b\Db^b\p\sigma_a
+(\p s\ind{^e_a}\p h^{bc}\p R\ind{^d_b_c_e} - \p K^{dbc}\p K_{abc})\p\sigma_d,
\label{eqn:LchiKa}
\end{align}
and again the $L_{\hat{\p\tau}} \p K_a$ 
terms are simply $\lie_{\p\tau}\p K_a$.  The remaining parts of 
$\delta \p K_a$ involving the spacetime metric are denoted $X^*\kappa_a$, 
and since $\kappa_a = h_{bc}\kappa\ind{_a^b^c} + K\ind{_a^b^c}\delta g_{bc}$, it is expressed
using (\ref{eqn:kappa}) as 
\beq\label{eqn:kappaa}
\kappa_a = \frac12 \Eth_a(h^{bc}\delta g_{bc}) -\tilde\Dt_b(h^{bd}s\ind{^c_a}\delta g_{bc}).
\eeq
The full variation of $\p K_a$ is then given by 
the sum 
$\delta\p K_a = X^*\kappa_a
+ L_{\hat{\p \sigma}} \p K_a
+ \lie_{\p\tau} \p K_a$
involving (\ref{eqn:LchiKa}), (\ref{eqn:kappaa}), and the tangential piece. 

\subsubsection{Jacobi equation} \label{sec:Jacobi}
An immediate application of the variational formula (\ref{eqn:LchiKa}) for $\p K_a$ is 
a derivation of the Jacobi equation, which describes deformations of extremal
surfaces which leave them extremal.  
These surfaces are described by an embedding map $X$ that is a stationary point
of the volume functional,
\beq
V = \int_{\p\Sigma} \p\mu,
\eeq
where $\p\Sigma$ is a fixed $(d-p)$-dimensional surface in the reference space with normal
form $\p\nu$.  Using (\ref{eqn:delmu2}) for $\delta\p\mu$, the variation of this functional is 
\beq
\delta V = \int_{\p\Sigma} \delta\p\mu = \frac12\int_{\p\Sigma} \p h^{cd}\delta \p g_{cd}\p\mu
=\int_{\p \Sigma} \p\sigma^e\p K_e\p\mu +\int_{\p\partial\p\Sigma} \p\tau^e\p\mu_e
+ \frac12 \int_{\p\Sigma}X^*(h^{cd}\delta g_{cd}\, \mu)
\eeq
where in the last expression the variation has been split into a contribution from the spacetime
metric variation $\delta g_{cd}$, and the contributions from the normal and tangential
components of $\p\rchi^a$.  Hence, for fixed spacetime metric, the embedding must satisfy
$\p K_e = 0$ to be a stationary point of $V$.\footnote{Additionally, boundary conditions
should be imposed on the tangential component so that $\p\tau^e\p\mu_e\big|_{\p\partial\p\Sigma}
=0$.}

The Jacobi equation involves perturbations to nearby stationary surfaces, 
so it comes from demanding that $\p K_e$ remains $0$ after varying the embedding.  
For a specific choice of the deformation described by the vector $\p \xi^a$, this
is just the statement that $L_{\hat{\p\xi}} \p K_e =  0$.  
The tangential component $\p\zeta^a = \p h\ind{^a_b}\p \xi^b$ does not contribute to this 
equation since it is just given by $\lie_{\p \zeta} \p K_e$, which vanishes since $\p K_e = 0$
everywhere on the surface.  The normal contribution $\p\eta^a = \p s\ind{^a_b}\p\eta^b$ 
is nontrivial, and  equation 
(\ref{eqn:LchiKa}) shows that it must satisfy
\beq \label{eqn:Jacobi}
\Db_b\Db^b \p \eta^a + ( \p K\ind{^a_b_c} \p K\ind{_d^b^c} - \p s^{ea}\p h^{bc}\p R_{dbce})
\p \eta^d = 0,
\eeq
which is the Jacobi equation for the vector field $\p\eta^a$.  Note that since $\p K_e = 0$
on the surface, the extrinsic curvatures appearing in this equation can be taken to be 
traceless.  This equation was first derived for embedded surfaces of arbitrary codimension
in \cite{Simons1968}.

It is interesting to express this equation using the modified normal bundle
connection from equation (\ref{eqn:Dtilde}), which gives
\beq\label{eqn:Dtildelong}
\tilde{\Db}_b\tilde{\Db}^b \p \eta^a + 2 \p L\ind{^b_c^a}\tilde\Db_b\p\eta^c
+\left(\Db_b\p L\ind{^b_d^a} +\p L\ind{_b_d_e}\p L\ind{^b^e^a} 
+\p K\ind{^a_b_c} \p K\ind{_d^b^c} - \p s^{ea}\p h^{bc}\p R_{dbce}\right) \p\eta^d = 0.
\eeq
After applying the identity (\ref{eqn:shsR}), this simplifies to
\beq\label{eqn:DtildeLap}
\tilde{\Db}_b\tilde{\Db}^b \p \eta^a + 2 \p L\ind{^b_c^a}\tilde\Db_b\p\eta^c
-\Ethb_d\p K^a \p\eta^d = 0.
\eeq
An advantage of using the modified connection is that $\tilde\Db_b\tilde\Db^b$ acts like 
a scalar Laplacian as opposed to a vector one.  This is because $\tilde \Db_b$ annihilates 
the normal basis vectors $\p w_A^a$, defined in  appendix \ref{app:coordexpr},
equation (\ref{eqn:wbasis}), so by expressing
the deformation vector in terms of its components $\p\eta^a = \p \eta^A\p w_A^a$, 
this equation becomes
\beq
\Db_i \Db^i \p\eta^A + 2 \p L\ind{^i_B^A}\p\partial_i\p\eta^B - (\Ethb_B\p K^A) \p\eta^B=0,
\eeq
with $\Db_i \Db^i$ the scalar Laplacian, and $\p\eta^A$ are now viewed as a collection of 
scalar functions.  
The drawback of (\ref{eqn:DtildeLap}) is that  $\Ethb_d\p K^a$ appears, which involves
derivatives of $\p K^a$ away from the surface.  These derivatives were not specified 
by demanding that $\p K^a$ vanish on the surface, so it seems the only way to make
sense of this object is to replace it with the equivalent expression  in (\ref{eqn:Dtildelong}).  

\subsection{Normal extrinsic curvature}
The last set of variational formulas we consider are for $\p L\ind{^a_b_c}$.  Using 
its definition (\ref{eqn:Labc}) and applying (\ref{eqn:delmixed}), we find
\begin{align}
\delta \p L\ind{^a_b_c} &= 
\delta(\p h\ind{^a_f} \p s\ind{^d_b} \p s\ind{^e_c}\p\nabla_d\p h\ind{^f_e}) 
\nonumber \\
&=
\delta \p g_{de}\p s\ind{^d_b}\p K\ind{_c^e^a} 
+\p h\ind{^a_m}\p s\ind{^d_b} \p s\ind{^e_c}
\left(-\delta\p\Gamma^m_{de}
+\p\nabla_d(\p h^{mf} \p s\ind{^n_e}\delta\p g_{mn})  \right)
\nonumber \\
&=
\delta\p g_{de}\left(\p s\ind{^d_b} \p K\ind{_c^e^a} +\p s\ind{^d_c}\p L\ind{^a_b^e}
-\p h^{da}\p L\ind{^e_b_c}  \right)
+\p h^{ae}\p s\ind{^d_b}\p s_{fc}\delta\p \Gamma^f_{de}.
\label{eqn:delL}
\end{align}
This can be simplified by decomposing the metric variation into its normal and tangential
components according to (\ref{eqn:delgdecomp}). 
First the $\delta \p s_{de}$ terms are 
\beq \label{eqn:Dtildedels}
\delta \p s_{ce}\p L\ind{^a_b^e} + \frac12\left(\Db^a\delta\p s_{bc} -\p L\ind{^a_b^e}\delta \p s_{ce}
+\p L\ind{^a_c^e}\delta\p s_{be}\right)  = \frac12\tilde\Db^a \delta \p s_{bc}.
\eeq
Next, the purely tangential contribution involving $\delta \p h^{mn}$ is
\beq\label{eqn:delhA}
\delta \p h^{ae}\p L_{ebc}+ \frac12 \left(-\p L_{ebc}\delta \p h^{ea} 
+ \p L_{ecb}\delta \p h^{ea}\right)
 = \delta \p h^{ae} \p A_{ebc}.
\eeq
Lastly, the normal-tangential terms involving $\delta\p h\ind{^m_e}$ are
\beq\label{eqn:Ethtildedelh}
\delta \p h\ind{^e_b}\p K\ind{_c_e^a} + \frac12\left(
\Ethb_b\delta\p h\ind{^a_c}- \Ethb_c\delta \p h\ind{^a_b}-\p K\ind{_b^a_e}\delta \p h\ind{^e_c}
-\p K\ind{_c^a_e}\delta \p h\ind{^e_b}\right)
 = \tilde\Ethb\ind{_{\un b}}\delta \p h\ind{^a_{\un c}}
\eeq
The expressions in (\ref{eqn:Dtildedels}) and (\ref{eqn:delhA}) are manifestly symmetric on 
$b$ and $c$, and hence contribute only to $\delta \p A\ind{^a_b_c}$, while (\ref{eqn:Ethtildedelh})
is antisymmetric in $b$ and $c$ and hence contributes to only $\delta \p F\ind{^a_b_c}$. 
This then leads to expressions for the contribution of $\p \sigma^a$ to the variations
\begin{align}
L_{\hat{\p \sigma}}\p A\ind{^a_b_c} &= \tilde\Db^a
\Ethb_{\overline b} \p\sigma_{\overline c}
-2\p A_{ebc}\p K^{dea}\p\sigma_d \\
L_{\hat{\p \sigma}}\p F\ind{^a_b_c} &= 2\tilde\Ethb_{\un c}\left(\Db^a\p\sigma_{\un b}
-\p L\ind{^a_{\un b}_e}\p\sigma^e\right).
\end{align}
The effects of a pure metric variation also can be read off from (\ref{eqn:Dtildedels}), 
(\ref{eqn:delhA}), and (\ref{eqn:Ethtildedelh}).
These are captured by the spacetime variational tensors
\begin{align}
\alpha\ind{^a_b_c} &= \frac12\tilde\Dt^a(s\ind{^d_b}s\ind{^e_c}\delta g_{de}) 
- A\ind{^e_b_c}h^{ad}\delta g_{de} \\
\digamma\ind{^a_b_c}&=2\tilde \Eth_{\un c}\left(h^{ad}s\ind{^e_{\un b}} \delta g_{de}\right)\\
\lambda\ind{^a_b_c} &= \alpha\ind{^a_b_c}-\frac12\digamma\ind{^a_b_c}.
\end{align}
with $X^*\alpha\ind{^a_b_c}$ giving the pure metric variation in $\delta \p A\ind{^a_b_c}$, 
$X^*\digamma\ind{^a_b_c}$ the contribution in $\delta\p F\ind{^a_b_c}$, 
and $X^*\lambda\ind{^a_b_c}$ the contribution in $\delta \p L\ind{^a_b_c}$.  
As before, the full variation satisfies $\delta \p A\ind{^a_b_c} = X^*\alpha\ind{^a_b_c}
+L_{\hat{\p\sigma}} \p A\ind{^a_b_c} + \lie_{\p\tau} \p A\ind{^a_b_c}$, and similarly
for $\delta \p F\ind{^a_b_c}$ and $\delta \p L\ind{^a_b_c}$.  

Finally, according to the discussion of section \ref{sec:derivs}, $\p L\ind{_a_b^c}$ has an 
interpretation as the connection coefficients of the normal bundle connection,
and hence one might expect its variation to be an invariant tensor of the surface,
even though $\p L\ind{_a_b^c}$ is not.  This is the case, provided that its first 
index is projected tangentially and last index normally.  To see this, note that 
the expression (\ref{eqn:delL}) leads straightforwardly to
\beq
\delta \p L\ind{_a_b^c} = \delta\p g_{de}\big(\p s\ind{^d_b}\p K\ind{^c^e_a}
+\p h^{dc}\p L\ind{_a_b^e}
+ \p s\ind{^d_a}\p L\ind{^e_b^c}\big)
+\p h\ind{^e_a}\p s\ind{^d_b} \p s\ind{^c_f}\delta \p \Gamma^f_{de},
\eeq
and the projection gives
\beq \label{eqn:hsdelL}
\p h\ind{^a_m} \p s\ind{^n_c}\delta\p L\ind{_a_b^c}  
= \delta\p g_{de} \p s\ind{^d_b}\p K\ind{^n^e_m}
+ \p h\ind{^e_m}\p s\ind{^d_b}\p s\ind{^n_f}\delta\p \Gamma^f_{de}.
\eeq
Since $\delta\p \Gamma^f_{de}$ is expressed solely in terms of the spacetime metric variation 
and its 
covariant derivative, it is an invariant tensor of the surface by the arguments of  
section \ref{sec:invten}.  $\p K\ind{^n^e_m}$, $\p s\ind{^a_b}$, $\p h\ind{^a_b}$,
are also all invariant, and so every 
term in (\ref{eqn:hsdelL}) is invariant.  

\section{Special Cases} \label{sec:special}

This section applies the formalism developed above to special choices for the codimension
of the surfaces.  The codimension-1 hypersurface is presented first, since this case is likely
the most familiar and allows an easy comparison between the geometric quantities defined
above and the usual quantities associated with a hypersurface.  Next, one-dimensional
submanifolds are treated, which is the case of a congruence of curves, and again 
comparisons are made between the constructions in this work and the expansion, shear, twist, and 
acceleration usually associated with such a congruence.  Finally, we treat the case of 
codimension-2, which exhibits some special features over the generic case.  

\subsection{Codimension 1}

The normal form for a codimension-1 hypersurface is simply a one-form $\nu_a$, and the 
unit normal is given by
\beq
n_a = N \nu_a.
\eeq
We will take the hypersurfaces to be spacelike, so that $\sg = -1$.  A common situation 
where a foliation by hypersurfaces is employed is the $3+1$ split used in the canonical
analysis of general relativity. There, the normal is $\nu = -\dd T$, with 
$T$  the time function labelling the hypersurfaces, and $N$ is the lapse.  The 
normal and tangential metrics are
\beq
s_{ab}= - n_a n_b,\qquad h_{ab} = g_{ab}+ n_a n_b.
\eeq
Since the normal space is one dimensional, the unit normal $n_a$ provides a natural vielbein 
for this space, as the equation for $s_{ab}$ shows.  

The one dimensional normal space also means that the extrinsic curvature tensor $K\ind{^a_b_c}$
is entirely determined by its contraction with $n_a$ on its normal index.  This is seen 
explicitly by applying its definition (\ref{eqn:Kabc}),
\beq
K\ind{^a_b_c} = - h\ind{^d_b}h\ind{^e_c}\nabla_d(n^an_e) = - n^ah\ind{^d_b}h\ind{^e_c}
\nabla_d n_e = - n^a K_{bc},
\eeq
where the last equality involves the usual extrinsic curvature tensor of the hypersurface,
\beq
K_{bc} = n_a K\ind{^a_b_c} = h\ind{^d_b}h\ind{^e_c}\nabla_d n_e.
\eeq
For the normal extrinsic curvature tensor $L\ind{^a_b_c}$, it is completely determined by 
its contraction with $n^b$ on both of its normal indices.  It cannot have an antisymmetric piece,
so $F\ind{^a_b_c}$ vanishes identically, and the symmetric piece is pure trace, $A\ind{^a_b_c}
 = A^as_{bc}$.  The tangential vector $A^a$ can be shown to be the acceleration of $n^a$,
\beq\label{eqn:Aa}
A^a = - n^b n^c A\ind{^a_b_c} = -n^b n^c \nabla_b(n^a n_c) = n^b\nabla_b n^a.
\eeq
The outer curvature tensor $\out_{abcd}$ from (\ref{eqn:outabcd}) 
vanishes identically since it is normal and antisymmetric
on $a$ and $b$, and consequently all terms in the Ricci-Voss equation (\ref{eqn:RVmain})
vanish separately.  Hence, this equation has no content when considering hypersurfaces.  

According to the discussion of section \ref{sec:invten}, if one is free to choose how to 
extend the foliation away from an initial hypersurface $\Sigma$, there exists a choice
that sets $A^a = 0$.  From equation (\ref{eqn:Aa}), this choice simply corresponds to extending
the surfaces along geodesics in the normal direction.  This choice defines a 
Gaussian normal coordinate system, in which the lapse $N$ is constant, and can therefore be 
set to $1$, and the shift vector $N^i = N^i_T$, discussed in appendix \ref{app:coordexpr}, 
vanishes.  
Applying the formulas in section \ref{app:christoffel}, we see that the Christoffel symbols
with two or three $T$ components, $\Gamma^T_{T i}$, $\Gamma^i_{TT}$ and $\Gamma^T_{TT}$,
all vanish in this coordinate system.

The Jacobi equation 
for perturbations between maximal volume slices takes a simple form
due to the one-dimensionality of the normal space.  The normal deformation vector field 
must be proportional to the normal vector, so by writing $\p\eta^a = \p H\p n^a$, the 
Jacobi equation (\ref{eqn:Jacobi}) 
becomes a scalar equation for the function $\p H$.  Recalling that 
$\Db_a \p n_b = 0$, the Jacobi equation reduces to 
\beq
\Db_a \Db^a \p H - \left(\p K_{ab}\p K^{ab} + \p n^a \p n^b \p R_{ab}\right) \p H = 0.
\eeq

\subsection{Codimension $(d-1)$} \label{sec:codimd-1}

A congruence of curves is a foliation by one-dimensional submanifolds, and this gives another
special case where a number of simplifications occur.   This case is relevant to many applications
on global properties of spacetime \cite{Hawking1975}, and is also important
for relativistic fluid mechanics, where the curves define the world lines of the fluid elements.  
We  take the curves to be timelike, so $\sg = +1$.  The normal metric $s_{ab}$ is given 
by (\ref{eqn:sabdef}), and has rank $d-1$.  On the other hand, the tangential 
metric is rank $1$, and hence is given by 
\beq
h_{ab} = - u_a u_b,
\eeq
where $u^a$ is the unit tangent vector to the curves.  

Since the tangential space is one-dimensional, $K\ind{^a_b_c}$ is totally determined by 
its contraction with $u^b$ on both its tangential indices.  This means it must be pure 
trace, $K\ind{^a_b_c} = K^a h_{bc}$, and the normal vector $K^a$ is simply the 
acceleration of the curves,
\beq
K^a = -u^b u^c K\ind{^a_b_c} = u^b u^c\nabla_b h\ind{^a_c} = u^b\nabla_b u^a.
\eeq
The normal extrinsic curvature $L\ind{^a_b_c}$ contains the remaining geometric information
about the congruence.  It is totally determined by its contraction with $u_a$ on its tangential index,
and takes the form $L\ind{^a_b_c} = - u^a L_{bc}$, where 
\beq
L_{bc} = - u_a s\ind{^d_b}s\ind{^e_c}\nabla_d(u^a u_e) = s\ind{^d_b}s\ind{^e_c}\nabla_d u_e.
\eeq
The twist and acceleration tensors defined in (\ref{eqn:F}) and (\ref{eqn:A}) are similarly determined
by their contraction with $u_a$, which determine the tensors $F_{bc} = u_a F\ind{^a_b_c}
= -2 L_{\un{bc}}$ and 
$A_{bc} = u_a A\ind{^a_b_c} = L_{\overline{bc}}$. These are then related to the usual twist 
$\omega_{bc}$,  expansion $\theta$, and shear $\sigma_{bc}$ of the congruence by
\begin{align}
\omega_{bc} &= -\frac12 F_{bc} \\
\theta &= s^{bc} A_{bc} \\
\sigma_{bc} &= \tilde A_{bc} = A_{bc} - \frac{1}{d-1} \theta s_{bc}.
\end{align}

The outer curvature $\out_{abcd}$ of (\ref{eqn:outabcd})
again vanishes, since it is antisymmetric and tangential
on $c$ and $d$, and again all terms in the Ricci-Voss equation (\ref{eqn:RVmain}) are 
identically zero.  Similarly, the intrinsic curvature of the curves also vanishes since they are 
one-dimensional, and the Gauss equation  (\ref{eqn:Gauss}) is also trivial.  However,
as discussed in appendix  \ref{app:normalcurv}, there is an analog of the Gauss relation
for the 
normal connection $\Eth_a$, and 
gives an identity (\ref{eqn:Cat}) that relates the normal curvature tensor $\cat_{abcd}$ defined by
equation (\ref{eqn:Ccurv}) to the spacetime Riemann tensor and $L\ind{_b_c}$
\cite{Massa1974, Boersma1995c}. 

In the situation where only a single curve is determined and the foliation away from the curve
is freely specifiable, the discussion of section \ref{sec:invten} shows that $A_{bc}$ can 
be chosen to vanish.  Furthermore, the additional freedom in changing the foliation after
$A_{bc}$ has been set to zero is enough to also set $F_{bc}$ to zero.  This is just the 
axial gauge choice  $u_aF\ind{^a_b_c}=0$ discussed in section \ref{sec:invten}.  
Hence, when the congruence is extended according to this prescription, the expansion, shear, and 
twist all vanish along the central curve.  Since $F_{bc}$ was set to zero, one can pick a basis
of normal vectors which are all commuting, and hence can serve as coordinate basis vectors.  
This then leads to the notion of Fermi normal coordinates along the curve.  In terms of the 
coordinate expressions discussed in appendix \ref{app:coordexpr}, these coordinates set the 
normal metric $s_{AB}$ along the curve to a constant,  and set its first transverse derivative
to zero;  additionally, they set the shift $N_A^0$ and its first transverse derivative to zero
(here $0$ denotes the coordinate along the curve).
Applying the formulas in section \ref{app:christoffel}, the only nonvanishing Christoffel
symbols along the curve in these coordinates are $\Gamma^A_{00}$ and $\Gamma^0_{A0}$, and if 
the curve is a geodesic, these vanish as well.

The covariant derivative $\Dt_a$ is also determined 
by its contraction with $u^a$, and this contraction $\Dt_F = u^a\Dt_a$ is referred to as 
the Fermi derivative along the curve \cite{Hawking1975}.  
From the fact that $\Dt_F u^a = 0$, the Jacobi
equation (\ref{eqn:Jacobi}) for a normal vector $\p \eta^a$ that defines a variation to 
a nearby geodesic reads
\beq
\Db_F^2 \p\eta^a + \p s^{ea} \p u^b \p u^c \p R_{dbce} \p \eta^d = 0,
\eeq
which is also known as the equation of geodesic deviation.

\subsection{Codimension 2}

For codimension-2 submanifolds, there are certain simplifications that occur in the expressions 
for the twist tensor $F\ind{^a_b_c}$ and the outer curvature tensor $\out_{abcd}$.  
This is because they contain normal, antisymmetric pairs of indices, which, since the 
normal space is two-dimensional, can be simplified by contracting with the unit normal
$n\ind{^a^b}$.  Hence, the twist tensor can reduced to a single tangential vector,
\beq
F^a = \frac12 n\ind{^b^c}F\ind{^a_b_c}, \qquad F\ind{^a_b_c} = \sg F^a n_{bc},
\eeq
and similarly the outer curvature reduces to a tangential 2-form,
\beq
\out_{cd} = \frac12 n^{ab}\out_{abcd}, \qquad \out_{abcd} = \sg n_{ab}\out_{cd}.
\eeq
This then leads to a simplification in the expression of $\out_{cd}$ in terms of $F_a$ and 
$A\ind{^a_b_c}$, which follows from (\ref{eqn:outdFAA}),
\beq \label{eqn:OcdA}
\out_{cd} = \Dt_{\un c} F_{\un d} + n^{ab}\tilde A\ind{_{\un c} _b^e} \tilde A\ind{_{\un d}_e_a},
\eeq
which no longer involves a term quadratic in $F\ind{^a_b_c}$.  Further, when the extension of the 
foliation away from the initial surface is allowed to be chosen to set $A\ind{_a_b^c}$ to zero,
the outer curvature restricted to the surface is an exact form,
\beq \label{eqn:Ocd}
\out_{cd}  = \Dt_{\un c} F_{\un d}.
\eeq
This reflects the fact that the orthogonal group acting on the normal bundle is abelian
for codimension-2 surface.  One can then form topological invariants  of 
the surface by wedging $\out_{cd}$ together to form a top form (when the submanifold is 
even dimensional), and then integrating it over the surface \cite{Carter1992}.  
This will result in the Euler 
number
of the normal bundle.  
Note that if a global choice of tensor $m\ind{_a^b}$ can be found that transforms $A\ind{_a_b^c}$
to zero everywhere, as discussed in \ref{sec:invten}, equation (\ref{eqn:Ocd}) says that 
the outer curvature is globally exact on the surface.  
This would imply that the Euler number vanishes, being the integral of an exact form.  
When $A\ind{_a_b^c}$ cannot be set to zero everywhere and equation (\ref{eqn:OcdA})
is used to compute the Euler number, all terms involving $F_d$ are exact and drop
out of the integral.  Hence, the acceleration tensor completely determines the Euler
number in this case, and, conversely, the Euler number represents an obstruction
to setting $A\ind{_a_b^c}$ to zero everywhere on the surface. 

\section{Boundary term in gravitational Hamiltonian} \label{sec:hamiltonian}

One application of this formalism is in analyzing boundary terms of Hamiltonians 
that arise in the 
covariant canonical analysis of general relativity. 
Given a finite subregion defined by a hypersurface $\Sigma$ with boundary $\partial \Sigma$,
one can form a symplectic form $\Omega$ associated with the subregion by integrating 
the symplectic current $(d-1)$-form $\omega$ over the surface,
\beq
\Omega = \int_\Sigma\omega.
\eeq
The symplectic current is a $2$-form on field space, so that $\delta g_{ab}$ appears quadratically,
and it is constructed from the field space exterior derivative of a symplectic potential $\theta$,
a spacetime $(d-1)$-form.  
This potential arises as the boundary term in the variation of the Einstein-Hilbert Lagrangian
for general relativity, $L = \frac{1}{16\pi G}  \ep R$,  through the equation
\beq
\delta L = E^{ab}\delta g_{ab} + \dd \theta,
\eeq
where $E^{ab}$ are the field equations, and the expression for $\theta$ is
\cite{Iyer1994a} 
\beq \label{eqn:thetadef}
\theta = 2\ep_a E^{abcd}\nabla_d \delta g_{bc},
\qquad E^{abcd} = \frac{1}{32\pi G}(g^{ac}g^{bd} - g^{ad} g^{bc}).
\eeq

Time evolution along the flow of a vector field $\xi^a$ should be generated by a Hamiltonian
$H_\xi$ on the phase space.  This Hamiltonian is required to satisfy \
\beq
\delta H_\xi = \int_{\partial \Sigma}(\delta Q_\xi  - i_\xi \theta),
\eeq
where $Q_\xi$ is the Noether charge \cite{Iyer1994a}, and $i_\xi$ denotes 
contraction of the vector $\xi^a$ into a differential form.
This formula assumes that the vector field $\xi^a$ is independent of the dynamical fields,
$\delta \xi^a = 0$, and 
we also assume that the embedding $X$ is fixed, so there are no contributions
from $\rchi^a$ (relaxing these constraints will be discussed later).  
In general, if $\xi^a$ has a transverse component to $\partial\Sigma$, 
this equation has no solutions since 
$\theta$ is not a total variation \cite{Wald2000b}.  
However, one can look for boundary conditions to impose on the 
fields so that $i_\xi\theta = \delta B_\xi$ for some $B_\xi$.  In this case the Hamiltonian is equal
to 
$\int_{\partial\Sigma} (Q_\xi - B_\xi)$,
up to a constant.  

To classify the possible boundary conditions, it is helpful to decompose 
$i_\xi \theta\big|_{\partial\Sigma}$ in terms of the geometric quantities associated with the 
surface $\partial\Sigma$.  First, note that only the normal component of $\xi^a$ is relevant,
since any tangential component contracting with $\theta$ will lead to a form that vanishes 
when restricted to $\partial\Sigma$.  Then applying $\ep = - n\wedge \mu$ 
 to the formula (\ref{eqn:thetadef}) gives
\beq \label{eqn:ixitheta}
i_\xi\theta\big|_{\partial\Sigma} = - \frac{\mu}{16\pi G} \xi^e n_{ae} (s^{ac}g^{bd}-s^{ad}g^{bc})
\nabla_d\delta g_{bc}.
\eeq
It is useful now to use the decomposition of the metric variation (\ref{eqn:delgdecomp}),
which can be applied to $\delta g_{ab}$ as opposed to the pullback $\delta \p g_{ab}$
because the embedding is held fixed. First, the terms involving 
$\delta s_{ab}$ are
\begin{align}
n\ind{^c_e}\left(s^{bd} (\Eth_d  \delta s_{bc} - \Eth_c\delta s_{bd}) 
+ K^b\delta s_{bc} \right) = n^{cd}\Eth_d\delta s_{ec} + n\ind{^c_e}K^b\delta s_{bc}.
\end{align}
Next, the contribution involving $\nabla_d(h_{mb}\delta h\ind{^m_c} + h_{mc}\delta h\ind{^m_b})$
is computed noting that the contraction with $g^{bc}$ gives zero since
$\delta h\ind{^m_c}$ is normal on its lower index.  The remaining terms involving 
$\delta h\ind{^m_c}$ are
\beq
n\ind{^c_e}\left(A_m \delta h\ind{^m_c} + L\ind{_m^b_c}\delta h\ind{^m_b} + \Dt_d\delta h\ind{^d_c}
\right)
\eeq
Finally, the $\delta h^{mn}$ terms are 
\begin{align}
n\ind{^c_e}\left(K_{cmn}\delta h^{mn} + \Eth_c(h_{mn} \delta h^{mn})\right)
= n\ind{^c_e}\left(K_{cbd}\delta h^{bd} - 2 \delta K_c -2 \Dt_b\delta h\ind{^b_c} 
-2L\ind{_m_c^b}\delta h\ind{^m_b}\right),
\end{align}
where the second expression employed the formula (\ref{eqn:kappa}) for the variation
of $K_c$.  
Combining these expressions in (\ref{eqn:ixitheta}) leads to
\begin{align}
i_\xi\theta\big|_{\partial\Sigma} = 
\frac{\mu}{16\pi G}\xi^en\ind{^c_e}\Big[  \;&
\delta K_c + s_{mc}h^{bd}\delta K\ind{^m_b_d}
+ \Dt_b\delta h\ind{^b_c} 
 \nonumber \\
& 
-s^{bd}(\Eth_d\delta s_{bc}-\Eth_c\delta s_{bd}) 
+ 2 L\ind{_d_c^a} \delta h\ind{^d_a} 
- A_b\delta h\ind{^b_c} - L\ind{_b^d_c}\delta h\ind{^b_d}
\Big]
\label{eqn:ixithetadSigma}
\end{align}
Each term on the first line of this expression consists of tensors 
that are invariant with respect to changes in the foliation away from $\partial\Sigma$,
according to the discussion in section \ref{sec:invten}.  The terms on the second line are not 
individually invariant, due to the appearance of normal covariant derivatives $\Eth_b$ and 
normal extrinsic curvatures $L\ind{_d_c^e}$;  however, together they form an invariant
object, since 
\beq
-s^{bd}(\Eth_d\delta s_{bc}-\Eth_c\delta s_{bd}) 
+ 2 L\ind{_d_c^e} \delta h\ind{^d_e} 
- A_b\delta h\ind{^b_c} - L\ind{_b^d_c}\delta h\ind{^b_d}
= -s^{bd}(\nabla_d\delta g_{bc} - \nabla_c\delta g_{bd}).
\eeq

The expression (\ref{eqn:ixithetadSigma}) can be organized into a somewhat simpler
form by introducing $\delta \gimel^a_{bc}$, the variation of the connection coefficients
for $\Eth_a$.  $\gimel^a_{bc}$ is defined in appendix 
\ref{app:conncoefs} through equation (\ref{eqn:gimeldef}),
and, in a coordinate system adapted to the foliation, it is given by the normal projection of the 
Christoffel symbols, as in equation (\ref{eqn:gimelGamma}).  Its variation satisfies
\beq
s\ind{^d_a}\delta\gimel^a_{bc} = -\delta h\ind{^e_b}L\ind{_e_c^d} 
- \delta h\ind{^e_c}L\ind{_e_b^d} +s\ind{^d_a} s\ind{^e_b}s\ind{^f_c}\delta\Gamma^a_{ef},
\eeq
where the relation (\ref{eqn:GammaL}) was used.  One can then calculate that
\beq
n\ind{^b_a}\delta\gimel^a_{be} = n\ind{^c_e}
\left[-s^{bd}(\Eth_d\delta s_{bc} - \Eth_c\delta s_{bd})
+2A\ind{_d_c^a}\delta h\ind{^d_a} - 2 A_b\delta h\ind{^b_d} \right]
\eeq
so that (\ref{eqn:ixithetadSigma}) becomes
\beq \label{eqn:ixithetasimp}
i_\xi\theta\big|_{\partial\Sigma} = \frac{\mu}{16\pi G} \xi^e\Big[
n\ind{^c_e}\left(\delta K_c + s_{mc}h^{bd}\delta K\ind{^m_b_d} + \Dt_b\delta h\ind{^b_c}
+A_d\delta h\ind{^d_c} - L\ind{_b^d_c}\delta h\ind{^b_d}\right)
+n\ind{^b_a}\delta\gimel^a_{be} +s\ind{^d_e}\delta F_d 
\Big]
\eeq

Since $\theta$ is defined as the boundary term obtained when varying the Lagrangian, 
it is ambiguous by the addition of an exact form, $\theta\rightarrow \theta + \dd\beta$.  
The presence of the divergence term $\Dt_b\delta h\ind{^b_c}$ in (\ref{eqn:ixithetadSigma})
suggest that some terms may be canceled by the appropriate choice of $\beta$.  This is 
indeed the case, and coincides with the natural choice for this ambiguity suggested in
\cite{Speranza2018}:
\beq \label{eqn:beta}
\beta_{e_2\ldots e_{d-1}} = \frac1{16\pi G}\ep\ind{^c_b_{e_2\ldots e_{d-1}} }\delta h\ind{^b_c}
\eeq 
gives
\beq
(\dd\beta)_{e_1\ldots e_{d-1}} = \frac1{16\pi G}\left[
\ep\ind{^c_b_{e_2\ldots e_{d-1}}} \nabla_{e_1}\delta h\ind{^b_c}
-(d-2)\ep\ind{^c_b_{e_1}_{\un{e_3\ldots e_{d-1}}}} 
\nabla\ind{_{\un{e_2\vphantom{e_{d-1}} }}} \delta h\ind{^b_c}
\right].
\eeq
Contracting with the normal component of 
$\xi^{e_1}$ and restricting the form to $\partial\Sigma$ forces
the indices $e_2,\ldots, e_{d-1}$ to be tangential.  In the second term above, the only nonzero
contribution then comes from $c$ normal and $b$ tangential in 
$\ep\ind{^c_b_{e_1}_{\un{e_3\ldots e_{d-1}}}}$, and the expression simplifies to
\begin{align}
i_\xi \dd \beta\big|_{\partial\Sigma} &= \frac{\mu}{16\pi G} \xi^e
\left[-n\ind{^c_e}\Dt_b\delta h\ind{^b_c} 
-n^{cb}L_{deb}\delta h\ind{^d_c}\right] \nonumber \\
&= %
\frac{\mu}{16\pi G} \xi^e n\ind{^c_e}
\left[  
-\Dt_b\delta h^b_c - A_d\delta h\ind{^d_c}+L\ind{_d^b_c}\delta h\ind{^d_b}
\right].
\label{eqn:ixidbeta}
\end{align}
Adding this term to (\ref{eqn:ixithetasimp}) then gives
\beq \label{eqn:ixithetadbeta}
i_\xi(\theta+\dd\beta)\big|_{\partial\Sigma} = \frac{\mu}{16\pi G} \xi^e
\Big[
n\ind{^c_e}\left( \delta K_c +s_{mc}h^{bd}\delta K\ind{^m_b_d} \right)
+ n\ind{^b_a}\delta\gimel^a_{be} +s\ind{^d_e}\delta F_d
\Big]
\eeq

This gives a fairy simple expression for the additional boundary contribution
to the Hamiltonian associated with the flow along $\xi^e$; however, a word
of caution is in order.  Although $\beta$ defined in (\ref{eqn:beta}) is an invariant form
on the surface, its spacetime exterior derivative is not, since it involves normal derivatives.  
The expression (\ref{eqn:ixidbeta}) is similarly not invariant, which is easily seen due to the 
explicit appearance of $L_{deb}$ in the first line.  Since the original boundary term 
(\ref{eqn:ixithetadSigma}) is invariant, the modified one 
(\ref{eqn:ixithetadbeta}) necessarily is not.  Although simpler to analyze than
(\ref{eqn:ixithetadSigma}), one should keep in mind that 
quantities in equation (\ref{eqn:ixithetadbeta}) depend on how the foliation is extended 
away from $\partial\Sigma$. Other choices for the ambiguity term $\beta$ generally share this
feature of breaking refoliation-invariance of $i_\xi\theta$, unless $\beta$ is constructed
solely from spacetime-covariant fields, such as $g_{ab}$ and $\delta g_{ab}$, as opposed
to tensors associated with the surface, such as $\delta h\ind{^a_b}$.  
For pure general relativity, there are no such choices for $\beta$ without using derivatives
of $g_{ab}$ and $\delta g_{ab}$, hence the expression (\ref{eqn:ixithetadSigma}) is the unique
refoliation-invariant
choice where all terms involve only one derivative
of the metric or its variation.

From here, one could classify the possible boundary conditions that allow (\ref{eqn:ixithetadSigma})
or (\ref{eqn:ixithetadbeta}) to be written as a total variation.  We will not attempt this 
general analysis here, although this problem has been partially considered before.  In particular,
\cite{Camps2019} found the necessary boundary conditions under the assumption of fixed 
normal metric $\delta s_{ab} = 0$ and a partial fixing of the the tangential projector variation 
$\delta h\ind{^a_b}$.   These boundary conditions turn out to be quite stringent:
they require either a fixed volume form $\delta\mu = 0$ or $K_c = 0$, and also 
fixing the traceless extrinsic curvature $\delta\tilde K\ind{^a_b_c} = 0$ or  
fixing the conformal class of the induced metric 
$\delta h_{ab} - \frac{1}{d-2} h_{ab}h^{cd}\delta h_{cd} = 0$.  Similar classifications 
of boundary conditions have appeared in \cite{Jafferis2016}.

For a finite subregion, it can appear overly restrictive to try to demand that this Hamiltonian
be integrable for a vector field that generates a diffeomorphism transverse to the surface.  This
is because one would expect symplectic flux to leak out if the surface is moved to a different
location.  However, another application for the above analysis is in the  considerations of 
diffeomorphism edge modes, which characterize the gauge degrees of freedom that become
physical in the presence of the fixed surface $\partial\Sigma$ \cite{Donnelly2016F}.  
In order to 
build up a larger phase space by assembling phase spaces associated with subregions, 
these edge modes are necessary in order to implement gauge constraints in the larger space
through a symplectic reduction procedure.  This procedure requires a symmetry algebra to 
act on the edge modes as Hamiltonian transformations in the local phase space, and hence
it is important that integrable Hamiltonians can be found for these transformations, including
the diffeomorphisms transverse to $\partial\Sigma$.  

The edge mode degrees of freedom are contained in the embedding fields $X$, 
and the Hamiltonian for a diffeomorphism in the reference space must satisfy
\beq
\delta \p H_{\p\xi} = \int_{\partial\p\Sigma} (\delta \p Q_{\p \xi} - i_{\p\xi}\p\theta)
\eeq
where $\p\theta$ is now a function of the pulled back metric variation $\delta \p g_{ab}$.  
This case now reduces to the same analysis as before, and the conclusion that possibly
overly strong boundary conditions are necessary.  However, one can generalize the allowed
symmetry transformations by letting the generator $\p \xi^a$  depend on the dynamical
fields, so that $\delta\p\xi^a \neq 0$. For such a field-dependent generator, the 
Hamiltonian variation is instead
\beq
\delta \p H_{\p \xi} = \int_{\partial\Sigma} (\delta\p Q_{\p \xi} - \p Q_{\delta \p\xi} - i_{\p\xi}\p\theta),
\eeq
and the extra freedom in $\delta\p\xi^a$ makes it plausible that this equation has solutions 
without overly restrictive boundary conditions on the fields.  The algebra satisfied by these 
field-dependent generators is modified from the Lie bracket of vector fields on spacetime to
the Lie bracket of the associated vector fields on field space, given in
\cite{Barnich2010, Barnich2010a, Gomes2019}.
Additionally, when the algebra is represented with Poisson brackets of the Hamiltonians
$\p H_{\p\xi}$, it can acquire central extensions.  
Examples of field-dependent generators leading to central charges include
the Brown-Henneaux analysis of asymptotically $\text{AdS}_3$ gravity \cite{Brown1986b},
Carlip's work on near horizon symmetry algebras of black holes \cite{Carlip1999a,
Carlip2017},
and the Barnich-Troessart analysis of the extended BMS algebras of asymptotically 
flat space \cite{Barnich2010, Barnich2010b}.  
We leave further analysis of such
field-dependent generators and their algebras to future work.

\section{Discussion} \label{sec:conclusion} 

We conclude with a discussion of possible generalizations of the above constructions with 
embedding fields and foliations, and point to additional applications for this formalism.  

\subsection{Null surfaces}\label{sec:null}

A notable deficiency in the formalism developed in this work is that it cannot 
be applied unmodified to null surfaces.  This is unfortunate, since numerous recent
works on asymptotic symmetries \cite{Strominger2014a, 
Barnich2010, Barnich2010b, Flanagan2017},
 actions for local subregions \cite{Brown2016, Lehner2016,Jubb2016}, 
 and degrees of freedom on null Cauchy surfaces \cite{Reisenberger2018,
 Hopfmuller2017, Hopfmuller2018}, among many others, all utilize null surfaces or null foliations.
There are a number of obstacles in trying to adapt the submanifold calculus of
this paper to null surfaces.
First, although a null surface can still be defined by a normal form $\nu$, it is no longer
possible to form a unit normal as in equation (\ref{eqn:nNnu}), 
since nullness means that $\nu$ has norm zero. One can still work with $\nu$ directly
as an unnormalized normal form, but it should not be considered an invariant tensor
of the folaition, since rescaling $\nu$ defines the same foliation.   

The ability to normalize $\nu$ led to expressions for normal and tangential projectors,
$s\ind{^a_b}$ and $h\ind{^a_b}$, which defined  canonical decompositions of all vectors and 
covectors into normal and tangential parts.  With a null surface $\Sigma$, no such projectors
are available, and the decompositions of the tangent and cotangent spaces are more complicated.
For vectors, the subspace of tangential vectors, $T\Sigma$, defined as all vectors which
annihilate $\nu$ upon contraction, is well-defined and independent of the metric.  
Since $\Sigma$ is a null surface, there is a one-dimensional subspace in $T\Sigma$ 
consisting of all null vectors tangent to $\Sigma$.  Call this the  lightlike subspace $L\Sigma$, 
and let $k^a$ denote a generic null vector in this space.  The metric can be used to define
the orthogonal complement $L\Sigma^\perp$, the subspace of all vectors with zero inner
product with $k^a$.  These spaces are related to each other by the following inclusions
\beq
L\Sigma\subseteq T\Sigma \subseteq L\Sigma^\perp
\eeq
The second inclusion reflects the fact that $k_a$ is a normal one-form, so that $\nu = k\wedge
\upsilon$ for some $(p-1)$-form $\upsilon$, and hence all tangential vectors must annihilate
$k_a$.  When dealing with a congruence of null curves ($p=d-1$), the first inclusion
is an equality, whereas for a null hypersurface ($p = 1$), the second inclusion is an equality.

$L\Sigma^\perp$ is the largest subspace in this discussion, and is always $(d-1)$-dimensional.
Hence, there are still vectors that do not lie in any of these subspaces, and there is 
no canonical subspace to associate with the vectors transverse to $L\Sigma^\perp$.  
Instead, one must work with quotients of vector spaces.  Two quotient spaces that are particularly
relevant are the transverse space $TR = TM/T\Sigma$, where two vectors are equivalent if they
differ by a tangential vector, and the spatial space $S = T\Sigma/L\Sigma$, consisting of 
tangential vectors to $\Sigma$ defined modulo arbitrary multiples of $k^a$.  

Similar subtleties exist for the decomposition of the cotangent space $T^*M$.  There is a 
natural one-dimensional lightlike subspace $L^*\Sigma$ generated by $k_a$, which is contained 
in the normal subspace $N^*\Sigma$ defined by $\nu$.  
Finally, there is $(d-1)$-dimensional horizontal subspace $H^*$ consisting of all covectors
that annihilate $k^a$ (which implies $H^* = L^*\Sigma^\perp$), and these
subspaces satisfy the inclusions
\beq
L^*\Sigma\subseteq N^*\Sigma \subseteq H^*.
\eeq
Equality occurs in the first inclusion for the case of a null hypersurface, and in the second in the 
case of a null congruence of curves.  Again, one must form quotients when performing 
general decompositions of covectors.  Important quotients are the tangential space $T^*\Sigma
 = T^*M/N^*\Sigma$, consisting of covectors defined modulo normal covectors
 which should serve as the cotangent bundle for the null surface, and 
 the transverse space $H^*/L^*\Sigma$, which is used in discussions of geodesic deviation
 for null congruences.  

Quotient spaces complicate the description of geometric quantities on the surface, since tensors
defined in the quotient space do not give rise to well-defined tensors on $M$.  
Instead, there are conditions a spacetime tensor 
must satisfy in order for it to define an unambiguous
tensor on the quotient space of interest
\cite[pg.~222]{Wald1984}.  

Another issue related to the lack of a projector is that not all of the subspaces or 
quotient spaces come equipped with a metric.  For example, the 
spacetime metric $g_{ab}$ does not induce a metric on the 
subspace $L\Sigma$ generated
by the null vector $k^a$ on $\Sigma$, since the only possible inner product vanishes, 
$k^a k^b g_{ab}=0$.  There is a metric induced on $T\Sigma$, but it is degenerate, which 
means, for one, that a metric-compatible connection on $T\Sigma$ is not uniquely defined, 
and
in fact  a torsion-free one does not exist in general.  
Without such an object, it is more difficult to discuss intrinsic  and extrinsic curvatures
for the null surface and to formulate analogs of the Gauss, Ricci-Voss, and Codazzi identities.  

In addition to these obstructions to standard geometric analyses of null surfaces, there 
are also problems related to the constraint nullness puts on metric variations.  For  
timelike or spacelike surfaces, one is free to vary the spacetime metric and the embedding
fields independently, since the norm of $\p\nu$ is not fixed, but merely required to be positive
or negative.  On the other hand,  null surfaces require the 
norm of $\p\nu$ to be zero, which, after imposing that
$\p\nu$ does not vary, necessarily restricts the allowed variations of the metric.  
In the case of a hypersurface, this condition is  $\delta \p g^{ab} \p\nu_a \p\nu_b
= X^*(\delta g^{ab} + \lie_\rchi g^{ab}) \p \nu_a \p\nu_b = 0$, which relates some components
of the metric variation to the variation of the embedding
described by $\rchi^a$.  Although 
one could choose to work only with embedding maps $X$ and metrics $g_{ab}$ that maintain
nullness of $\p\nu$, it is somewhat contrary to the perspective of this work where the two
objects are meant to be varied independently.  

Imposing nullness by hand can also lead to 
more severe restrictions on the class of metrics under consideration.  For example, 
null hypersurfaces are  generated by geodesics, and therefore 
generically develop caustics and crossovers.  These 
would  be reflected in a singularity
in either the embedding or  the normal form $\p\nu$.  By not allowing for such singularities, 
one is making a strong assumption about the metric, in that it admits a complete, caustic-free,
null hypersurface.

Clearly a different formalism is required in order to discuss the geometry and dynamics of 
null surfaces and foliations.  This section concludes by offering a number of suggestions on how
the various issues outlined above might be addressed, and leaves full analysis of these
possibilities to future work.  

A common way of dealing with the necessary appearance of quotient spaces is to 
introduce additional structures that allow these quotients to be canonically identified
with subspaces of the spacetime tangent and cotangent bundles.  One way of doing 
this for the null surface is to choose an arbitrary slicing of the null surface by
spatial sections.  Doing so essentially converts the problem to the analysis 
of a foliation by spatial submanifolds with codimension one higher than the null
surface.  One can then define the intrinsic and extrinsic geometry of this spatial foliation,
and many quantities computed in this way can be argued to produce well-defined tensors
on the quotient spaces.  For many applications, this construction is sufficient,
although the issue of caustics and the assumed regularity of the null surface is still present.  

A related approach is to define a Carroll structure on the null surface
\cite{Duval2014}.
This is simply a choice of preferred spatial subspace to identify with the spatial quotient
space by way of an Ehresmann connection \cite{Leigh2019}.  The Carroll structure is more general than the choice of spatial slicing 
since the spatial subspaces are not required to be integrable.  A related construction
parameterizes the transverse space to $\Sigma$ by choosing an auxiliary 
null vector $\ell^a$ that satisfies $\ell^a k_a = -1$, in terms of which the metric
can be fully decomposed (see e.g.\ \cite{Poisson2007}).  An interesting question in these 
approaches is whether these additional structures can be used to fix a unique connection 
$\Dt_a$ on $\Sigma$.  Note that it is 
generically not possible to require that $\Dt_a$ be compatible with the degenerate 
induced metric on the surface.\footnote{I thank Rafael Sorkin for pointing this out.} 
It seems possible that once a preferred $k^a$ and additional
Carroll or transverse structure has been chosen, a connection could be fixed by imposing
 compatibility conditions with these structures (see, e.g.\ \cite{Hartong2015}).  
 
The issue of nullness overconstraining the metric variations could possibly be addressed 
by only imposing that $\p\nu$ is null in the background, and not requiring that variations
preserve nullness.
This approach was advocated for recently in \cite{Jafari2019}, which provided 
a physical interpretation for metric variations that do not preserve nullness of the surface.  
Note there are still challenges in this case related to quotient spaces and the inability to construct
projectors.  Furthermore, variations of the metric could now produce a surface whose
induced metric changes signature, which can cause the intrinsic geometry to look singular.
Such problems would have to be addressed when pursuing this construction, perhaps
along the lines of \cite{Mars2001}.

Finally, a possible way forward with the problem of caustics and crossovers would be 
to consider embeddings into the cotangent bundle $T^*M$ or a tensor product thereof, 
rather than into the manifold 
$M$ \cite{Friedrich1983}.  
The image of the embedding  specifies both the location of the surface and its 
normal form.  Hence, surfaces with caustics or crossovers can look singular when embedded 
in $M$, but are perfectly smooth as surfaces in $T^*M$.  There are drawbacks in this approach
in that it allows for immersed surfaces in $M$ as opposed to embeddings, since the surfaces
can now intersect themselves.  While this raises additional complications, it seems 
to be a promising way forward in analyzing the geometry of generic null surfaces.

\subsection{Hamiltonians for field-dependent generators and asymptotic symmetries}

Section \ref{sec:hamiltonian} examined boundary terms that appear when
defining a Hamiltonian in general relativity
for transformations that move the boundary of a finite subregion.  Although the 
conditions for which this Hamiltonian is integrable were not rigorously analyzed,
it was clear that the necessary conditions are likely overconstraining in that most
metric components must be held fixed.  An alternative was suggested 
that  allows for the vector field generating the flow to be field-dependent,
$\delta\p\xi^a\neq0$, to allow
for more freedom in imposing integrability of the Hamiltonian. Many analyses of 
asymptotic symmetries employ such field-dependent generators, since they are 
necessary when attempting to preserve a privileged asymptotic structure.
Note that field-dependent generators can lead to algebras that differ from the Lie algebra
of the vector fields on $M$.  This occurs for two reasons.  First, the Lie bracket is 
explicitly modified due to the field dependence \cite{Barnich2010}.  Second,
representing the algebra as a Poisson bracket of Hamiltonians can give rise to 
central extensions, as found in, e.g., \cite{Brown1986b,Carlip1999a,
Carlip2017,Barnich2010, Barnich2010b}.  
Together, these modifications of the algebra can have important implications 
when one considers quantizations of the theory in question.  The case of asymptotic
symmetries of $\text{AdS}_3$ gravity is a renowned example, where the central 
charge of the asymptotic Virasoro algebra is related to the entropy of 
black holes in the theory \cite{Strominger1998}.

\subsection{Edge modes and entanglement entropy}

Another reason to consider field-dependent generators comes from the application
of embedding fields to the problem of entanglement entropy in a gravitational theory.  
The $X$ field encapsulates the extra edge mode degrees of freedom that must be 
incorporated into a finite subregion in order to properly implement gauge-invariance 
when gluing to an adjacent subregion \cite{Donnelly2016F}.  Variations of the embedding 
transverse to the entangling surface are 
among the degrees of freedom encoded in $X$.  In order to analyze the contribution
of the edge modes to the entanglement entropy, one must have a handle on their quantization,
for which their symmetry algebra plays an important role.  It may be necessary
to include the surface deformations in the symmetry algebra
to account for the full diffeomorphism invariance, and such a symmetry algebra
will likely involve field-dependent generators to avoid overconstraining boundary conditions
on the metric.  

\subsection{Boundary terms in gravitational actions}

An application of the general geometric framework developed in this paper is 
to the problem of finding boundary terms in the action of higher curvature gravitational theories.  
These boundary terms are added to ensure that the action is stationary only for 
field variations that satisfy the appropriate boundary conditions.   For general relativity,
the Gibbons-Hawking boundary term is added on spacelike or timelike codimension-1 boundaries
to implement a Dirichlet condition on the induced metric at the boundary \cite{Gibbons1977b}.
Additional contributions must be added when
the region has null components or corners of higher codimension, as 
were recently analyzed in detail in \cite{Lehner2016, Jubb2016}.  For higher curvature
theories,  less is known about the necessary boundary terms, although see
\cite{Myers1987, Dyer2008, Cremonini2010, Bueno2016} for some results in this area.  
One reason these theories 
are more complicated is that the presence of higher derivatives in the 
field equations  entails
additional boundary conditions on derivatives of the fields 
in the variation principal.  Nevertheless, a general analysis of 
boundary terms and boundary conditions in these theories is lacking at present, and 
the systematic decompositions described in this work may lend themselves to 
addressing this problem in the future.  

\subsection{Perturbations of RT surfaces}

RT surfaces are important objects in holography since they bound subregions in the 
bulk that are dual to corresponding subregions in the boundary \cite{Dong2016a}.  
Perturbations of these surfaces in response to a change in the boundary subregion or 
the bulk dynamical fields are of interest, since these can be used in
perturbative calculations of entanglement entropy \cite{Hung2011b, Casini2016}, proofs of 
CFT energy
conditions \cite{Kelly2014, Koeller2016, 
Akers2017, 
Leichenauer2018}, and considerations of bulk reconstruction 
\cite{Lewkowycz2018a, Bao2019}.  The embedding fields provide a covariant description of these 
perturbations, and perturbations to nearby extremal surfaces are controlled by 
solutions to the Jacobi equation, discussed in section \ref{sec:Jacobi} (see
\cite{Mosk2018, Ghosh2018} for a perturbative analyses of solutions to this equation).
Recently, \cite{Engelhardt2019} used a similar submanifold formalism to the one developed in 
this paper to 
 recast and extend some of these holographic proofs involving RT surfaces
in a covariant language.  
Further application of the covariant formalism
may lead to deeper understanding
of the mechanisms at play in these holographic constructions,
and suggest possible generalizations.

\subsection{Magnetohydrodynamics and force-free electrodynamics}

A final application of the embedding field formalism is to the theory of relativistic
magnetohydrodynamics (MHD) and force-free electrodynamics (FFE).  This application is similar to 
the use of embedding fields in fluid dynamics.  It differs, however, in that the 
foliation is by two-dimensional timelike manifolds, as opposed to the one-dimensional flow
lines.  MHD and FFE are thus theories of a string fluid.  

This foliation arises since the abundance of free charges tends to short out any electric
fields in the rest frame of the fluid, implying that the electromagnetic field tensor satisfies
$F_{ab} u^b = 0$, where $u^b$ is the fluid velocity.  Since $u^a$ is a zero eigenvector
of $F_{ab}$, the field strength must have rank 2, and is also closed, $\dd F = 0$.  Hence,
$F_{ab}$ is normal to a foliation of two-dimensional manifolds according
to the discussion of \ref{sec:foliations}, which comprise the strings
of the fluid description.
In fact, the entire theory can be recast as the theory of this foliation using embedding 
fields and a fixed two-form in a reference space $M_0$ \cite{Gralla2014}.  
  An intriguing relation geometrical relation holds for the 
electromagnetic stress tensor, which for a degenerate field takes the form \cite{Gralla2014}
\beq
T^{\text{EM}}_{ab} = \frac14 F^{cd}F_{cd}\big( s_{ab} - h_{ab}\big),
\eeq
being proportional to the difference of the normal and tangential metrics of the foliation. 
It could be useful to express other geometrical tensors associated with the foliation in 
terms of the electromagnetic field strength, which could lead to deeper intuition 
for properties of MHD and FFE solutions \cite{Compere2016a}.

The string fluid is also the starting point for the effect field theory of MHD
and FFE, where the foliation 
 has an interpretation in terms of the generalized global 
symmetries \cite{Gaiotto2015}.  
The one-form symmetry of electromagnetism has a conserved charge that is integrated over 
codimension-2 surfaces, and this flux counts the number of strings passing through
the surface, which are the fundamental charged paritcles for the one-form symmetry
\cite{Grozdanov2017}.  This is analogous to the viewpoint that the charge of an ordinary
symmetry is the integral over a spacelike hypersuface, and the flux counts the number 
of charged particle worldlines piercing the surface.  The effective field theory for the 
one-form symmetry of electromagnetism should arise from the most general action
consistent with this symmetry, and the string fluid and embedding fields provide
an efficient way of writing terms that are manifestly symmetric
\cite{Grozdanov2017, Gralla2018, Armas2018}.

Force-free electrodynamics is known to be dynamically well-posed if the field strength
is magnetically dominated, $B^2> E^2$, or equivalently, $F_{ab}F^{ab} > 0$.  
There is considerable interest in the question
of how and when this condition breaks down, signaling the onset of an instability
which can have observable astrophysical consequences.
At saturation, $B^2=E^2$, the foliation becomes null, implying that a singularity develops 
in its intrinsic geometry.  Additionally, magnetic reconnection events are associated with 
crossing field lines, which can be interpreted as the development
of a caustic from the foliation perspective.  In order to better characterize how these
breakdowns occur, it would be helpful to develop singularity and focusing theorems for the 
intrinsic and extrinsic geometry of the field sheets.  A possible useful tool
would be a generalization of the Raychaudhuri equation that applies to the two
dimensional foliation.  The starting point for such a generalization would be 
equation (\ref{eqn:mixedCodazzi}), which relates the tangential derivative 
of $L\ind{_c_a_d}$ to other quantities on the surface.  Some initial analysis of 
this expression, interpreted as a generalized Raychaudhuri equation, has 
appeared in \cite{Capovilla1995a}.  Further analysis of the force-free 
equations using the geometric tools developed in this work could lead interesting
and useful results.  

\section*{Acknowledgments}

I would like to thank Joan Camps, William Donnelly, Netta Engelhardt, Sebastian Fischetti, 
Laurent Freidel, 
Stephen Green, 
Theo Johnson-Freyd, Rob Leigh, 
Aitor Lewkowycz, Rafael Sorkin, and Wolfgang Wieland
for helpful discussions.  
This research was supported in part by Perimeter Institute for Theoretical Physics. Research at 
Perimeter Institute is supported by the Government of Canada through the Department of 
Innovation, Science and Economic Development Canada and by the Province of Ontario through 
the Ministry of Economic Development, Job Creation and Trade. 
\appendix

\section{Curvature identities for $\Db_a$, $\Eth_a$}
\label{app:curvature}

As discussed in section \ref{sec:gauss}, there are curvature tensors $\mathcal{R}_{abcd}$ and 
$\out_{abcd}$ associated with  the commutator of two tangential
covariant derivatives $[\Dt_c , \Dt_d]$.  These are 
related to the spacetime Riemann curvature and 
extrinsic curvature tensors through the Gauss and Ricci-Voss equations.  Additionally, the
Codazzi equation arises from the requirement that $[\Dt_c, \Dt_d]V^a$ remain tangential on 
$a$ if $V^a$ is itself tangential.  These three equations relate certain components 
of $R_{abcd}$ to intrinsic and extrinsic geometrical quantities at the surface.  The specific 
components of $R_{abcd}$ covered by these identities are those with all indices tangential, 
those with three indices tangential and one normal, and those with the first two indices normal
and second two tangential (or equivalently, second two normal and first two tangential).  

A natural generalization of these equations comes from the commutator 
of two normal covariant derivatives $[\Eth_a ,\Eth_b]$.  This case is slightly more subtle 
due to the presence of torsion-like contributions; however, one can find a tensorial expression 
that can be called the curvature of the normal connection, along with the analog of the 
outer curvature.  These are then related to the purely normal components of $R_{abcd}$ and 
its tangent-tangent-normal-normal components.  An analog of the Codazzi equation
gives a relation between $L_{abc}$ and the components of $R_{abcd}$ with three normal and 
one tangential index.  

The remaining components of $R_{abcd}$ that are not constrained by these equations are 
normal on $a$ and $c$ and tangential on $b$ and $d$.  The desired relation on these components
is obtained by considering the commutator of one tangential and one normal covariant derivative,
$[\Eth_c, \Dt_d]$.  Mixed curvature tensors associated with this commutator can be constructed,
which are related to $K_{abc}$ and $L_{abc}$.  Finally, by examining the Codazzi
identity for this commutator, a number of additional relations between 
the various curvature quantities can be derived, one of which relates the remaining components of 
$R_{abcd}$ to $K_{abc}$ and $L_{abc}$.  

Carrying out the above program leads  to  six curvature tensors associated with the 
normal and tangential covariant derivatives, which are related in various ways to 
$R_{abcd}$, $K_{abc}$ and $L_{abc}$.  In addition, three Codazzi identities lead to further 
relationships between $R_{abcd}$, $K_{abc}$ and $L_{abc}$.  This appendix  systematically
derives these identities and works out some of their consequences.  
An early treatment of many of the concepts presented in this appendix was given
by Schouten in \cite[Ch. V, Sec. 7]{Schouten1954}.

\subsection{Tangential curvatures} \label{app:tangcurv}

The intrinsic tangential curvature $\mathcal{R}\indices{^a_b_c_d}$ 
is defined by $[\Dt_c, \Dt_d]$ acting on a tangential vector $V^a$
by the equation
\beq \label{eqn:DDV}
(\Dt_c \Dt_d -\Dt_d \Dt_c) V^a = \mathcal{R}\indices{^a_b_c_d}V^b.
\eeq
The Gauss equation for this curvature can be derived by simply writing out the 
definition for the tangential covariant derivative,
\begin{align}
\Dt_c \Dt_d V^a 
&= %
 h\indices{^p_c} h\indices{^q_d} h\indices{^a_m} \nabla_p(\Dt_q V^m)
\nonumber \\
&= %
h\indices{^p_c} h\indices{^q_d} h\indices{^a_m}\nabla_p(h\indices{^e_q}\nabla_eV^m
+ K\indices{^m_q_b} V^b ) \nonumber \\
&= %
-K\indices{^e_c_d} \Eth_e V^a
+ h\indices{^p_c}h\indices{^q_d} h\indices{^a_m}\nabla_p\nabla_q
V^m + K\indices{_m_c^a} K\indices{^m_d_b}V^b.
\label{eqn:DDVdeldelV}
\end{align}
Subtracting $\Dt_d \Dt_c V^a$ from this expression and recalling that $K\indices{^e_{\un c}_{\un d}}
=0$ and that $(\nabla_p\nabla_q - \nabla_q \nabla_p)V^m = R\indices{^m_n_p_q}V^n$, 
produces the Gauss equation,
\beq
\mathcal{R}_{abcd} = h\indices{^m_a} h\indices{^n_b} h\indices{^p_c} h\indices{^q_d}R_{mnpq}
+ K\indices{_e_c_a}K\indices{^e_d_b} - K\indices{_e_d_a}K\indices{^e_c_b}.
\eeq

The tangential outer curvature $\out\indices{^a_b_c_d}$ is defined by $[\Dt_c,\Dt_d]$
acting on a normal vector $W^a$ through the equation
\beq \label{eqn:Oabcd}
(\Dt_c \Dt_d-\Dt_d\Dt_c)W^a = \out\indices{^a_b_c_d}W^b.
\eeq
To derive the Ricci-Voss equation, we explicitly compute the action of the two derivatives acting
on $W^a$,
\begin{align}
\Dt_c \Dt_d W^a 
& = %
 h\indices{^p_c}h\indices{^q_d} s\indices{^a_m} \nabla_p(\Dt_q W^m) \nonumber\\
 &= %
h\indices{^p_c}h\indices{^q_d} s\indices{^a_m} \nabla_p(h\indices{^e_q}\nabla_e W^m 
- K\indices{_b_q^m}W^b) \nonumber \\
&=%
-K\indices{^e_c_d} \Eth_e W^a + h\indices{^p_c} h\indices{^q_d} s\indices{^a_m}
\nabla_p\nabla_q W^m + K\indices{^a_c_m} K \indices{_b_d^m} W^b
\label{eqn:DDWdeldelW}
\end{align}
Subtracting the expression with $c$ and $d$ reversed yields the Ricci-Voss equation,
\beq\label{eqn:RV}
\out_{abcd} = s\indices{^m_a}s\indices{^n_b}h\indices{^p_c}h\indices{^q_d}R_{mnpq}
+K\indices{_a_c_e}K\indices{_b_d^e} - K\indices{_a_d_e}K\indices{_b_c^e}.
\eeq

Finally, the Codazzi equation arises by requiring that $[\Dt_a,\Dt_b]V^q$ is tangential
on the $q$ index.  This leads to
\begin{align}
0&=%
s\indices{^d_q}\Dt_a \Dt_b V^q = s\indices{^d_q}h\indices{^m_a}h\indices{^n_b} h\indices{^q_e}
\nabla_m(\Dt_n V^e) \nonumber\\
&=%
s\indices{^d_q}h\indices{^m_a}h\indices{^n_b}\nabla_m(h\indices{^p_n}\nabla_p V^q +
K\indices{^q_n_c} V^c) + K\indices{^d_a_e}\Dt_b V^e \nonumber \\
&=%
-K\indices{^p_a_b} L\ind{_q_p^d} V^q + s\indices{^d_q} h\indices{^m_a} h\indices{^n_b}
\nabla_m\nabla_n V^q + \Dt\ind{_a}K\indices{^d_b_c}V^c+ K\indices{^d_b_e}\Dt_a V^e +
K\indices{^d_a_e}\Dt_b V^e
\label{eqn:sDDV}
\end{align}
Subtracting the expression with $a$ and $b$ reversed and rearranging terms yields 
the Codazzi equation
\beq \label{eqn:Codazzi}
h\ind{^m_a}h\ind{^n_b} h\ind{^p_c}s\ind{^q_d}R_{mnpq} = \Dt\ind{_a}K\ind{_d_b_c} -
\Dt\ind{_b}K\ind{_d_a_c}.
\eeq
Note that the identity obtained from requiring that $[\Dt_a, \Dt_b]W^q$ is normal on $q$
for $W^q$ normal is equivalent to equation (\ref{eqn:Codazzi}).  

\subsection{Normal curvatures} \label{app:normalcurv}
Next we look to define a tensorial quantity to associate with the curvature of the normal
connection $\Eth_a$.  This case is more subtle than the tangential curvature since, as discussed
around equation (\ref{eqn:Ethtors}), the twist tensor $F\indices{^e_a_b}$ imbues the connection
$\Eth_a$ with a property similar to torsion.  This causes the commutator
$[\Eth_c, \Eth_d]W^a$ that to not be tensorial, and an additional term of the form
$F\indices{^e_c_d}\Dt_e W^a$ must be subtracted in order to define a tensor.  
We take this modified commutator as the definition of the curvature of the normal connection,
and in the process derive the analogue of the Gauss equation for this curvature.  Taking $W^a$
to be normal, we start by
computing
\begin{align}
\Eth_c \Eth_d W^a &= %
s\indices{^p_c}s\indices{^q_d} s\indices{^a_m}\nabla_p(\Eth_q W^m) \nonumber\\
&=%
s\indices{^p_c}s\ind{^q_d}s\ind{^a_m}\nabla_p(s\ind{^e_q}\nabla_e W^m + L\ind{^m_q_b} W^b)
\nonumber\\
&=%
-L\ind{^e_c_d}\Dt_e W^a+ s\ind{^p_c}s\ind{^q_d}s\ind{^a_m}\nabla_p\nabla_q W^m 
+ L\ind{_m_c^a} L\ind{^m_d_b}W^b \label{eqn:dbardbarW}
\end{align}
When we subtract $\Eth_d \Eth_c W^a$ from this expression, the right hand side still contains 
the term $F\indices{^e_c_d}\Dt_e W^a$, which prevents the commutator from being tensorial.
However, we can simply move this term involving a derivative of $W^a$ to the left hand side, and 
take the definition of the normal curvature $\cat\ind{^a_b_c_d}$ to be\footnote{The letter
$\cat$ is employed for this curvature tensor in honor of Cattaneo-Gasparini, who
appears to have been the first to consider this quantity in the special case 
of a foliation by one-dimensional curves \cite{C-G1961, C-G1963}. } 
\beq \label{eqn:Ccurv}
(\Eth_c \Eth_d - \Eth_d \Eth_c - F\ind{^e_c_d}\Dt_e)W^a = \cat\ind{^a_b_c_d}W^b 
\eeq
This equation for the curvature further justifies the interpretation of $-F\ind{^e_c_d}$ as a type 
of torsion for the connection $\Eth_a$.  For comparison, the definition of the curvature 
$\tilde{R}\ind{^a_b_c_d}$ for an affine connection $\tilde\nabla_a$ with torsion 
$\tilde{T}\ind{^e_a_b}$
is 
\beq
(\tilde\nabla_c\tilde\nabla_d - \tilde\nabla_d \tilde\nabla_c + T\ind{^e_c_d}\tilde\nabla_e)V^a
 = \tilde R\ind{^a_b_c_d}V^b.
\eeq
Combining equation (\ref{eqn:dbardbarW}) with the definition (\ref{eqn:Ccurv}) 
leads to the Gauss equation for the normal curvature,
\beq \label{eqn:Cat}
\cat_{abcd} = s\ind{^m_a}s\ind{^n_b}s\ind{^p_c}s\ind{^q_d}R_{mnpq} + L\ind{_e_c_a}L\ind{^e_d_b}
- L\ind{_e_d_a}L\ind{^e_c_b}
\eeq
This expression shows that the normal curvature enjoys the symmetries $\cat_{abcd} = 
-\cat_{bacd} = -\cat_{abdc}$; however, it does not satisfy either of the remaining symmetries
associated with a torsionless curvature tensor, since $\cat_{abcd} \neq \cat_{cdab}$ and 
$\cat_{\un{abcd}}\neq 0$. 

We can also define a normal outer curvature tensor $\nout\indices{^a_b_c_d}$ by acting with 
the derivative operator in (\ref{eqn:Ccurv}) on a tangential vector $V^a$,
\beq \label{eqn:Pdef}
(\Eth_c \Eth_d - \Eth_d \Eth_c - F\ind{^e_c_d}\Dt_e)V^a = \nout\ind{^a_b_c_d}V^b 
\eeq
Explicitly expanding out the derivatives, we find
\begin{align}
\Eth_c \Eth_d V^a 
&= %
s\indices{^p_c}s\ind{^q_d}h\ind{^a_m}\nabla_p(\Eth_q V^m) \nonumber \\
&=%
s\ind{^p_c}s\ind{^q_d}h\ind{^a_m}\nabla_p(s\ind{^e_q}\nabla_e V^m - L\ind{_b_q^m}V^b) 
\nonumber \\
&=%
-L\ind{^e_c_d}\Dt_e V^a+ s\ind{^p_c}s\ind{^q_d}h\ind{^a_m}\nabla_p\nabla_q V^m
+L\ind{^a_c_m}L\ind{_b_d^m}V^b
\end{align}
Using this to form the combination of derivatives in (\ref{eqn:Pdef}), we derive the Ricci-Voss
equation for the normal outer curvature,
\begin{align}
\nout_{abcd} &= %
h\ind{^m_a}h\ind{^n_b}s\ind{^p_c}s\ind{^q_d}R_{mnpq} + L\ind{_a_c_e}L\ind{_b_d^e}
- L\ind{_a_d_e}L\ind{_b_c^e}   
\label{eqn:nRV} \\
&= \out_{cdab} -K\ind{_c_a_e}K\ind{_d_b^e} + K\ind{_c_b_e}K\ind{_d_a^e}
+ L\ind{_a_c_e}L\ind{_b_d^e}
- L\ind{_a_d_e}L\ind{_b_c^e}, 
\end{align}
where the second expression was obtained by applying the tangential Ricci-Voss equation,
(\ref{eqn:RV}).  Hence we see that the outer curvature for $\Eth_a$ is expressible in terms 
of the outer curvature for $\Dt_a$ and $K_{abc}$ and $L_{abc}$. 

Finally, we can derive a Codazzi identity for the normal connection by requiring that 
$[\Eth_a, \Eth_b]W^q$ is normal on $q$.  We have
\begin{align}
0&=%
h\ind{^d_q}\Eth_a\Eth_bW^q = h\ind{^d_q}s\ind{^m_a}s\ind{^n_b}s\ind{^q_e}\nabla_m(\Eth_n W^e)
\nonumber \\
&=%
h\ind{^d_q} s\ind{^m_a}s\ind{^n_b}\nabla_m(s\ind{^p_n}\nabla_p W^q+ L\ind{^q_n_c}W^c)
+L\ind{^d_a_e}\Eth_b W^e \nonumber \\
&=%
-L\ind{^p_a_b}K\ind{_q_p^d} W^q + h\ind{^d_q}s\ind{^m_a}s\ind{^n_b}\nabla_m\nabla_n W^q
+ \Eth_aL\ind{^d_b_c}W^c + L\ind{^d_b_c}\Eth_a W^c + L\ind{^d_a_e}\Eth_b W^e.
\end{align}
Subtracting the expression with $a$ and $b$ reversed leads to the normal Codazzi
identity,
\beq\label{eqn:nCodazzi}
s\ind{^m_a}s\ind{^n_b}s\ind{^p_c}h\ind{^q_d}R_{mnpq} = \Eth_a L_{dbc} - \Eth_b L_{dac} 
+ F\ind{^e_a_b}K\ind{_c_e_d}.
\eeq

\subsection{Mixed curvatures} \label{app:mixedcurv}
Additional relations arise from considering  
the mixed commutator of the two derivatives, $[\Eth_c, \Dt_d]V^a$.
Taking $V^a$ tangent, we compute
\begin{align}
\Eth_c \Dt_d V^a &= s\ind{^p_c} h\ind{^q_d}h\ind{^a_m} \nabla_p(\Dt_q V^m) \nonumber \\
&= %
s\ind{^p_c}h\ind{^q_d}h\ind{^a_m}\nabla_p
(h\ind{^e_q}\nabla_eV^m +K\ind{^m_q_b}V^b) \nonumber\\
&=%
L\ind{_d_c^e}\Eth_e V^a + s\ind{^p_c}h\ind{^q_d}h\ind{^a_m}\nabla_p\nabla_q V^m
-L\ind{^a_c_m}K\ind{^m_d_b}V^b.
\end{align}
We also need to separately calculate the expression with the derivatives in the opposite order,
\begin{align}
\Dt_d \Eth_c V^a 
&= %
h\ind{^q_d}s\ind{^p_c}h\ind{^a_m}\nabla_q(\Eth_p V^m) \nonumber \\
&= %
h\ind{^q_d}s\ind{^p_c}h\ind{^a_m}\nabla_q(s\ind{^e_p}\nabla_e V^m - L\ind{_b_p^m}V^b) 
\nonumber \\
&=%
K\ind{_c_d^e}\Dt_e V^a + h\ind{^q_d}s\ind{^p_c}h\ind{^a_m}\nabla_q\nabla_p V^m 
-K\ind{_m_d^a}L\ind{_b_c^m}V^b.
\end{align}
By subtracting these two expressions, we can form a tensor by considering the following
combination of derivatives of $V^a$,
\beq \label{eqn:mixdef}
(\Eth_c \Dt_d - \Dt_d \Eth_c + K\ind{_c_d^e}\Dt_e - L\ind{_d_c^e}\Eth_e)V^a = 
\mix\ind{^a_b_c_d}V^b.
\eeq
The tangential mixed curvature tensor $\mix\ind{_a_b_c_d}$ then satisfies the identity
\begin{align}
\mix\ind{_a_b_c_d} &= h\ind{^m_a}h\ind{^n_b} s\ind{^p_c}h\ind{^q_d}R_{mnpq} 
+L\ind{_b_c_e}K\ind{^e_d_a} - L\ind{_a_c_e}K\ind{^e_d_b}  \\
&= \label{eqn:Mabcd}%
\Dt_b K_{cad} -\Dt_a K_{cbd} +L\ind{_b_c_e}K\ind{^e_d_a} - L\ind{_a_c_e}K\ind{^e_d_b},
\end{align}
where the second line is obtained by employing the tangential Codazzi identity (\ref{eqn:Codazzi}).
Note that the mixed curvature possesses the following symmetries: $\mix_{abcd} = -\mix_{bacd}$,
and $\mix_{\un{ab} c \un{d}} = 0$.

We can also form a normal mixed curvature tensor $\nix\ind{^a_b_c_d}$ through a definition
similar to (\ref{eqn:mixdef}), except acting on a normal vector $W^a$,
\beq
(\Eth_c \Dt_d - \Dt_d \Eth_c + K\ind{_c_d^e}\Dt_e - L\ind{_d_c^e}\Eth_e)W^a = 
\nix\ind{^a_b_c_d}W^b.
\eeq
Again expanding out the derivatives that appear in this expression, we find
\begin{align}
\Eth_c \Dt_d W^a &= s\ind{^p_c} h\ind{^q_d}s\ind{^a_m} \nabla_p(\Dt_q W^m) \nonumber \\
&= %
s\ind{^p_c}h\ind{^q_d}s\ind{^a_m}\nabla_p
(h\ind{^e_q}\nabla_eW^m -K\ind{_b_q^m}W^b) \nonumber\\
&=%
L\ind{_d_c^e}\Eth_e W^a + s\ind{^p_c}h\ind{^q_d}s\ind{^a_m}\nabla_p\nabla_q W^m
-L\ind{_m_c^a}K\ind{_b_d^m}W^b. \\
\Dt_d \Eth_c W^a 
&= %
h\ind{^q_d}s\ind{^p_c}s\ind{^a_m}\nabla_q(\Eth_p W^m) \nonumber \\
&= %
h\ind{^q_d}s\ind{^p_c}s\ind{^a_m}\nabla_q(s\ind{^e_p}\nabla_e W^m + L\ind{^m_p_b}W^b) 
\nonumber \\
&=%
K\ind{_c_d^e}\Dt_e W^a + h\ind{^q_d}s\ind{^p_c}s\ind{^a_m}\nabla_q\nabla_p W^m 
-K\ind{^a_d_m}L\ind{^m_c_b}W^b.
\end{align}
These expressions then lead to the identity satisfied by the normal mixed curvature,
\begin{align}
\nix_{abcd} &=%
s\ind{^m_a}s\ind{^n_b}s\ind{^p_c}h\ind{^q_d}R_{mnpq} +K\ind{_a_d_e}L\ind{^e_c_b}
-K\ind{_b_d_e}L\ind{^e_c_a} \\
&= \label{eqn:Nabcd}%
\Eth_a L_{dbc} - \Eth_b L_{dac} + F\ind{^e_a_b}K\ind{_c_e_d} +K\ind{_a_d_e}L\ind{^e_c_b}
-K\ind{_b_d_e}L\ind{^e_c_a},
\end{align}
where the normal Codazzi identity (\ref{eqn:nCodazzi}) was used to obtain the second line. 
The only symmetry that the normal mixed curvature possesses is $\nix_{abcd} = -\nix_{bacd}$,
so in general $\nix_{\un{abc}d}\neq0$. 

Finally, we can examine the mixed Codazzi identity, which will lead to a number of useful relations
between various curvature quantities.  These arise from requiring $[\Eth_a,\Dt_b]V^q$
is tangential on $q$.  Computing in a similar manner as before, we find
\begin{align}
0&=%
s\ind{^d_q}\Eth_a\Dt_b V^q = s\ind{^d_q}s\ind{^m_a}h\ind{^n_b}h\ind{^q_e}\nabla_m(\Dt_n V^e)
\nonumber\\
&=%
s\ind{^d_q} s\ind{^m_a}h\ind{^n_b}\nabla_m(h\ind{^p_n}\nabla_p V^q + K\ind{^q_n_c}V^c)
-L\ind{_e_a^d}\Dt_b V^e \nonumber \\
&=%
L\ind{_b_a^p}L\ind{_q_p^d}V^q + s\ind{^d_q}s\ind{^m_a}h\ind{^n_b}\nabla_m\nabla_n V^q
+ \Eth_aK\ind{^d_b_c}V^c + K\ind{^d_b_c}\Eth_a V^c - L\ind{_e_a^d}\Dt_b V^e. \\
0&= %
s\ind{^d_q}\Dt_b\Eth_a V^q = s\ind{^d_q}h\ind{^n_b}s\ind{^m_a} h\ind{^q_e}\nabla_n(\Eth_m V^e)
\nonumber \\
&=%
s\ind{^d_q}h\ind{^n_b}s\ind{^m_a}\nabla_n(s\ind{^p_m}\nabla_p V^q - L\ind{_c_m^q}V^c)
+K\ind{^d_b_e}\Eth_a V^e \nonumber \\
&=%
-K\ind{_a_b^p}K\ind{^d_p_q}V^q + s\ind{^d_q}h\ind{^n_b}s\ind{^m_a}\nabla_n\nabla_mV^q
-\Dt_bL\ind{_c_a^d}V^c - L\ind{_c_a^d}\Dt_b V^c +K\ind{^d_b_e}\Eth_a V^e.
\end{align}
Subtracting these two expressions produces the mixed Codazzi identity,
\beq \label{eqn:mixedCodazzi}
s\ind{^m_a}h\ind{^n_b}s\ind{^q_d} h\ind{^p_c}R_{qpmn}
=-\Dt_b L_{cad} -L\ind{_b_a^e}L\ind{_c_e_d} - \Eth_a K_{dbc}-K\ind{_a_b^e}K\ind{_d_e_c}.
\eeq
Note that this relation appeared previously in \cite{Capovilla1995a}, equation 3.1, where 
it was presented as a generalization of the Raychaudhuri equation for foliations 
with arbitrary dimension for the leaves.  This interpretation takes the submanifolds
to be timelike, and views (\ref{eqn:mixedCodazzi}) as an evolution equation for $L_{cad}$.
In the case of one-dimensional manifolds, the resulting equation
captures the evolution of the expansion, shear, and twist, using the identifications
between these quantities and $L_{abc}$ described in section \ref{sec:codimd-1}.

(\ref{eqn:mixedCodazzi}) can be unpackaged somewhat by symmetrizing or antisymmetrizing on 
various indices.  First, consider the antisymmetrization on $b$ and $c$, which
leads to
\beq
s\ind{^q_d} s\ind{^m_a}h\ind{^p_c} h\ind{^n_b} R_{qmpn}=\Dt_c L_{bad}-\Dt_bL_{cad}
+L\ind{_c_a^e}L\ind{_b_e_d} - L\ind{_b_a^e}L\ind{_c_e_d} +K\ind{_a_c^e}K\ind{_d_e_b}
-K\ind{_a_b^e}K\ind{_d_e_c}.
\eeq
The Riemann tensor component in this equation is the same as the one appearing in the 
Ricci-Voss equation (\ref{eqn:RV}), and so we can substitute in the outer curvature tensor
to this equation.  This leads to an alternative expression for $\out_{abcd}$,
\beq \label{eqn:ODL}
\out_{abcd} = \Dt_c L_{dba} - \Dt_{d}L_{cba} +L\ind{_c_b^e}L\ind{_d_e_a}
 - L\ind{_d_b^e}L\ind{_c_e_a}.
\eeq
Similarly, considering the antisymmetric part of (\ref{eqn:mixedCodazzi}) on $a$ and $d$
and comparing to the normal Ricci-Voss equation for $\nout_{abcd}$ (\ref{eqn:nRV}) gives
an alternative expression for this tensor,
\beq \label{eqn:PEthK}
\nout_{abcd} = \Eth_cK_{dab}-\Eth_dK_{cab}+K\ind{_c_b^e}K\ind{_d_e_a} 
- K\ind{_d_b^e}K\ind{_c_e_a} +\Dt_bF_{adc}+L\ind{_b_d^e}F\ind{_a_e_c}
- L\ind{_b_c^e}F\ind{_a_e_d}.
\eeq

The symmetric part of (\ref{eqn:mixedCodazzi}) gives a relation on the remaining components
of the spacetime Riemann tensor that have not appeared in any of the previous identities.  
Two identities are thus obtained by symmetrizing on either $b$ and $c$, or on $a$ and $d$, and 
these read
\begin{align}
s\ind{^m_a}h\ind{^n_b }  h\ind{^p_c }  s\ind{^q_d} R_{q\overline{p}\overline{n} m } &= 
\Eth_a K_{dbc} + K\ind{_a_{\overline{b}}^e}K\ind{_d_e_{\overline{c}}}
+\Dt_{\overline{b}} L_{\overline{c} ad} + L\ind{_{\overline{b}}_a^e}L\ind{_{\overline{c}}_e_d} 
\label{eqn:symbc}\\
s\ind{^m_a}h\ind{^n_b }  h\ind{^p_c }  s\ind{^q_d} R_{\overline{q} pn \overline{m} } &= 
\Dt_b A_{cad} + L\ind{_b_{\overline{a}}^e}L\ind{_c_e_{\overline{d}} }
+\Eth_{\overline{a}} K_{\overline{d} bc} + K\ind{_{\overline{a}}_b^e}K\ind{_{\overline{d}}_e_c}.
\end{align}
After symmetrizing on one pair of indices as in the above expressions, the Riemann tensors 
appearing are automatically symmetric in their second pair of indices, since
$R_{q\overline{pn}m} - R_{m\overline{pn}q} = R_{\overline{np}qm} = 0$. 
Hence, antisymmetrizing the above expressions on the remaining index pair will lead to 
differential identities involving only the $K_{abc}$ and $L_{abc}$ tensors.  These are
\begin{align}
\Eth_aK_{dbc} - \Eth_d K_{abc} &= \Dt_{\overline{b}} F_{\overline{c} ad} 
+ L\indices{_{\overline{b}}_d^e}L\indices{_{\overline{c}} _e_a}
- L\indices{_{\overline{b}}_a^e}L\indices{_{\overline{c}}_e_d} 
\label{eqn:EthK}\\
\Dt_b A_{cad} -\Dt_c A_{bad} &= L\ind{_c_{\overline{a}}^e} L\ind{_b_e_{\overline{d}}}
-L\ind{_b_{\overline{a}}^e} L\ind{_c_e_{\overline{d}}}.
\label{eqn:DA}
\end{align}
which simplify using the modified connection $\tilde\Dt_a$ from (\ref{eqn:Dtilde}) to 
\begin{align}
\Eth_aK_{dbc} - \Eth_d K_{abc} &= \tilde\Dt_{\overline{b}} F_{\overline{c} ad} \\
\tilde \Dt_b A_{cad} -\tilde\Dt_c A_{bad} &= 0.
\end{align}
It is also useful to consider the traces of the above identities, since, for example,
the trace of (\ref{eqn:symbc}) on $b$ and $c$ contains an expression that appears 
in the Jacobi equation for perturbations of extremal surfaces.  Hence, the traced identities
read
\begin{align}
s\ind{^m_a}h\ind{^b^c} s\ind{^q_d}R_{qcbm} &= 
\Eth_a K_d+ K\ind{_a^b^c}K\ind{_d_c_b}+ \Dt_b L\ind{^b_a_d} + L\ind{^b_a^c}L\ind{_b_c_d} 
\label{eqn:shsR}
\\
s\ind{^a^d}h\ind{^n_b}h\ind{^p_c}R_{apnd} &= 
\Dt_b A_c + L\ind{_b^a^e}L\ind{_c_e_a} + \Eth_a K\ind{^a_b_c} + K\ind{^a^e_b}K\ind{_a_e_c}
\\
\Eth_a K_d - \Eth_d K_a &= \Dt_b F\ind{^b_a_d}
+L\ind{^b_d^e}L\ind{_b_e_a} - L\ind{^b_a^e}L\ind{_b_e_d} \\
\Dt_b A_c - \Dt_c A_b &= 0.
\end{align}
Finally, equations (\ref{eqn:EthK}) and (\ref{eqn:DA}) can be used in the expressions 
(\ref{eqn:ODL}) and (\ref{eqn:PEthK}) for the outer curvature tensors to further 
simplify their expressions and make manifest some of the index symmetries.  
The resulting equations are 
\begin{align}
\label{eqn:Ocoordexpr}
\out_{abcd} &= 
\Dt_{\un c} F_{\un{d}ab}+\frac12 F\ind{_{\un c}_b^e}F\ind{_{\un d} _e_a} 
+ 2 A\ind{_{\un c}_b^e} A\ind{_{\un d}_e_a}
\\
\label{eqn:Pcoordexpr}
\nout_{abcd} &= \Dt_{\un a} F_{\un{b} cd} + L\ind{_{\un a} _c^e} F\ind{_{\un b} _e_d}
-L\ind{_{\un a} _d^e} F\ind{_{\un b} _e_c} +2 K\ind{_c_{\un b}^e}K\ind{_d_e_{\un a}} 
\\
&= \tilde\Dt_{\un a} F_{\un{b} cd} + 2 K\ind{_c_{\un b}^e} K\ind{_d_e_{\un a}}
\end{align}

\subsection{Modified outer curvatures} \label{app:modifiedcurv}

The modified connections $\tilde\Dt_a$ and $\tilde\Eth_a$ 
defined in (\ref{eqn:Dtilde}) and (\ref{eqn:Ethtilde}) also have curvature identities 
associated with them.  We can define the outer curvatures associated with these connections
in analogy with equations (\ref{eqn:Oabcd}) and (\ref{eqn:Pdef}),
\begin{align}
(\tilde \Dt_c \tilde \Dt_d - \tilde \Dt_d \tilde \Dt_c)W^a &= \tilde{\out}\ind{^a_b_c_d} W^b\\
(\tilde \Eth_c \tilde\Eth_d -\tilde\Eth_d\tilde\Eth_c -F\ind{^e_c_d}\Dt_e)V^a 
&= \tilde{\nout}\ind{^a_b_c_d}V^b. \label{eqn:tildeP}
\end{align}
These curvatures can be related to $\out\ind{^a_b_c_d}$ and $\nout\ind{^a_b_c_d}$
straightforwardly.  Beginning with $\tilde\Dt_a$, we have 
\begin{align}
\tilde\Dt_c\tilde \Dt_d W^a &= 
\tilde\Dt_c(\Dt_d W^a - L\ind{_d_b^a}W^b) \\
&= \Dt_c \Dt_d W^a -L\ind{_c_b^a}\Dt_dW^b -\Dt_c L\ind{_d_b^a}W^b-L\ind{_d_b^a}\Dt_c W^b
+ L\ind{_c_e^a}L\ind{_d_b^e}W^b.
\end{align}
Antisymmetrization on $c$ and $d$ then leads to 
\beq
\tilde\out\ind{^a_b_c_d} = \out\ind{^a_b_c_d} -2\left(\Dt\ind{_{\un c}}L\ind{_{\un d}_b^a}
+L\ind{_{\un d}_e^a}L\ind{_{\un c}_b^e} 
\right) = 0
\eeq
by equation (\ref{eqn:ODL}).  Hence, $\tilde{\Dt}_a$ is a flat connection on the normal bundle. 

The computation for $\tilde\Eth_a$ proceeds similarly,
\begin{align}
\tilde\Eth_c\tilde\Eth_d V^a &= \tilde \Eth_c(\Eth_d -K\ind{_d_b^a}V^b) \\
&= \Eth_c \Eth_d V^a -K\ind{_c_b^a}\Eth_c V^b - \Eth_c K\ind{_d_b^a} V^b 
-K\ind{_d_b^a}\Eth_c V^b + K\ind{_c_e^a}K\ind{_d_b^e} V^b.
\end{align}
Forming the combination of derivatives appearing in (\ref{eqn:tildeP}) then gives the relation
\beq
\tilde\nout\ind{^a_b_c_d} = \nout\ind{^a_b_c_d} -2\left(\Eth\ind{_{\un c}} K\ind{_{\un d}_b^a}
+K\ind{_{\un d}_e^a}K\ind{_{\un c}_b^e}\right) = \tilde\Dt\ind{_b} F\ind{^a_d_c},
\eeq
using (\ref{eqn:PEthK}).

\section{Coordinate expressions} \label{app:coordexpr}

It is often useful when performing computations 
to have coordinate expressions for the various curvature quantities 
defined for a foliation.  In this appendix we will derive the relevant quantities for 
a foliation-adapted coordinate system.  This adapted coordinate system  
splits the spacetime coordinates $y^\mu$ into $p$ normal coordinates $y^A$, $A = 0, \ldots, p-1$,
and $(d-p)$ tangential coordinates $y^i$, $i = p,\ldots, d-1$, where the normal coordinates 
are required to be constant on each surface of the foliation, i.e.\ $\nabla_a y^A$ are normal
forms.  This means the coordinate basis vectors $\partial_i^a$ are tangential, and so $y^i$ 
define an intrinsic coordinate system on surface. The remaining coordinate basis 
vectors $\partial_A^a$ are transverse to the surface, but in general are not normal, since
determining if a vector is normal to the surface requires a metric.  
Once a metric has been specified, we can determine the tangential piece of $\partial_A^a$ and 
subtract it off to form a normal vector $v_A^a = \partial_A^a + N_A^i\partial_i^a$.  This 
equation determines the shift vectors   $N_A^a = N_A^i\partial_i^a$ as simply the tangential 
vector that must be added to $\partial_A^a$ to produce a normal vector.   
It is then straightforward to see that although $\nabla_a y^i$ are not tangential forms to the 
surface, the normal piece is also determined by the shift, so instead 
the covectors $w^i_a = \nabla_a y^i - N^i_A \nabla_a y^A$ are tangential.  
This leads us to define a basis $v_\mu^a$ for the tangent space,
\beq
v_\mu^a = 
\begin{cases}
\partial_\mu^a + N_\mu^i\partial_i^a& \text{for } \mu=0,\ldots, p-1 \\
\partial_\mu^a  & \text{for }\mu = p,\ldots, d-1
\end{cases}
\eeq
so that $v_A^a$ are a basis for the normal subspace while $v_i^a$ are a basis for the 
tangential subspace.  Similarly, a basis $w^\mu_a$ for the cotangent space is given by
\beq \label{eqn:wbasis}
w^\mu_a = 
\begin{cases}
\nabla_a y^\mu& \text{for } \mu=0,\ldots, p-1 \\
\nabla_a y^\mu - N_A^\mu \nabla_a y^A  & \text{for }\mu = p,\ldots, d-1
\end{cases}
\eeq
with $w_a^A$ a basis for the normal subspace and $w^i_a$ a basis for the tangential subspace.  

\subsection{Metrics}
The coordinate expressions for the normal and tangential
metrics can then be expressed in terms of these bases,
\begin{align}
\label{eqn:sab} 
s_{ab} &= s_{AB} w_a^A w_b^B = s_{AB} \nabla_a y^A \nabla_b y^B\\
\label{eqn:hab} 
h_{ab} &= h_{ij}w_a^i w_b^j = h_{ij}(\nabla_a y^i - N_A^i \nabla_a y^A)(\nabla_b y^j - N_B^j 
\nabla_b y^B),
\end{align}
where $s_{AB}$ and $h_{ij}$ are  the 
components of the normal and tangential metric.
Since the spacetime metric is given by the sum $g_{ab} = s_{ab} + h_{ab}$, (\ref{eqn:sab}) and 
(\ref{eqn:hab}) lead to the following expressions for the line element,
\begin{align}
ds^2 &= s_{AB} dy^A dy^B + h_{ij} (dy^i - N_A^i dy^A)(dy^j - N_B^j dy^B) \\
&= (s_{AB} + h_{ij}N_A^i N_B^j)dy^A dy^B - 2 h_{ij}N^j_A dy^A dy^i + h_{ij}dy^i dy^j 
\label{eqn:lineelt}
\end{align}
The mixed-index projectors can be constructed directly from the basis vectors and covectors,
and depend only on the shift vectors,
\begin{align}
\label{eqn:smixed}
s\indices{^a_b} &= v^a_A w^A_b = (\partial_A^a + N^i_A \partial_i^a)\nabla_b y^A \\
\label{eqn:hmixed}
h\indices{^a_b} &= v^a_i w^i_b = (\nabla_b y^i - N^i_A \nabla_b y^A) \partial_i^a.
\end{align}
The inverse metrics are constructed in a similar manner,
\begin{align}
s^{ab} &= s^{AB} v_A^a v_B^b = 
s^{AB}(\partial_A^a + N^i_A \partial_i^a)(\partial_B^b+N_B^j\partial_j^b) \\
h^{ab} &= h^{ij} v_i^a v_j^b = h^{ij}\partial_i^a \partial_j^b.
\end{align}
Here, $s^{AB}$ must the the matrix inverse of $s_{BC}$, which follows from the 
requirement  that $s^{ab}s_{bc} = s\indices{^a_c}$, using the expressions 
(\ref{eqn:sab}) and (\ref{eqn:smixed}).  Unsurprisingly, $h^{ij}$ must the the matrix
inverse of $h_{jk}$ by an analogous argument.  
The decomposition of the line element in (\ref{eqn:lineelt}) shows that for a fixed
foliation, the metric variation can be parameterized by variations of $s_{AB}$, $N_B^i$ and 
$h^{ij}$.  These match onto the respective terms in the covariant decomposition of 
$\delta \p g_{ab}$ given in (\ref{eqn:delgdecomp}).

The expression for the unit normal form is\footnote{The minus sign is conventional when
the normal is timelike, and reflects the fact that $-\nabla^a y^0$ is a future
pointing timelike vector, when $y^0$ is a timelike coordinate.  It can be omitted for 
timelike foliations with a spacelike normal.}
\beq \label{eqn:unitnorm}
n_{a_1\ldots a_p} = -\frac{\sqrt{\sg s} }{p!} 
\, \vep_{A_1 \ldots A_p} w^{A_1}_{a_1} \ldots w^{A_p}_{a_p}
=-\sqrt{\sg s} (dy^0 \wedge \ldots \wedge dy^{p-1})_{a_1 \ldots a_p},
\eeq
where $s = \det(s_{AB}) $,
and $\vep_{A_1\ldots A_p}$ is the antisymmetric symbol with $\vep_{01\ldots(p-1)} = 1$.
The prefactor $\sqrt{\sg s}$ can be derived by applying the normalization condition
(\ref{eqn:normnorm}) to $n_{a_1\ldots a_p}$. The induced volume form $\mu_{a_1 \ldots a_{d-p}}$
on the surfaces has a similar expression,
\begin{align}
\mu_{a_1 \ldots a_{d-p}} &= \frac{\sqrt{-\sg h} }{(d-p)!} \, \vep_{i_1\ldots i_{d-p}} w^{i_1}_{a_1} \ldots 
w^{i_{d-p}}_{a_{d-p}} \\
\label{eqn:indvol}
&=\sqrt{-\sg h}\left[ (dy^p -N_A^p dy^A)\wedge\ldots\wedge
(dy^{d-1}-N_B^{d-1}dy^B)\right]_{a_1\ldots a_{d-p}}
\end{align}
where again $h = \det(h_{ij})$.  Finally, since the spacetime volume form $\ep$ is related to
the unit normal form and induced volume form by $\ep = -n\wedge \mu$, and since it also
has the coordinate expression $\ep = \sqrt{-g}\, dy^0\wedge\ldots\wedge dy^{d-1}$ with
$g = \det g_{\mu\nu}$, 
from (\ref{eqn:unitnorm}) and (\ref{eqn:indvol}) we find the relation
\beq
\sqrt{-g} = \sqrt{\sg s} \,\sqrt{-\sg h}.
\eeq

\subsection{Connection coefficients} \label{app:conncoefs}
The next objects to consider are the connection coefficients for the tangential and normal 
covariant derivatives, $\Dt_a$ and $\Eth_a$.  Beginning with $\Dt_a$ acting on a tangent
vector $V^a = V^j\partial_j^a$, the connection coefficients $\gamma^a_{bc}$ are defined by the equation
\beq\label{eqn:DV}
\Dt_b V^a =h\indices{^a_d} h\indices{^c_b}\partial_c V^d + \gamma^a_{bc}V^c
= (\partial_i V^j + \gamma^i_{jk}V^k)v_j^a w_b^i
\eeq
where $\partial_c$ is the coordinate derivative operator associated with the 
surface-adapted coordinate system $y^\mu$.  Its action extends to covectors and multi-index
vectors in the usual way.  From the relation that $\Dt_a h_{bc}=0$, we conclude that $\Dt_a$
is the unique
metric-compatible, torsion-free connection on the surface, and hence the connection
coefficients have the usual expression in terms of derivatives of the intrinsic metric components,
\beq
\gamma^i_{jk} = \frac12 h^{il}(\partial_j h_{lk} + \partial_k h_{lj} -  \partial_l h_{jk}).
\eeq

For $\Dt_b$ acting on a normal vector $W^a = W^Av_A^a$, the connection coefficients can be
obtained by writing
\beq
\Dt_b W^a = h\ind{^d_b}s\ind{^a_c}(\partial_d W^c + \Gamma^c_{de}W^e)
= h\ind{^d_b}s\ind{^a_c}\partial_dW^c +L\ind{_d_e^c}W^e
=(\partial_i W^A +L\ind{_i_B^A}W^B)v^a_A w^i_b,
\eeq
where the second equality employs the relation (\ref{eqn:GammaL}).  This shows that 
the tensor
$L\ind{_d_e^c}$ gives the connection coefficients for $\Dt_b$ acting on the normal bundle.

The normal connection coefficients $\cn^a_{bc}$
can be defined by the action on a normal covector $W_c
= W_A\nabla_c y^A$, 
\beq\label{eqn:gimeldef}
\Eth_b W_c = s\ind{^d_b} s\ind{^a_c}  \partial_d W_a - \cn^a_{bc}W_a
= (\partial_B W_A+ N_B^i\partial_iW_A - \cn^A_{BC} W_A)w_a^B w_b^C
\eeq
which involves both transverse $\partial_A$ and tangential $\partial_i$ derivatives, due 
to the form of the normal projector from (\ref{eqn:smixed}).  It is convenient to define a modified 
transverse derivative 
\beq
\partial_{*A} = \partial_A + N^i_A \partial_i,
\eeq 
which will appear often 
in the following formulas.  The coordinate expression for $\cn^A_{BC}$ again follows from 
the fact that $\Eth_a s_{bc}=0$. However, since the derivatives appearing in this expression
are $\partial_{*A}$ rather than $\partial_A$, the formula for the connection coefficients will 
be the usual expression except with derivatives of $s_{AB}$ with respect to $\partial_{*C}$ 
instead of $\partial_C$,
\beq
\cn^{A}_{BC} = \frac12 s^{AD}(\partial_{*B}s_{DC} + \partial_{*C}s_{DB} - \partial_{*D} s_{BC}).
\eeq

The connection coefficients for $\Eth_b$ acting on a tangent vector $V^a = V^iv_i^a$ 
can be obtained applying (\ref{eqn:Kconnect}),
\beq
\Eth_bV^a = s\ind{^d_b} h\ind{^a_c}(\partial_d V^c + \Gamma^c_{de})V^e = \left(\partial_{*A}
V^i +(K\ind{_A_j^i}-\partial_j N_A^i) V^j \right) v_i^a w^A_b,
\eeq
so that $K\ind{_A_j^i}-\partial_j N_A^i$ are the connection coefficients.

\subsection{Extrinsic curvatures} \label{app:extrinsic}
The extrinsic curvatures are the next items to consider.  To derive the coordinate expression
for the tangential extrinsic curvature, 
it is useful to start with the interpretation of $K\ind{^a_b_c}$ as measuring the change 
in the induced metric on the surface under a flow in the normal direction.  Specifically,
the projected Lie derivative of $h_{ab}$ along a normal vector $W^c$ satisfies
\beq \label{eqn:liehab}
\frac12 h\ind{^c_a}h\ind{^d_b} \lie_{W}h_{cd} = 
\frac12 h\ind{^c_a}h\ind{^d_b} (W^e\nabla_e h_{ab} + \nabla_a W^e h_{eb} + \nabla_b W^e
h_{ae}) = W^e K_{eab}.
\eeq
This formula can then be converted straightforwardly to a coordinate expression by writing
\begin{align} \label{eqn:KAij}
K\ind{^A_i_j} = w_d^A s^{da}K_{abc}v^b_i v^c_j 
= s^{AB} v_B^a K_{abc} v_i^b v_j^c
=\frac12 s^{AB} (\lie_{\partial_B + N_B^k\partial_k}
h_{bc})v_i^b v_j^c.
\end{align}
The first term $\lie_{\partial_A} h_{bc}$ will just produce the partial derivative of the components
$h_{ij}$ of the induced metric.  For the second term, note that $N_B^k\partial_k^e$ is a tangent
vector, and hence we can evaluate this Lie derivative using the metric-compatible 
derivative $\Dt_a$.  This results in 
\beq \label{eqn:Kdh}
K\ind{^A_i_j} =\frac12 s^{AB}(\partial_B h_{ij} + \tilde{\Dt}_i N_{jB} + \tilde{\Dt}_j N_{iB}),
\eeq 
where the modified normal connection of (\ref{eqn:Dtilde}) 
is used, given by
$\tilde{\Dt}_i N_{jB}  = \partial_i N_{jB}-\gamma^k_{ij}N_{kB}$, and $N_{jB} = h_{jl}N^l_B$. 
In particular, the 
tilde means to treat $N_{j B}$ as a collection of covectors indexed by $B$, as opposed to 
a two-index tensor with one tangential index $j$ and one normal index $B$.  

A slight modification of this argument leads to the coordinate expression for $A\ind{^i_{AB}}$.
It is straightforward to see that this object measures the change in the normal metric
$s_{ab}$ when flowing in a tangential direction, since it satisfies the analog of equation 
(\ref{eqn:liehab}) for a tangent vector $V^c$,
\beq
\frac12 s\ind{^c_a} s\ind{^d_b} \lie_V s_{cd} = V^e A_{eab}.
\eeq
Performing the analogous calculation to equation (\ref{eqn:KAij}) leads to
\beq\label{eqn:AiAB}
A\ind{^i_A_B}  = \frac12 q^{ij}\partial_j s_{AB}
\eeq
which involves no shift vectors since the tangential basis vectors $v^a_i$ are defined independent
of $N_A^i$. 

The coordinate expression for the twist tensor $F\ind{^a_b_c}$ can be obtained using the 
expression (\ref{eqn:smixed}) for the normal projector and the direct definition of 
the twist (\ref{eqn:F}).  This leads to 
\begin{align}
F\ind{^i_A_B} &= w_a^i v_A^b v_B^c F\ind{^a_b_c} 
= 2 w_a^i v_A^b v_B^c\nabla\ind{_{\un b}}s\ind{^a_{\un c}}
=2 w_a^i v_A^b v_B^c\left[(\partial_{\un b} N_C^j) \partial_j^a \nabla_{\un c}y^C
+ \Gamma^a_{{\un b} d} s\ind{^d_{\un c}} -\Gamma^d_{\un{bc}} s\ind{^a_d} \right] \nonumber \\
&= \partial\ind{_{*A}}N_B^i - \partial\ind{_{*B}}N_A^i. \label{eqn:FiAB}
\end{align}
Note this is consistent with with equation (3.62) of reference \cite{Donnelly2016F}
for the twist tensor.  

\subsection{Intrinsic curvatures}

The coordinate expressions for the curvature tensors are derived directly from their
defining formulas in terms of commutators of derivative operators, which were presented in
appendix \ref{app:curvature}.  Equation (\ref{eqn:DDV}) defines the intrinsic curvature 
$\mathcal{R}\ind{^a_b_c_d}$ of the 
tangential connection $\Dt_a$.  To compute its coordinate expression, we apply 
the formula (\ref{eqn:DV}) for the coordinate expression of $\Dt_b V^c$, along with
its generalization to covectors and multi-index tensors.  This leads to
\begin{align}
v_k^a v_l^b\Dt_a \Dt_b V^c &= %
v_k^a v_l^b \Dt_a\left[  (\partial_m V^i + \gamma^i_{mj}V^j)v_i^c w_b^m \right] \nonumber \\
&= %
\left[ \partial_k \partial_l V^i + \partial\ind{_k} \gamma^i_{lj} V^j+ \gamma^i_{lj}\partial_k V^j
+\gamma^i_{km}(\partial_l V^m +\gamma^m_{lj}V^j) - \gamma^m_{kl}(\partial_m V^i
+\gamma^i_{mj}V^j) \right]v_i^c 
\end{align}
Subtracting the expression with the derivative order reversed retains only the antisymmetric
piece on $k$ and $l$, and this leads to the expected formula for the intrinsic curvature 
components,
\beq
\mathcal{R}\ind{^i_j_k_l} = \partial\ind{_k} \gamma^i_{lj} - \partial\ind{_l}\gamma^i_{kj}
+\gamma^i_{km}\gamma^m_{lj} - \gamma^i_{lm}\gamma^m_{kj}.
\eeq

For the intrinsic normal curvature $\cat\ind{^a_b_c_d}$, we utilize the definition (\ref{eqn:Ccurv})
along with the coordinate expression $\Eth_b W^c = (\partial_{*D}W^A +\cn^A_{DB}W^B)w_b^D 
v^c_A$ and its generalization to covectors and tensors.  This gives
\begin{align}
v_C^a v_D^b \Eth_a\Eth_b W^c  &=%
v_C^a v_D^b \Eth_a\left[  (\partial_{*E}W^A +\cn^A_{EB} W^B) w_b^E v_A^c \right] \nonumber \\
&=%
\big[\partial_{*C}\partial_{*D} W^A + \partial\ind{_{*C}}\cn^A_{DB}W^B
+\cn^A_{DB}\partial_{*C}W^B \nonumber\\
&\qquad
+\cn^A_{CE}(\partial_{*D}W^E + \cn^E_{DB}W^B) 
-\cn_{CD}^E (\partial_{*E}W^A +\cn^A_{EB}W^B) \big] \label{eqn:EthEthW}
\end{align}
This time when subtracting the expression with the derivatives reversed, the term 
involving two derivatives of $W^A$ is nonzero due to the shift
vectors present in the starred derivatives. Instead, we find using (\ref{eqn:FiAB})
\beq
(\partial_{*C}\partial_{*D}-\partial_{*D}\partial_{*C})W^A = F\ind{^i_C_D}\partial_i W^A
=F\ind{^i_C_D} (D_i W^A -L\ind{_i_B^A}W^B).
\eeq
The first term here is exactly the piece that must be subtracted from the $[\Eth_a,\Eth_b]$ 
commutator to form the curvature according to (\ref{eqn:Ccurv}). Adding the second term
to the antisymmetric pieces in (\ref{eqn:EthEthW}) on $C$ and $D$ 
then yields the coordinate expression for the normal curvature,
\beq
\cat\ind{^A_B_C_D} = \partial\ind{_{*C}}\cn^A_{DB}-\partial\ind{_{*D}} \cn ^A_{CB}
+\cn^A_{CE}\cn^E_{DB}-\cn^A_{DE}\cn^E_{CB} - F\ind{^i_C_D}L\ind{_i _B^A},
\eeq
and the components $L\ind{_i_B^A}$ come from (\ref{eqn:AiAB}) and (\ref{eqn:FiAB})
using $L\ind{_m_b^a} = A\ind{_m_b^a}-\frac12 F\ind{_m_b^a}$.  
This coordinate expression for the normal curvature has appeared previously in the special 
case of a one-dimensional foliation in \cite{C-G1961, Massa1974, Boersma1995c}.

For the remaining curvature tensors, the formulas derived in appendix \ref{app:mixedcurv}
lead directly to the coordinate expressions.  For the outer curvature tensors 
$\out_{ABkl}$ and $\nout_{ijCD}$ 
one simply uses (\ref{eqn:Ocoordexpr}) and (\ref{eqn:Pcoordexpr}) along with 
the expressions (\ref{eqn:Kdh}), (\ref{eqn:AiAB}) and (\ref{eqn:FiAB}) for 
$K\ind{^A_i_j}$, $A\ind{^i_A_B}$ and $F\ind{^i_A_B}$.  Similarly, equations 
(\ref{eqn:Mabcd}) and (\ref{eqn:Nabcd}) lead to coordinate expressions for 
the mixed curvatures $\mix_{abcd}$ and $\nix_{abcd}$.

\subsection{Spacetime Christoffel symbols} \label{app:christoffel}

The final set of formulas will relate the spacetime Christoffel symbols for the metric 
defined by (\ref{eqn:lineelt}) to the intrinsic connection coefficients $\gamma^i_{jk}$, $\cn^A_{BC}$,
extrinsic curvatures $K\ind{^A_i_j}$, $L\ind{^i_A_B}$, and shift vectors $N_B^i$.  First, note that 
for the tangential covariant derivative,
\beq
\Dt_a V^b = h\ind{^c_a}h\ind{^b_d}(\partial_c V^d +\Gamma^d_{ce}V^e) 
=  h\ind{^c_a}h\ind{^b_d}(\partial_c V^d +\gamma^d_{ce}V^e),
\eeq 
leading to the conclusion
\beq \label{eqn:gammaGamma}
\gamma^a_{bc} = h\ind{^a_d}h\ind{^e_b}h\ind{^f_c}\Gamma^d_{ef}.
\eeq
A similar argument applies to the normal connection,
\beq
\Eth_a W^b = s\ind{^c_a}s\ind{^b_d}(\partial_c W^d +\Gamma^d_{ce}W^e) 
=  s\ind{^c_a}s\ind{^b_d}(\partial_c W^d +\cn^d_{ce}V^e),
\eeq
which implies
\beq \label{eqn:gimelGamma}
\cn^a_{bc} = s\ind{^a_e}s\ind{^e_b}s\ind{^f_c}\Gamma^d_{ef}.
\eeq
The remaining components of the Christoffel symbols can be arrived at by examining
the covariant derivative of $s\ind{^a_b}$. From its coordinate expression (\ref{eqn:smixed}),
we find
\beq
\nabla_b s\ind{^a_c} = (\nabla_b N_A^i)\partial_i^a \nabla_c y^A + \Gamma^a_{bd}s\ind{^d_c}
-\Gamma^m_{bc}s\ind{^a_m},
\eeq
and projecting onto different components yields the following expressions,
\begin{align}
s\ind{^a_d}h\ind{^e_b}h\ind{^f_c}\Gamma^d_{ef} &= -K\ind{^a_b_c} \\
\label{eqn:Kconnect}
h\ind{^a_d}h\ind{^e_b}s\ind{^f_c} \Gamma^d_{ef} &= K\ind{_c_b^a}-
(h\ind{^e_b}\nabla_eN_A^i)\partial_i^a \nabla_c y^A \\
\label{eqn:GammaL}
s\ind{^a_d}s\ind{^e_b}h\ind{^f_c}\Gamma^d_{ef} &= L\ind{_c_b^a} \\
h\ind{^a_d}s\ind{^e_b}s\ind{^f_c}\Gamma^d_{ef} &= -(s\ind{^e_b}\nabla_e N_A^i)\partial_i^a
\nabla_c y^A-L\ind{^a_b_c}
\end{align}
These four relations along with (\ref{eqn:gammaGamma}) and (\ref{eqn:gimelGamma}) 
lead to coordinate expressions for all components of the Christoffel symbols:
\begin{align}
\Gamma^i_{jk} &= \gamma^i_{jk} - N_A^i K^A_{jk} \\
\Gamma^A_{ij} &= - K\ind{^A_i_j} \\
\label{eqn:GammaiAj}
\Gamma^i_{Aj} &= K\ind{_A_j^i}-\Dt_j N_A^i +N_A^k N_B^i K\ind{^B_k_j} \\
\Gamma^A_{Bi} &= L\ind{_i_B^A} + N_B^j K\ind{^A_j_i} \\
\label{eqn:GammaiAB}
\Gamma^i_{AB} &= -A^i_{AB}- \Eth\ind{_{\overline{A}}}N_{\overline{B}}^i  
-N_{\overline{A}}^j K\ind{_{\overline{B}}_j^i}
+2 N_{\overline{A}}^j \Dt_j N_{\overline{B}}^i
-\gamma^i_{jk}N_A^j N_B^k
-N_A^j N_B^k N_C^i K\ind{^C_j_k}
\\
\Gamma^A_{BC} &= \cn^A_{BC} -N_B^j L\ind{_j_C^A}- N_C^jL\ind{_j_B^A}
-N_B^i N_C^j K\ind{^A_i_j}.
\end{align}
The explicit expressions for the covariant derivatives of the shift vectors appearing in 
(\ref{eqn:GammaiAj}) and (\ref{eqn:GammaiAB}) are
\begin{align}
\Dt_j N^i_A &= \partial_j N^i_A +\gamma^i_{jk}N^k_A - L\ind{_j_A^B}N_B^k \\
\Eth_AN_B^i &= \partial_{*A} N_B^i - \cn^C_{AB}N_C^i +K\ind{_A_j^i}N_B^j 
-(\partial_j N_A^i)N_B^j \\
\Eth_{\overline A}N_{\overline B}^i &= \partial_{\overline A}  N_{\overline B}^i - \cn^C_{AB}N_C^i 
+K\ind{_{\overline A}_j^i}N_{\overline B}^j.
\end{align}

\bibliographystyle{JHEPthesis}
\bibliography{embed}

\end{document}